\documentclass[aps,prd,showpacs,showkeywords,preprintnumbers]{revtex4-1}
\usepackage{amsmath,amssymb,amsthm,bm,bbm,color,hyperref}
\usepackage{graphicx,psfrag,subfigure,rotating}

\begin{document} 

\title{Center Vortices and Chiral Symmetry Breaking in $SU(2)$ Lattice Gauge Theory}

\author{Roman H\"ollwieser}
\email{hroman@kph.tuwien.ac.at}
\affiliation{Institute of Atomic and Subatomic Physics, Vienna University of
Technology,\\ Operngasse 9, 1040 Vienna, Austria}
\affiliation{Department of Physics, New Mexico State University, PO Box 30001, Las Cruces, NM 88003-8001, USA}
\author{Thomas Schweigler}
\author{Manfried Faber}
\email{faber@kph.tuwien.ac.at}
\affiliation{Institute of Atomic and Subatomic Physics, Vienna University of
Technology,\\ Operngasse 9, 1040 Vienna, Austria}
\author{Urs M. Heller}
\affiliation{American Physical Society, One Research Road, Ridge, NY 11961, USA}

\date{\today}

\begin{abstract}
We investigate the chiral properties of near-zero modes for thick classical
center vortices in $SU(2)$ lattice gauge theory as examples of the phenomena
which may arise in a vortex vacuum. In particular we analyze the creation of
near-zero modes from would-be zero modes of various topological charge
contributions from center vortices. We show that classical colorful spherical vortex and instanton ensembles have almost identical Dirac spectra and the low-lying
eigenmodes from spherical vortices show all characteristic properties for chiral
symmetry breaking. We further show that also vortex intersections are able to
give rise to a finite density of near-zero modes, leading to chiral symmetry
breaking via the Banks-Casher formula. We discuss the mechanism by which center
vortex fluxes contribute to chiral symmetry breaking.
\end{abstract}

\pacs{11.15.Ha, 12.38.Gc}

\keywords{Center Vortices, Chiral Symmetry Breaking, Atiyah-Singer Index Theorem, Overlap Operator, Lattice Gauge Field Theory}

\maketitle

\newpage

\section{Introduction}
Quantum chromodynamics (QCD) at low energies is dominated by the
nonperturbative phenomena of quark confinement and spontaneous chiral
symmetry breaking ($\chi$SB). Presently, a rigorous treatment of them is
only possible in the lattice regularization. Many of the important features
of non-Abelian gauge theories are already present in SU(2), which simplifies
theoretical and numerical calculations.

The nonperturbative vacuum can be characterized by various kinds of
topological gauge field excitations. A well established theory of $\chi$SB
relies on
instantons~\cite{Belavin:1975fg,Actor:1979in,'tHooft:1976fv,Bernard:1979qt},
which are localized in space-time and carry a topological charge of modulus
$1$. According to the Atiyah-Singer index
theorem~\cite{Atiyah:1971rm,Schwarz:1977az,Brown:1977bj,Narayanan:1994gw},
a zero mode of the Dirac operator arises, which is concentrated at the
instanton core. In the instanton liquid
model~\cite{Ilgenfritz:1980vj,Diakonov:1984vw,Diakonov:1986tv} overlapping would-be zero
modes split into low-lying nonzero modes which create the chiral
condensate.

Center vortices~\cite{'tHooft:1977hy,Vinciarelli:1978kp,Yoneya:1978dt,Cornwall:1979hz,Mack:1978rq,Nielsen:1979xu},
on the other hand, are promising candidates for explaining confinement. They form closed magnetic flux tubes, whose flux is quantized, taking only
values in the center of the gauge group. These properties are the key
ingredients in the vortex model of confinement, which is theoretically
appealing and was also confirmed by a multitude of numerical calculations, both in lattice Yang-Mills theory and within a corresponding infrared effective
model, see {\it e.g.}~\cite{DelDebbio:1996mh,Langfeld:1997jx,DelDebbio:1997ke,Kovacs:1998xm,Engelhardt:1999wr,Bertle:2002mm,Engelhardt:2003wm}. Lattice simulations indicate
that vortices may be responsible for topological charge and $\chi$SB as
well~\cite{deForcrand:1999ms,Alexandrou:1999vx,Reinhardt:2000ck2,Engelhardt:2002qs,Bornyakov:2007fz,Hollwieser:2008tq},
and thus unify all nonperturbative phenomena in a common framework.  A
similar picture to the instanton liquid model exists insofar as lumps of
topological charge arise at the intersection and writhing points of
vortices.  The colorful, spherical SU(2) vortex as introduced in previous
articles of our group
\cite{Jordan:2007ff,Hollwieser:2010mj,Hollwieser:2012kb,Schweigler:2012ae},
may act as a prototype for this picture, as it contributes to the
topological charge by its color structure, attracting a zero mode like an
instanton. 

In this article we want to show how the interplay of various topological
structures from center vortices (and instantons) leads to near-zero modes, which
by the Banks-Casher relation~\cite{Banks:1979yr} are responsible for a finite chiral condensate.
Using the overlap and asqtad staggered Dirac operator, we compute a varying
number of the lowest-lying Dirac eigenfunctions, including the zero modes.
By visualizing the probability density, we compare the distribution of the
eigenmode density with the position of the vortices and the topological
charge created by intersection points and color structures. These results
manifest the importance of center vortices also for chiral symmetry
breaking.

\section{Topological charge from center vortices}
\label{sec:vortQ}

From the definition 
\begin{gather}
  Q \propto \epsilon_{\mu\nu\alpha\beta}F_{\mu\nu}F_{\alpha\beta} \, ,
\end{gather}
it is clear that topological charge contributions arise where two
perpendicular nontrivial plaquettes meet. On a center vortex, the vortex
sheet is locally orthogonal to the nontrivial plaquettes. Hence,
topological charge emerges at so-called singular points of center vortices,
where the set of tangent vectors to the vortex surface spans all four
space-time directions \cite{Reinhardt:2001kf}. There are two possibilities
where this can occur:
\begin{itemize}
  \item intersection points of two different surface patches: $Q=\pm1/2$ \, ,
  \item writhing points of a single surface patch: $|Q|<1/2$ \, .
\end{itemize}
In lattice language, intersection points have four plaquettes attached to
them extending in two particular space time directions ({\it e.g.} $xy$), and
another four plaquettes extending in the two orthogonal directions ({\it e.g.}
$zt$).  By contrast, at a writhing point one can start at any attached
plaquette and pass around over all others along a continuous path since
they are connected by common links.

In SU(2), a Wilson loop cannot distinguish between center fluxes of
opposite sign, since $\exp(i\pi)=\exp(-i\pi)=-1$. However, the sign of the
topological charge is sensitive to the direction of the flux. To preserve
this information for thin vortices, one can assign an orientation to the
vortex patches. Note that this quantity is not directly linked to the
geometry of the vortex. A non-orientable vortex need not be non-orientable
as a surface~\cite{Bertle:1999tw}.

Vortices always have closed surfaces. Furthermore, lattice studies show
that the major fraction of vortex patches belongs to a single large vortex
winding through the whole lattice. Intersections between different vortices
are therefore only of minor importance. But for a closed oriented surface
the total charge is zero. Charged vortices must therefore be globally
non-oriented and consist of differently oriented patches, which are
separated by monopole worldlines. The colorful spherical vortex is a
classical representation of such a vortex configuration, contributing to
topological charge only through its color structure. It will be discussed
shortly in Sec.~\ref{sec:spher}, but first let us explain the relation
between vortices and magnetic monopoles, thereby making precise the
definition of the orientation of a vortex.

On the lattice, monopoles are located by so-called Abelian projection.
First, one fixes the links up to a residual $U(1)$-symmetry, which
corresponds to an Abelian gauge theory. For example, we can use the maximal
Abelian gauge~\cite{Greensite:2003bk} to rotate the color vector of the
links as much as possible in, say, the $\sigma_3$-direction by maximizing
\begin{gather}
  R = \sum_{x,\mu} \mathrm{tr}[U_\mu(x)\sigma_3 U_\mu^\dagger(x) \sigma_3] \, .
\end{gather}
Afterwards, the link variables are replaced by their diagonal part
\begin{gather}
  U_\mu(x) = a_0 \mathbbm{1} + i\vec{a}\cdot\vec{\sigma} \rightarrow u_\mu(x)
 = \frac{1}{\sqrt{a_0^2+a_3^2}} [a_0 \mathbbm{1}+i a_3 \sigma_3] =
 \begin{pmatrix} \exp\{i\theta_\mu(x)\} & 0 \\ 0 & \exp\{-i\theta_\mu(x)\} \end{pmatrix} \, .
\end{gather}
Consider a 3-dimensional cube on the lattice. Normally, the total magnetic
flux out of the cube vanishes due to $\mathrm{div}\,\vec{B}=0$. If the cube
contains a monopole, there would be a nonzero net flux of the (nonphysical)
Dirac string, which compensates the physical monopole flux. The magnetic
charge inside the cube is defined by~\cite{Chernodub:2001ht}
\begin{gather}
  m = \frac{1}{2\pi} \sum_p \bar{\theta}_p \, .
\end{gather}
$\theta_p$
is the sum of the angles $\theta_\mu$ around one plaquette $p$; $\sum_p
\theta_p$ is always zero as a consequence of the Bianchi identity.
$\bar{\theta}_p$ is $\theta_p + 2\pi k$ with $k$ such that $\bar{\theta}_p$
falls into the range $[-\pi,\pi]$. If the absolute value of all plaquette
angles is smaller than $\pi$, $\bar{\theta}_p = \theta_p$ and $m=0$, as
usual.  Plaquettes which are greater than $\pi$ are pierced by the Dirac
string, at the end of which a monopole is located. $\bar{\theta}_p$
discards the flux caused by the Dirac string and a value $m\neq 0$ results.

Now let us return to center vortices. Traversing a thick vortex sheet, the
link variables change gradually, building up to a center element. Pictured
in group space of $SU(2) \simeq S^3$, we travel along a path from unity to
the antipode $-\mathbbm 1$. After Abelian projection, only one possible
direction for the path remains and this path will go either in $+\sigma_3$
or in $-\sigma_3$ direction. We can use this sign to allocate an
orientation to every patch of the vortex surface. The sign also corresponds
to the direction of the center flux bundled within the vortex, which is
quantized in units $\pm\pi$. Where regions of opposite orientation touch,
the flux jumps by $2\pi$, indicating the presence of a sink or source, {\it
i.e.}, a magnetic monopole carrying a quantized magnetic charge. To
summarize, a center vortex can be imagined as a chain of magnetic
monopoles, whose flux is bundled within the vortex surface. The monopole
worldlines divide the vortex sheet into patches of different orientations.

\subsection{Plane vortices}\label{sec:plane}

\begin{figure}[htb]
\psfrag{x}{$x$}
\psfrag{y}{$y$}
\psfrag{z}{$z$}
\psfrag{0}{\small $0$}
\psfrag{+Q}{\small $+Q$}
\psfrag{-Q}{\small $-Q$}
\begin{tabular}{ccc}
Parallel Vortices & Geometry & Anti-parallel Vortices\\
	a)\includegraphics[keepaspectratio,width=0.33\textwidth]{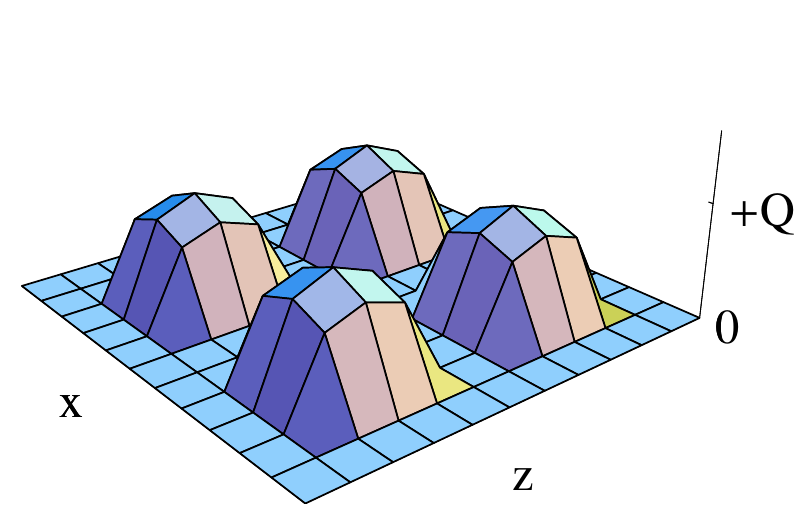} & 
	b)\includegraphics[keepaspectratio,width=0.24\textwidth]{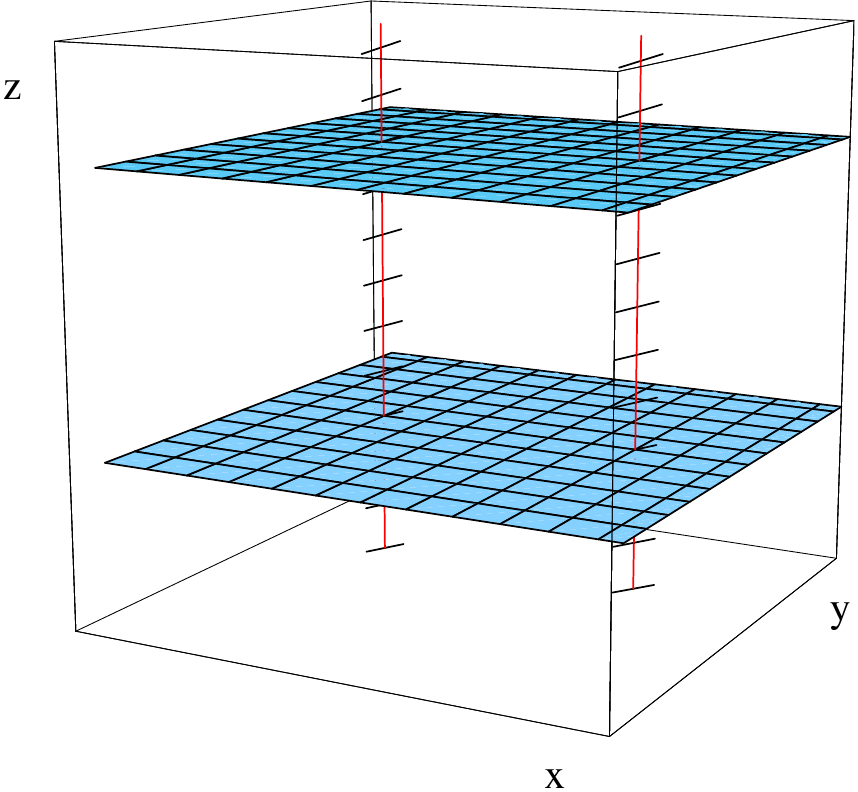} &
	c)\includegraphics[keepaspectratio,width=0.33\textwidth]{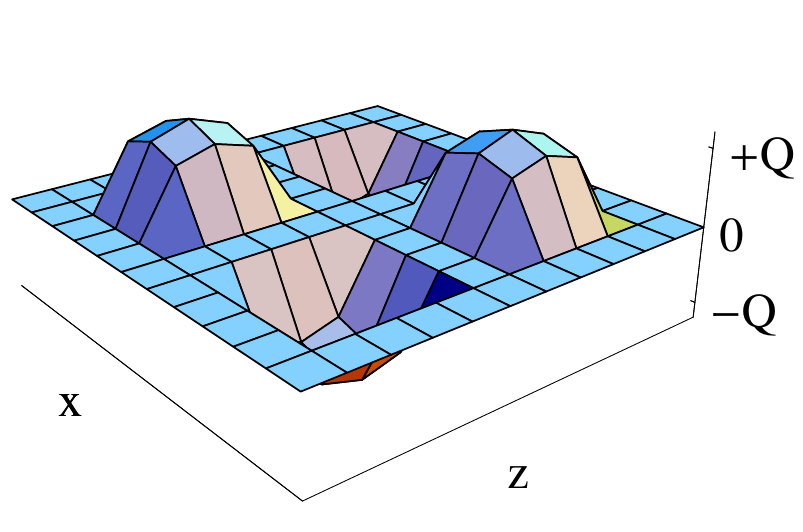}
\end{tabular}
\caption{A single time slice of a $12^4$-lattice with intersecting vortices (center). The horizontal planes are the $xy$-vortices, which exist only at this time. The vertical lines are the $zt$-vortices, which continue over the whole time axis. The vortices intersect in four points, giving topological charge $Q=2$ for parallel vortices (lhs) or $Q=0$ for anti-parallel vortices (rhs).}
\label{fig:vortcross3d}
\end{figure}

We define plane vortices parallel to two of the coordinate axes by links varying 
in a $U(1)$ subgroup of $SU(2)$. This $U(1)$ subgroup is generated by one of the 
Pauli matrices $\sigma_i$, {\it i.e.}, $U_\mu=\exp(\mathrm{i} \phi \sigma_i)$.  
The direction of the flux and the orientation of the vortices are determined by
the gradient of the angle $\phi$, which we choose as a piece wise linear
function of a coordinate perpendicular to the vortex.  The explicit
functions for $\phi$ are given in Eq. (2.1) of~\cite{Hollwieser:2011uj} (see
also Fig.1 therein).
Upon traversing a vortex sheet, the angle $\phi$ increases or decreases by
$\pi$ within a finite thickness of the vortex. Since we use periodic (untwisted) boundary conditions for the links, vortices occur in pairs of parallel sheets, 
each of which is closed by virtue of the lattice periodicity. We call vortex
pairs with the same vortex orientation parallel vortices and vortex pairs of
opposite flux direction anti-parallel. Of course, there are always two
coordinates perpendicular to a vortex surface, and vortices are always thin in
one of these directions. Their cross-sections thus strictly speaking do not
correspond to thickened tubes of magnetic flux, but rather thin strips. If the
thick, planar vortices intersect orthogonally, each intersection carries a
topological charge with modulus $|Q|=1/2$, whose sign depends on the relative
orientation of the vortex fluxes~\cite{Engelhardt:1999xw}, see
Fig.~\ref{fig:vortcross3d}. Fig.~\ref{fig:vortcross3d}b indicates the position
of the vortices after center projection, leading to (thin)
P-vortices at half the thickness~\cite{DelDebbio:1996mh}.

\subsection{The Colorful Spherical Vortex}\label{sec:spher}

\begin{figure}[h]
\psfrag{x}{$x$}
\psfrag{y}{$y$}
\psfrag{z}{$z$}
\centering
a)\includegraphics[keepaspectratio,width=0.3\textwidth]{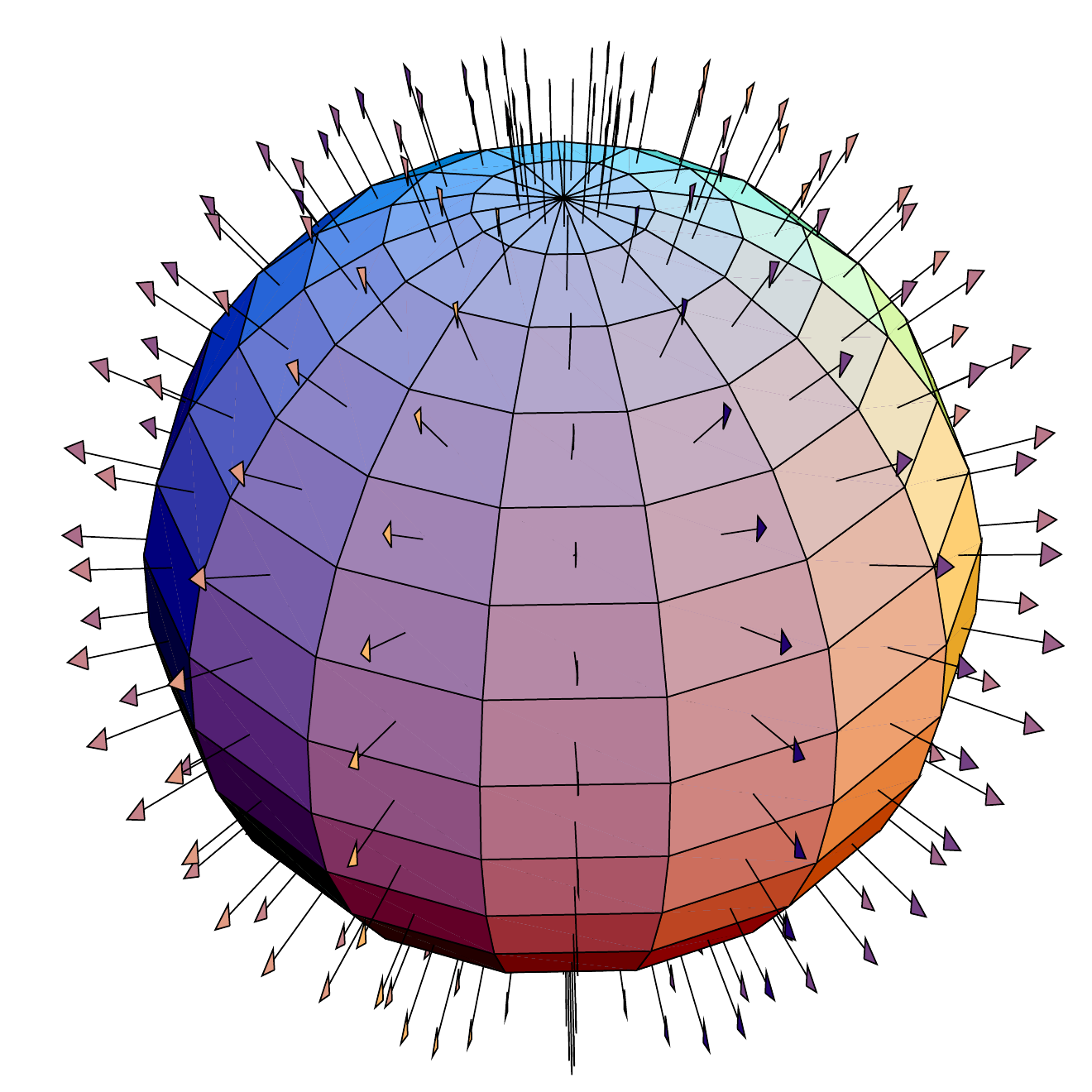}
b)\includegraphics[keepaspectratio,width=0.26\textwidth]{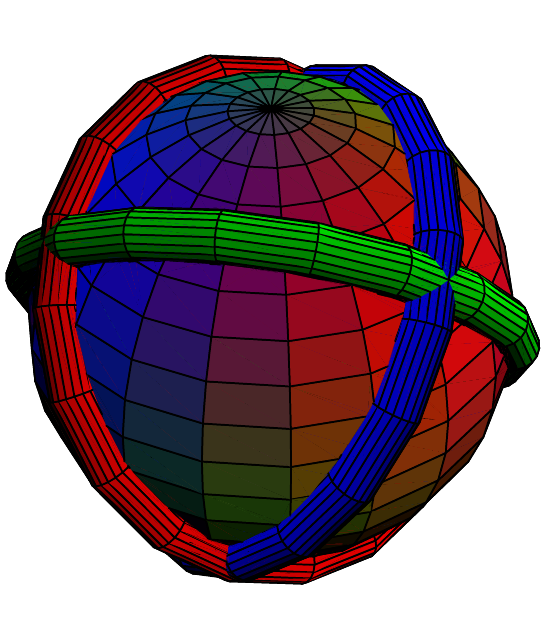}
c)\includegraphics[keepaspectratio,width=0.3\textwidth]{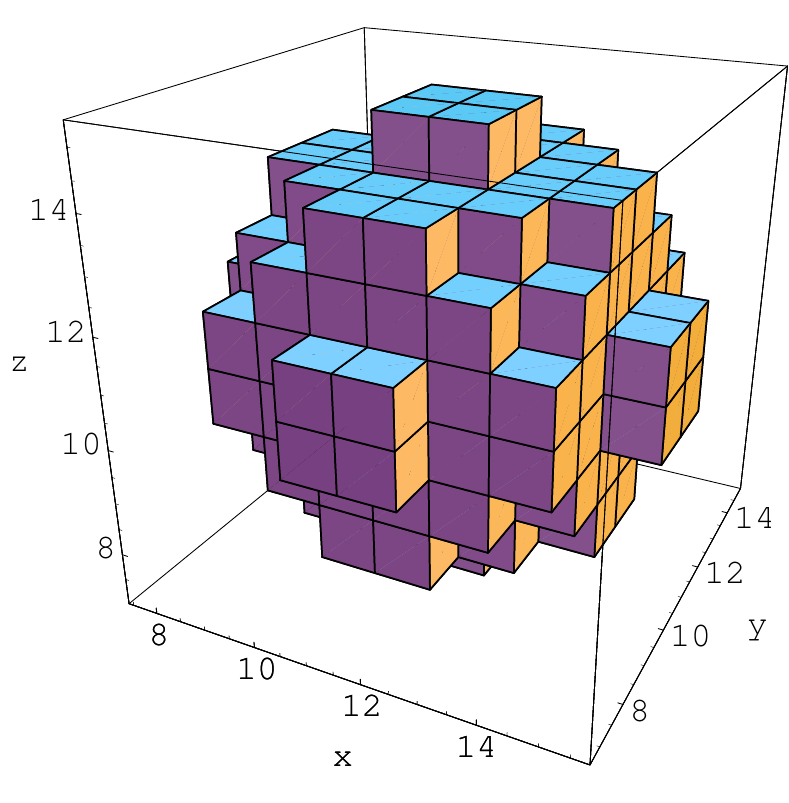}
\caption{a) The color structure of the vortex surface from the hedgehog
configuration leads to b) monopole lines after Abelian projection in the
maximal Abelian gauge. c) Lattice representation of a 3-sphere for our spherical
vortex after center projection in maximal center gauge.}
\label{fig:vortices}
\end{figure}

The non-orientable spherical vortex of radius $R$ and thickness $\Delta$ was
introduced in~\cite{Jordan:2007ff} and analyzed in more detail
in~\cite{Hollwieser:2010mj} and~\cite{Schweigler:2012ae}. It is constructed
with the following $SU(2)$ links:
\begin{gather}
U_\mu(x_\nu) = \begin{cases}
 \exp\left(\mathrm i\alpha(|\vec r-\vec r_0|)\vec n\cdot\vec\sigma\right)&t=t_i, \mu=4\\
 \mathbbm{1}&\mathrm{elsewhere}\end{cases}\quad\mbox{with}\quad
 \vec n(\vec r,t)=\frac{\vec r-\vec r_0}{|\vec r-\vec r_0|} \, ,
\label{eq:sphv}
\end{gather}
where $\vec r$ is the spatial part of $x_\nu$ and the profile function
$\alpha$ is either one of $\alpha_+, \alpha_-$, which are defined by
\begin{gather}
\alpha_-(r) = \begin{cases} \pi \\
     \frac{\pi}{2}\left( 1-\frac{r-R}{\frac{\Delta}{2}} \right) \\
                        0 
          \end{cases} \ldots\;
\alpha_+(r) = \begin{cases} \pi & r < R-\frac{\Delta}{2} \\
                          \frac{\pi}{2}\left( 3+\frac{r-R}{\frac{\Delta}{2}} \right) & R-\frac{\Delta}{2} < r < R+\frac{\Delta}{2} \\
                          2\pi & R+\frac{\Delta}{2} < r
          \end{cases}\label{eq:alpha} \, .
\end{gather}
This means that all links are equal to $\mathbbm 1$ except for the
$t$-links in a single time slice at fixed $t=t_i$. The phase changes from
$\pi$ to $0$ within a thickness $\Delta$ for $\alpha_-(r)$ (or from $\pi$ to $2\pi$ for $\alpha_+(r)$). The graph of $\alpha_-(r)$ is plotted in Fig. 2
in~\cite{Jordan:2007ff}, giving a hedgehog-like configuration, since  the
color vector $\vec n$ points in the ``radial'' direction $\vec r/r$ at the
vortex radius $R$, see Fig.~\ref{fig:vortices}a. The check that this
configuration is a vortex is done with maximal center gauge fixing and
center projection and results in a P-vortex forming a lattice
representation of a 3-sphere of radius $R$ at time slice $t_i$, see Fig.~\ref{fig:vortices}c. The color
structure of the thick vortex leads to a monopole loop on a great circle of
the P-vortex after fixing to maximal Abelian gauge and Abelian projection. The
direction of the loop depends on the $U(1)$ subgroup chosen as Abelian
degrees of freedom.  For the subgroup defined by the Pauli matrices
$\sigma_1, \sigma_2$ or $\sigma_3$ the monopole loops are in the $yz$-,
$zx$- and $xy$-plane, respectively. This is indicated schematically in
Fig.~\ref{fig:vortices}b with three colors.

The hedgehog-like structure is crucial for our analysis. The $t$-links of
the spherical vortices fix the holonomy of the time-like loops, defining a
map $U_t(\vec x,t=t_i)$ from the $xyz$-hyperplane at $t=t_i$ to $SU(2)$.
Because of the periodic boundary conditions, the time slice has the
topology of a 3-torus. But, actually, we can identify all points in the
exterior of the 3 dimensional sphere since the links there are trivial.
Thus the topology of the time slice is $\mathbbm R^3 \cup \{\infty\}$ which
is homeomorphic to $S^3$. A map $S^3 \to SU(2)$ is characterized by a
winding number
\begin{gather*}
  N = -\frac{1}{24\pi^2} \int d^3 x
  \,\epsilon_{ijk}\,\mbox{Tr}[(U^\dag\partial_i U)(U^\dag\partial_j
  U)(U^\dag\partial_k U)],\label{eq:winding}
\end{gather*}
resulting in $N=-1$ for positive and $N=+1$ for negative spherical
vortices. Obviously such windings, given by the holonomy of the time-like
loops of the spherical vortex, influence the Atiyah-Singer index theorem
\cite{Atiyah:1971rm,Schwarz:1977az,Brown:1977bj,Narayanan:1994gw,Nye:2000eg,Poppitz:2008hr},
giving a topological charge $Q=-1$ for positive and $Q=+1$ for negative
spherical vortices (anti-vortices). Hence, spherical vortices attract Dirac
zero modes similar to instantons. In~\cite{Schweigler:2012ae} we showed that
the spherical vortex is in fact a vacuum-to-vacuum transition in the
time direction which can even be regularized (smoothed out in the time
direction) to give the correct topological charge also from gluonic definitions (see also~\cite{Schweigler:2012} for more details).

\section{Dirac eigenmodes}
\label{sec:modes}

According to the Banks-Casher analysis \cite{Banks:1979yr}, chiral symmetry
breaking ($\chi$SB) is necessarily associated with a finite density of
near-zero eigenmodes of the chiral-invariant Dirac operator, resulting in a
finite chiral condensate, the order parameter of chiral symmetry breaking.
We compute the lowest-lying chiral eigenvectors $\chi_{R,L}$ and eigenvalues 
$|\lambda| \in [0,1]$ of the overlap Dirac operator 
\begin{equation}
D_{ov} = \frac{1}{2} \left[ 1 + \frac{D_W}{\sqrt{D_W^\dagger D_W}} \right]\label{eq:Dov} 
\end{equation}
with the kernel Wilson Dirac operator $D_W$~\cite{Edwards:1998yw}. Henceforth 
we will simply write $\lambda$ instead of $|\lambda|$ and assume it to be the 
absolute value of the two complex conjugate eigenvalues of $D_{ov}$ if $
\lambda\neq 0,1$. We will however distinguish between right- and left-handed 
zero modes of $D_{ov}$. When speaking of an eigenmode, we always mean both of 
the eigenvectors $\psi_\pm$ belonging to one value of $\lambda\neq 0$, which 
have the same scalar and chiral densities. For convenience, we will
number the eigenfunctions in ascending order of the eigenvalues. \#0+
denotes the right-handed zero mode, \#0-- the left-handed zero mode. \#1
labels the lowest nonzero mode, \#2 the second lowest etc. Their
eigenvalues are referred to as $\lambda$\#0+, $\lambda$\#0--, $\lambda$\#1,
etc., and their densities as $\rho$\#0+, etc. Finally, we will name
near-zero modes which emerge from would-be zero modes also by \#0 for good
reasons, which will be discussed later.

The corresponding eigenvectors are given by
\begin{gather}
  \psi_\pm = \frac{1}{\sqrt 2} (\chi_R \pm i\chi_L).
\end{gather}
They have scalar and chiral densities
\begin{gather}
  \rho = \psi^\dagger_\pm \psi_\pm = \frac{1}{2}(\chi_R^\dagger\chi_R + \chi_L^\dagger \chi_L) \, , \\
  \rho_+ = \psi^\dagger_\pm \frac{1}{2}(1+\gamma_5) \psi_\pm = \frac{1}{2} (\chi_R^\dagger\chi_R) \, , \\
  \rho_- = \psi^\dagger_\pm \frac{1}{2}(1-\gamma_5) \psi_\pm = \frac{1}{2} (\chi_L^\dagger\chi_L) \, , \\
  \rho_5 = \psi^\dagger_\pm\gamma_5\psi_\pm = \frac{1}{2}(\chi_R^\dagger\chi_R - \chi_L^\dagger \chi_L) =\rho_+-\rho_- \, .
\label{eq:densities}
\end{gather}

The chiral density $\rho_5$ is important to assess the local chirality
properties, in particular to test the notion that the near-zero modes arise
from the splitting of exact zero modes localized at lumps of topological
charge.

In a gauge field with topological charge $Q\neq 0$, $D_{ov}$ has $|Q|$
exact zero modes with chirality $-\mathrm{sign} Q$, and an equal number of
eigenvectors of opposite chirality and eigenvalue 1 (doubler modes). This
is required in order that $\mbox{Tr}\gamma_5=0$. The topological charge is
\begin{gather}
  Q = \mbox{Tr} (\gamma_5 D_{ov}) = n_- - n_+ = \text{ind} D_{ov} \, ,
\end{gather}
as for any Ginsparg-Wilson operator.

The plot titles of the density plots (see {\it e.g.} Fig.~\ref{fig:freeoverlap})
give the $x$- and $y$-coordinates of the shown $zt$-slice, the chirality
({\it i.e.}, "chi=0" means we plot $\rho_5$, "chi=1" would be $\rho_+$ and
"chi=-1" $\rho_-$), the numbers of plotted modes ("n=1-1" means we plot
$\rho\#1$, "n=1-2" would be $\rho\#1+\rho\#2$) and the maximal density in
the plotted area ("max=..." - some maxima are cut off in order to resolve other substructures). For instanton and spherical vortex
configurations we use $12^3\times24$-lattices and always plot $xt$-slices
at $y=z=6$ since they are symmetric in spatial directions around their
centers at $x=y=z=6.5$. For plane vortex configurations we use
$16^4$-lattices, the plotted slices vary.

\subsection{Eigenvalue Spectrum of the free Dirac operator}

In this section we will analytically calculate the eigenvalues of the
massless free Dirac operator. These eigenvalues and their multiplicity will
come handy later when we compare the eigenvalues for different gauge field
configurations to the eigenvalues for the free case. The free continuum
Dirac operator for massless fermions is given by $D = \gamma_{\mu}
\partial_{\mu}$, where $\gamma_{\mu}$ are the Euclidean Dirac matrices. We
want to solve the Dirac equation $D_{\alpha \beta} \  \psi_{\beta}(x) = 
\lambda \ \psi_{\alpha}(x)$. Using the well know ansatz of plane wave 
functions $\psi_{\alpha} (x) = u_{\alpha} \ \exp(i  p_{\mu} x_{\mu})$,
one gets the eigenvalues of the free continuum operator $\lambda = \pm i \sqrt
{p_{\mu} p_{\mu}}$. Identifying the eigenvalue $\lambda$ with the fermion 
mass $M$ makes clear, that this is simply the relativistic energy momentum 
relation in Euclidean space given by $M^2 = E^2 + p^2$, where we have denoted 
$p_4^2 $ by $E^2$ and $p_i p_i$ by $p^2$. Note that both the eigenvalue for 
the plus and the minus sign have a degeneracy of two, which gives in total 
four eigenvalues for the $4\times4$ Dirac matrix ($4\times4$ Dirac indices). 
Additional degeneracy comes from the color indices. The free Dirac operator 
is color blind. Therefore, the degeneracy of the different eigenvalues is 
multiplied by $n_{col}$, where $n_{col}$ is the range of the color indices. 
For the overlap Dirac operator the eigenvalues can also be calculated 
analytically for the free case using the lattice ansatz $\psi_{\alpha} (n) = 
u_{\alpha} \ \exp(i \ a \  p_{\mu} n_{\mu})$, with the lattice sites $n_\mu$ 
and yield the continuum solutions with corrections of orders of $a^2$:
\begin{gather}
D_{ov}(p_\mu) = \frac{1}{2} \left[ 1 + \frac{i \gamma_\mu \sin(p_\mu) 
- M(p_\mu)} {\sqrt{\sum_\mu \sin(p_\mu)^2 + M^2(p_\mu)}} \right]
\end{gather}
with $M(p_\mu) = M - \sum_\mu \left( 1 - \cos(p_\mu) \right)$. Here $a=1$ and 
$M=1$ in the free case. This gives ($H_{ov} = \gamma_5 D_{ov}$):
\begin{gather}
H^2_{ov}(p_\mu) = D_{ov}^\dagger D_{ov} = \frac{1}{2} \left[ 1 - \frac{
M(p_\mu)} {\sqrt{\sum_\mu \sin(p_\mu)^2 + M^2(p_\mu)}} \right] = \lambda^2
\end{gather}
with expected multiplicities ($H^2_{ov}(p_\mu)$ is a diagonal $4\times4$ 
matrix). For $p_\mu << 1$ we get 
\begin{gather}
\lambda^2 = \frac{p_\mu p_\mu}{4 M^2} ( 1 + O(p_\mu p_\mu) ) \, .
\end{gather}
With the normalization of Eq.~(\ref{eq:Dov}), which leads to $\lambda^2
\in [0,1]$, a wave function renormalization is needed to convert to the
usual continuum normalization. This compensates the factor $\frac{1}{4 M^2}$,
see e.g. Eq. (6) in~\cite {Edwards:1998wx}, giving the usual free Dirac
eigenvalues in the continuum limit.
Note that the plane wave ansatz is periodic, not only in $x_
\mu$ or lattice indices $n_{\mu}$, but also in $p_{\mu}$.
This means, that $p_{\mu}$ and $p_{\mu} + \frac{2 \pi z_{\mu}}{a}$ (with
$z_{\mu} \in \mathbb{Z}$) correspond to the same eigenfunction. Therefore,
in order to get the correct multiplicities for the eigenvalues, we have to
restrict the range of $p_{\mu}$ to
\begin{equation}
	- \frac{\pi}{a} < p_{\mu} \leq \frac{\pi}{a} \, .
\label{restriction_FB}
\end{equation}
For the usual periodic boundary conditions in the spatial directions and
periodic or anti-periodic boundary conditions in the temporal direction, the
allowed values for $p_{\mu}$ are
\begin{equation*}
	p_i = \frac{2 n \pi}{a N_{sp}}, \quad \quad p_4 = \begin{cases} \frac{2 n \pi}{a N_{t}} & \mbox{for periodic BC} \\ \frac{ (2 n + 1) \pi}{a N_{t}} & \mbox{for anti-periodic BC} \end{cases},
\quad \quad n \in \mathbb{Z} \, ,
\end{equation*}
where $N_{sp}$ is the spatial and $N_{t}$ the temporal extent of the
lattice. The total multiplicity of an eigenvalue $\lambda$ is given by
\begin{equation*}
n_{mult}(\lambda) = 2 \ n_{col} \ n_{p}(\lambda),	
\end{equation*}
where $n_{p}(\lambda)$ is the number of different $p_{\mu}$ corresponding
to a particular $\lambda$. The factor two comes from the Dirac indices as
discussed above. Let us now have a quick look at the multiplicity of the
eigenvalues for one particular lattice size, {\it i.e.}, $N_{sp} = 12$ and
$N_t = 24$, which we will use later for our lattice configurations. We
assume anti-periodic boundary conditions in the temporal direction, $a=1$ and a
$SU(2)$ gauge field theory ({\it i.e.}, $n_{col} = 2$). The momentum
vectors corresponding to the lowest eigenvalues are then given by $p_i = 0$
and $p_4 = \pm \pi / 24 $. Therefore, we have $n_{p}(\lambda\#1) = 2$ for
the lowest eigenvalues $\lambda\#1$. That means, we get a multiplicity
$n_{mult}(\lambda\#1) = 2 \ n_{col} \ n_{p}(\lambda\#1) = 8$. For the
second lowest eigenvalues $\lambda\#2$ we have $p_i = 0$ and $p_4 = \pm
3\pi / 24 =\pm\pi/8$ and therefore another degeneracy of eight. Then we get
$p_i = \pm \pi / 6$ (with $i \in \{ 1, 2, 3 \}$) and $p_4 = \pm \pi/24$,
{\it i.e.}, $n_p(\lambda\#3) = 12$. This gives a multiplicity
$n_{mult}(\lambda\#3) = 48$. 

\begin{figure}
	\centering
		\includegraphics[width=.32\columnwidth]{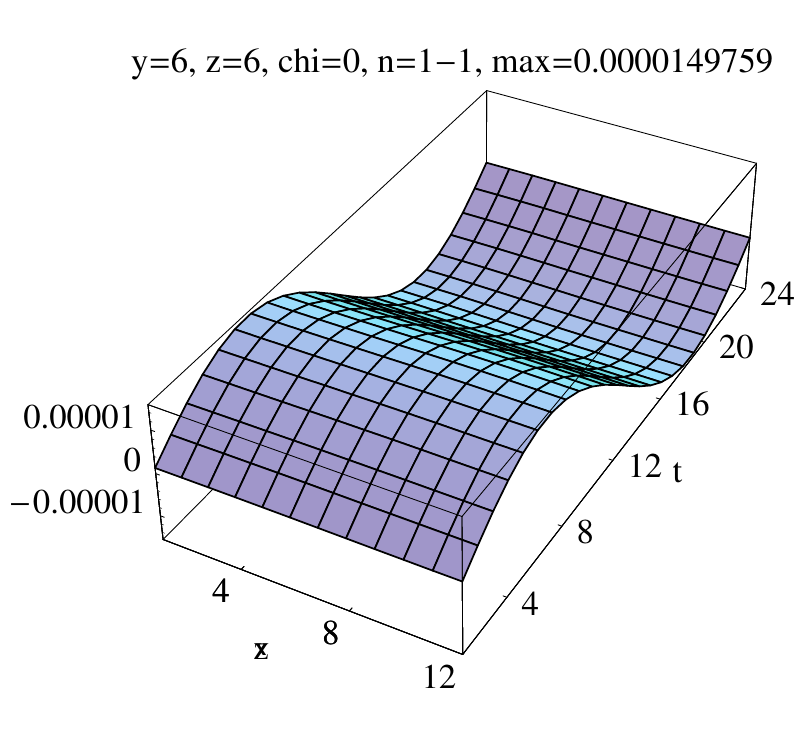}
		\includegraphics[width=.32\columnwidth]{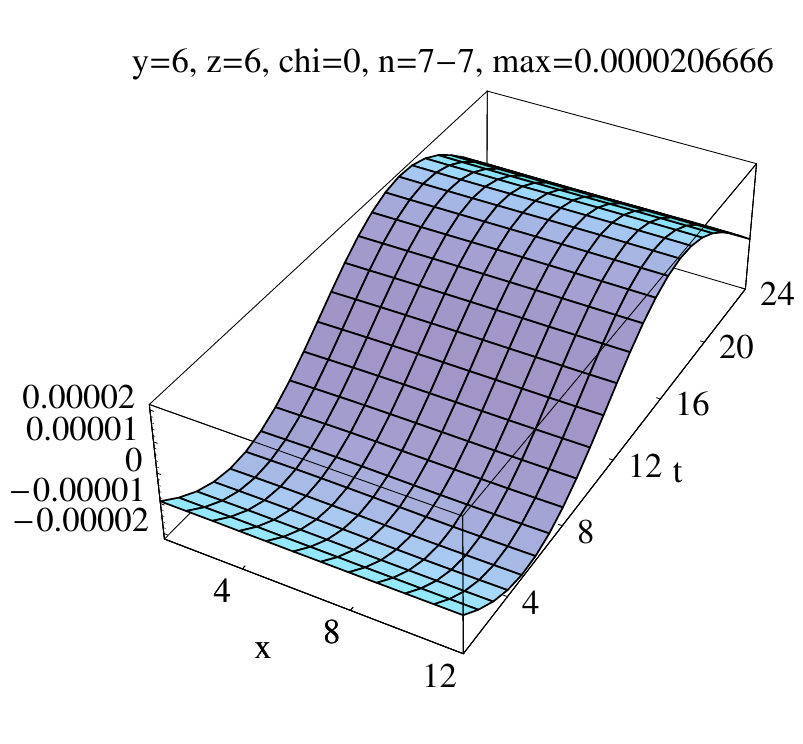}
		\includegraphics[width=.32\columnwidth]{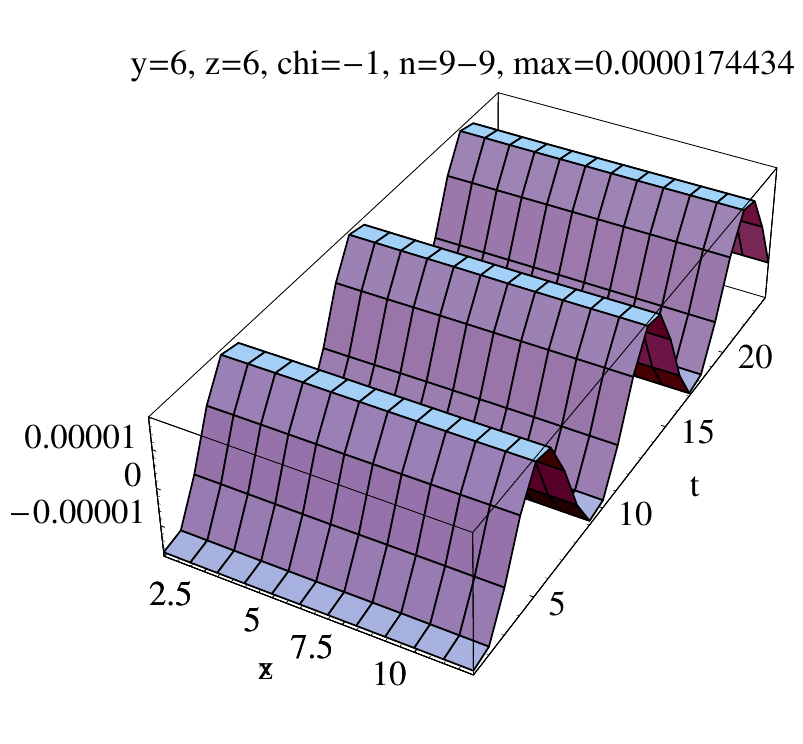}
	\caption{Chiral density of the low-lying eigenmodes of the free
	overlap Dirac operator: $\rho_5\#1$ (left), $\rho_5\#7$ (center)
	$\rho_5\#9$ (right). The modes clearly show the plane wave
behavior with oscillations of $2p_\mu$ (see text).}
	\label{fig:freeoverlap}
\end{figure}

Fig.~\ref{fig:freeoverlap} shows the chiral density of free overlap 
eigenmodes obtained numerically using the MILC code. The modes are found with 
the Ritz functional algorithm~\cite{Bunk:1994uv,Kalkreuter:1995mm} with 
random start and for degenerate eigenvalues the eigenmodes span a randomly 
oriented basis in the degenerate subspace.
Therefore the numerical modes presented in Fig.~\ref{fig:freeoverlap} are 
linear combinations of plane waves with $\pm p_\mu$ and show plane wave 
oscillations of $2p_\mu$ in the chiral density. The first eight degenerate 
modes consist of plane waves with $p_4=\pm\pi/24$, hence there is one sine 
(cosine) oscillation in time direction, the next eight have $p_4=\pm3\pi/24$, 
{\it i.e.}, three oscillations in the time direction. The oscillations of $
\chi_R$ and $\chi_L$ are separated by half an oscillation length, {\it i.e.}, 
the maxima of $\rho_+$ correspond to minima of $\rho_-$ and vice versa. 
Accordingly, the scalar density $\rho(x_\mu) = \frac{1}{2}(\chi_R^\dagger(x_
\mu)\chi_R(x_\mu) + \chi_L^\dagger(x_\mu)\chi_L(x_\mu))=1/N_V$ is constant 
($N_V\ldots$ lattice volume).

\subsection{Zero Modes and Instantons}\label{sec:zms:inst}

An instanton field gives rise to an exact zero mode (the precise expression
is given in {\it e.g.}~\cite{Diakonov:2002fq}). Its probability density
\begin{gather}
  \rho(x) = \frac{2}{\pi^2}\frac{R^2}{(x-x_0)^2+R^2}
\end{gather}
is localized at the instanton core $x_0$, with a half-radius of the
instanton parameter $R$. Since the zero mode is exactly chiral, its chiral
density $\psi^\dag \gamma_5 \psi$ equals $\pm\rho$.
In the instanton liquid model the near-zero modes originate from the
overlap of the would-be zero modes carried by individual instantons and
anti-instantons. If the overlap is not too large, one expects that the
resulting near-zero modes still exhibit definite chirality
\textit{locally}. This picture predicts characteristic properties of the
low-lying modes \cite{Horvath:2002gk}: 
\begin{itemize}
\item Their probability density should be clearly peaked, indicating the location of instantons. 
\item The local chirality at the peaks should match the sign of the topological charge and the size of the chiral lump should be correlated to the extension of the topological structure.
\item As an instanton and an anti-instanton approach each other, the eigenvalues should be shifted further away from zero and the localization and local chirality properties should fade.
\end{itemize}
Numerical evidence supporting these assumptions about the local chirality
structure of the low-lying Dirac modes are presented in {\it e.g.}
\cite{Edwards:1999dc,DeGrand:2000gq,Edwards:2001nd}. Here we want to analyze
these issues for center vortices.

\subsection{Zero Modes and Center Vortices}\label{sec:zms:vort}

Reinhardt et al. \cite{Reinhardt:2002cm} analytically calculate the exact
zero modes of the Dirac operator in the background of plane vortices, both
non-intersecting and perpendicularly intersecting ones, for an Abelian
$U(1)$ gauge field on $\mathbbm R^2, \mathbbm R^4, \mathbbm T^2$ and
$\mathbbm T^4$.
The first vortex field of interest consists of two (anti-)parallel fluxes
on a two-torus. Because of a two-dimensional translation symmetry we can
identify this with the four-dimensional configuration of a single vortex pair
presented in Sec.~\ref{sec:plane}. The second example contains four flat
vortices on a four-torus, which intersect orthogonally in four points. This
corresponds to the configuration shown in Fig.~\ref{fig:vortices}a.

We want to discuss the boundary conditions used in~\cite{Reinhardt:2002cm}.
The four-dimensional problem can be reduced to two two-dimensional ones
because of the translation invariance parallel to the vortex sheets. It
therefore suffices to deal with the $\mathbbm T^2$ case. The boundary conditions
for a gauge field on $\mathbbm T^2$ are specified by the two transition
functions $\Omega_x, \Omega_y$. In a suitable gauge, it is possible to set
$\Omega_x = 1$. The co-cycle condition then turns into a periodicity
condition
\begin{gather}
  \Omega_y(x+L_x)=\Omega_y(x)
\end{gather}
for $\Omega_y(x)$, which lives only at the boundary of the rectangle which
represents the torus. Therefore $\Omega_y$ defines a function $S^1 \to S^1$
belonging to a class of the homotopy group $\pi_1(S^1)\simeq \mathbbm Z$.
Its winding number $n$ determines the magnetic flux through the $xy$-plane
by
\begin{gather}
  \Phi := \oint A_\mu \text d x_\mu = 2\pi n \, .
\end{gather}
To prove this, we simply perform the line integral over the boundary of the
rectangle resulting from cutting up the torus. We set $\Omega_y(x) =
\exp\{i\chi(x)\}$. Since $\Omega_y$ is periodic, $\chi(x+L_x) = \chi(x) +
2\pi n$. Then
\begin{align}
  \Phi &= \int_0^{L_x} A_x(x,0) \text d x + \int_0^{L_y} A_y(L_x,y) \text d y - \int_0^{L_x} A_x(x,L_y)\text d x - \int_0^{L_y} A_y(0,y)\text d y \nonumber \\
     &= \int_0^{L_x} A_x(x,0) \text d x + \int_0^{L_y} A_y(0,y) \text d y - \int_0^{L_x} [A_x(x,0) - \partial_x\chi(x)]\text d x - \int_0^{L_y} A_y(0,y)\text d y \nonumber \\ &= \int_0^{L_x}
\partial_x \chi(x) \text d x = \chi(L_x) - \chi(0) = 2\pi n \, .
\end{align}
Consequently, the flux on $\mathbbm T^2$ is quantized. Note that this would
not hold for SU(2) since $\pi_1(S^3) = 1$.  To create two vortices which
carry a flux of $+\pi$ each, Reinhardt et al. use $\Omega_y=\exp\{2\pi i
x\}$. Returning finally to the fermion field, it obeys boundary conditions
consistent with gauge invariance,
\begin{gather}
  \psi(x,y+L_y) =\Omega_y(x)\psi(x,y) \, .
\end{gather}
In~\cite{Hollwieser:2011uj} we analyzed the zero modes for plane vortex
configurations and found good agreement with the results obtained by
Reinhardt {\it et al.}~\cite{Reinhardt:2002cm}. The discrepancies discussed in
our work originate in the finite thickness of our vortex configurations
because of finite lattice sizes.

\begin{figure}
	\centering
		a)\includegraphics[width=.48\columnwidth]{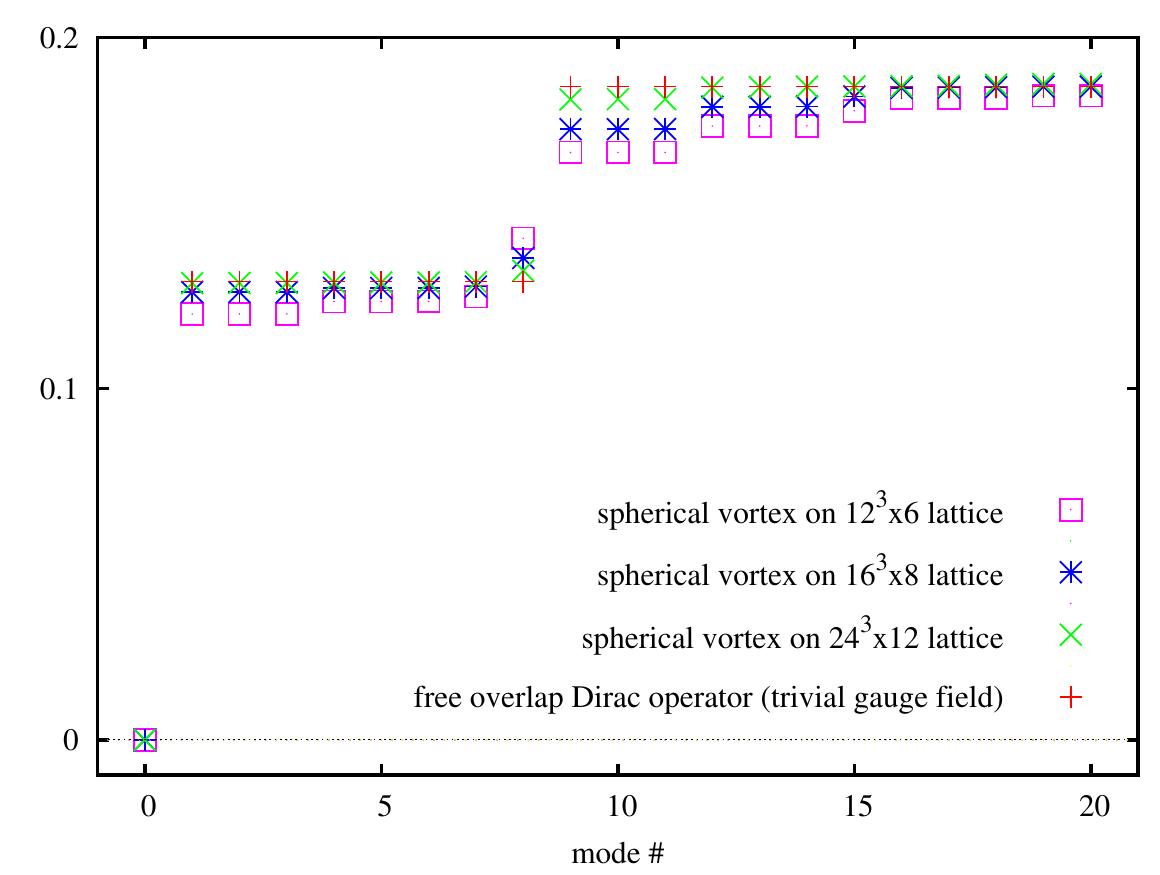}
		b)\includegraphics[width=.4\columnwidth]{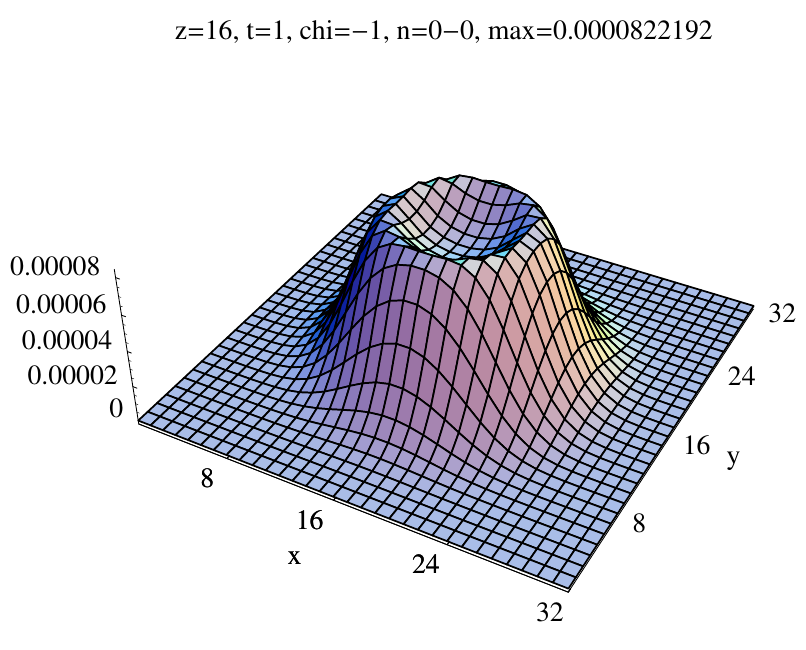}
	\caption{a) The lowest eigenvalues for the spherical vortex with
	$R = d = 3 a$ compared to the eigenvalues of the free Dirac operator
	(red crosses). Eigenvalues are calculated on a $12^3 \times 6$ lattice
	with relative lattice constant $a=2$ (magenta boxes), on a $16^3 \times 8$ lattice with $a=\frac{3}{2}$ (blue stars) and on a $24^3 \times 12$ lattice with
	$a=1$ (green crosses). The multiplicity of $\lambda\#1$ is eight,
	of $\lambda\#2$ 48. b) Spatial distribution of the zero mode,
	mainly located at the vortex core.}
	\label{fig:finitesizespher}
\end{figure}

We conclude this section with a quick discussion of the eigenvalues for the
spherical vortex. As already mentioned, there always occurs exactly one
zero mode, a positive chirality one for the spherical vortex and a negative
chirality zero mode for the anti-vortex. The lowest nonzero modes can be
seen as some sort of modified eigenmodes of the free Dirac operator. This
point of view is motivated by the results presented in
Fig.~\ref{fig:finitesizespher}a. In this diagram, one can see the
eigenvalues for configurations with vortices of the same size in lattice
units but different spatial $N_{sp}$ and temporal $N_t$ lattice extents. In
other words, the vortex gets smaller while the lattice gets finer. One sees
from the figures, that the nonzero eigenvalues converge to eigenvalues of
the free Dirac operator as the vortex gets smaller and smaller. For the
investigated eigenvalues, there seems to be a one to one correspondence
between the eigenvalues of the free Dirac operator and the nonzero
eigenvalues of the Dirac operator for the vortex configuration. The zero
mode is not a lowered nonzero mode, it occurs in addition to the low lying
modes. Because the total number of eigenmodes only depends on the lattice
size, the number of complex eigenvalues is lowered by two (because of the
zero and the doubler mode) for the vortex configuration in comparison to
the free case. Note that we still can have a one to one correspondence
between all the complex eigenvalues for the vortex and the free case. One
can see this by remembering that the correspondence is established in the
limit of an infinitely large lattice.

\section{Interactions between topological objects}\label{sec:inter}

We want to discuss the Dirac equation for a gauge field $\mathcal{A}_{\mu}=
\mathcal{A}_{1\mu} + \mathcal{A}_{2\mu}$ consisting of two fields
$\mathcal{A}_{1\mu}$ and $\mathcal{A}_{2\mu}$ which are separated in
Euclidean space and have non-vanishing topological charge $Q_i$. Therefore,
the Dirac operators $D_1$ for $\mathcal{A}_{1\mu}$ alone and $D_2$ for
$\mathcal{A}_{2\mu}$ alone would have at least one zero mode. In the
following, the zero modes of $D_1$ and $D_2$ will be called would-be zero
modes. Lets discuss the case in which $\mathcal{A}_{1\mu}$ has $Q_1=1$ and
$\mathcal{A}_{2\mu}$ has $Q_2=-1$. For simplicity we assume that $D_1$ has
only one left handed zero mode $\mid \psi_1 \rangle$ and $D_2$ only one
right handed zero mode $\mid \psi_2 \rangle$. Clearly, these two would-be
zero modes are orthogonal and can be part of an orthogonal basis. Let us
now have a look at the Dirac operator in this orthogonal basis. In
particular we are interested in the upper left $2 \times 2$ block
\begin{equation}
	\begin{pmatrix} \langle \psi_1 \mid D \mid \psi_1 \rangle & \langle \psi_1 \mid D \mid \psi_2 \rangle \\ \langle \psi_2 \mid D \mid \psi_1 \rangle & \langle \psi_2 \mid D \mid \psi_2 \rangle \end{pmatrix} \label{2x2_block_general}
\end{equation}
of the Dirac matrix. The continuum Dirac operator $D$ is given by
\begin{align}
\begin{split}
	D &= \gamma_{\mu} \ \left( \partial_{\mu} + \mbox{i} \mathcal{A}_{\mu} (x) \right) = \gamma_{\mu} \ \left( \partial_{\mu} + \mbox{i} \mathcal{A}_{1 \mu} (x) + \mbox{i} \mathcal{A}_{2 \mu} (x) \right) \\
	&= D_1 + \gamma_{\mu} \mbox{i} \mathcal{A}_{2 \mu} (x) = D_2 + \gamma_{\mu} \ \mbox{i} \mathcal{A}_{1 \mu} (x) \, . 
	\end{split} 
\end{align}
The first element of (\ref{2x2_block_general}) therefore evaluates to
\begin{equation}
	\langle \psi_1 \mid D \mid \psi_1 \rangle = \langle \psi_1 \mid D_1 \mid \psi_1 \rangle +
 \langle \psi_1 \mid \gamma_{\mu} \ \mbox{i} \mathcal{A}_{2 \mu} (x) \mid \psi_1 \rangle = 0 \, .
\label{first_diag}
\end{equation}
One can see that $\langle \psi_1 \mid \gamma_{\mu} \ \mbox{i}
\mathcal{A}_{2 \mu} (x) \mid \psi_1 \rangle$ vanishes from
\begin{align}
\begin{split}
	\langle \psi_1 \mid \gamma_{\mu} \ \mbox{i} \mathcal{A}_{2 \mu} (x) \mid \psi_1 \rangle &= \langle \psi_1 \mid \gamma_{\mu} \ \mbox{i} \mathcal{A}_{2 \mu} (x) \ {\gamma_5}^2 \mid \psi_1 \rangle \\
	&= - \ \langle \psi_1 \mid  \gamma_5 \ \gamma_{\mu} \ \mbox{i}
 \mathcal{A}_{2 \mu} (x) \ {\gamma_5} \mid \psi_1 \rangle = - \ \langle \psi_1 \mid  \gamma_{\mu} \ \mbox{i} \mathcal{A}_{2 \mu} (x) \mid \psi_1 \rangle \ ,.
\label{diag_sec_term_vanish}
	\end{split}
\end{align}
%
In the same way one can prove
\begin{equation}
	\langle \psi_2 \mid D \mid \psi_2 \rangle = 0 \, .
\label{el_22}
\end{equation}
%
Let us now calculate the off-diagonal terms. The first off-diagonal term
evaluates to
\begin{equation}
	\langle \psi_1 \mid D \mid \psi_2 \rangle = \langle \psi_1 \mid D_2 \mid \psi_2 \rangle +
 \langle \psi_1 \mid \gamma_{\mu} \ \mbox{i} \mathcal{A}_{1 \mu} (x) \mid \psi_2 \rangle = 0 + c = c \, .
\label{el_12}
\end{equation}
Here $c$ stands for the overlap integral $\langle \psi_1 \mid \gamma_{\mu}
\ \mbox{i} \mathcal{A}_{1 \mu} (x) \mid \psi_2 \rangle$. In general, $c$
will be large for eigenmodes that overlap a lot, and small for eigenmodes
that overlap only a little. Note that it is crucial that $\mid \psi_1
\rangle$ and $\mid \psi_2 \rangle$ have different chirality. Otherwise, by
the same argument as used in (\ref{diag_sec_term_vanish}), the overlap
integral would have to vanish. The second off-diagonal term of the
upper-left $2 \times 2$ block is
\begin{equation}
	\langle \psi_2 \mid D \mid \psi_1 \rangle = \left( \langle \psi_1 \mid D^{\dagger} \mid \psi_2 \rangle  \right)^*
 = \left( - \  \langle \psi_1 \mid D \mid \psi_2 \rangle  \right)^* = - c^* \, .
\label{el_21}
\end{equation}
Here it was used that the continuum Dirac Operator $D$ is anti-hermitian,
{\it i.e.}, $D^{\dagger} = - D$.

Combining (\ref{first_diag}), (\ref{el_22}), (\ref{el_12}) and
(\ref{el_21}), the upper left $2 \times 2$ block of the Dirac matrix reads
\begin{equation}
	\begin{pmatrix} 0 & c\\ -c^* & 0 \end{pmatrix} \, .
\label{2x2_1_-1}
\end{equation}
This $2 \times 2$ block can easily be diagonalized. The eigenvalues
$\lambda_{1,2}$ and the normalized eigenvectors $\psi'_{1,2}$ of
(\ref{2x2_1_-1}) are
\begin{equation}
	\lambda_{1,2} = \pm \mbox{i} \sqrt{c c^*} = \pm \mbox{i} | c |, \quad \psi'_{1,2} = \frac{1}{\sqrt{2}} \begin{pmatrix} \pm \mbox{i} \sqrt{\frac{c}{c^*}} \\ 1 \end{pmatrix} \, .
\label{eigval_eigvec_v_av}
\end{equation}
This means that the interaction transforms the two (would-be) zero modes
into two near-zero modes. Those near-zero modes also occur additionally to
the free Dirac eigenmodes and therefore we still enumerate them like zero
modes by \#0. The strength of the interaction quantified by the overlap
integral $c$ determines the size of the near-zero eigenvalue. Note that the
new near-zero modes consist in equal parts of the would-be zero modes $\mid
\psi_1 \rangle$ and $\mid \psi_2 \rangle$. Therefore the scalar and chiral
densities of the near-zero modes are simply an average of the densities of
the would-be zero modes.

So far we have ignored everything except the upper left $2 \times 2$ block
of the Dirac matrix. To get exact eigenmodes we clearly have to diagonalize
the whole Dirac matrix and not only this $2 \times 2$ block. However,
(\ref{eigval_eigvec_v_av}) represents a legitimate approximation to the
exact eigenvalues and eigenmodes if the elements of the form
\begin{equation*}
 \langle	\psi_j \mid D \mid \psi_{1,2} \rangle = \langle \psi_j \mid \gamma_{\mu} \ \mbox{i} \mathcal{A}_{2,1 \mu} (x) \mid \psi_{1,2} \rangle  \quad \mbox{with} \quad j > 2 
\end{equation*}
are a lot smaller than the overlap integral $c=\langle \psi_1 \mid
\gamma_{\mu} \ \mbox{i} \mathcal{A}_{1 \mu} (x) \mid \psi_2 \rangle$.
Usually this will be the case, because $\mid \psi_1 \rangle$ is localized
at $\mathcal{A}_{1 \mu}$ and $\mid \psi_j \rangle$ with $j > 2$ is not.
Let us also have a quick look at what happens when we have two would-be
zero modes with the same chirality. In this case also the off-diagonal
terms of the upper left $2 \times 2$ matrix vanish and the would-be zero
modes are actual zero modes.

Note that the mechanism discussed in this section is the basis for the
instanton liquid model of spontaneous chiral symmetry breaking (see
\cite{Diakonov:2002fq} for a detailed review). In the instanton liquid
model the QCD-vacuum consists of an ensemble of instantons and
anti-instantons whose would-be zero modes split into near-zero modes
because of interactions. Therefore, one gets a non-vanishing eigenmode
density around zero, which gives via the Banks-Casher relation a finite
chiral condensate and broken chiral symmetry. Clearly, one can construct
such a model also with other topological objects, as will be shown in the
next section for center vortices.

\section{Vortices and $\chi$SB}\label{sec:vortmodel:scsb}

In Sec.~\ref{sec:vortQ} we discussed the various possibilities of
vortices to create topological charge, {\it i.e.}, via writhing and
intersection points as well as through their color structure. In previous
works we presented results on the attraction of zero modes for
flat~\cite{Hollwieser:2011uj} and
spherical~\cite{Jordan:2007ff,Hollwieser:2010mj,Hollwieser:2012kb,Schweigler:2012ae}vortex
configurations. Here we present some results on how vortices form near-zero
modes from would-be zero modes through interactions.

\subsection{Spherical Vortices and Instantons}\label{sec:sphvort}

\begin{figure}
	\centering
		a)\includegraphics[width=.48\columnwidth]{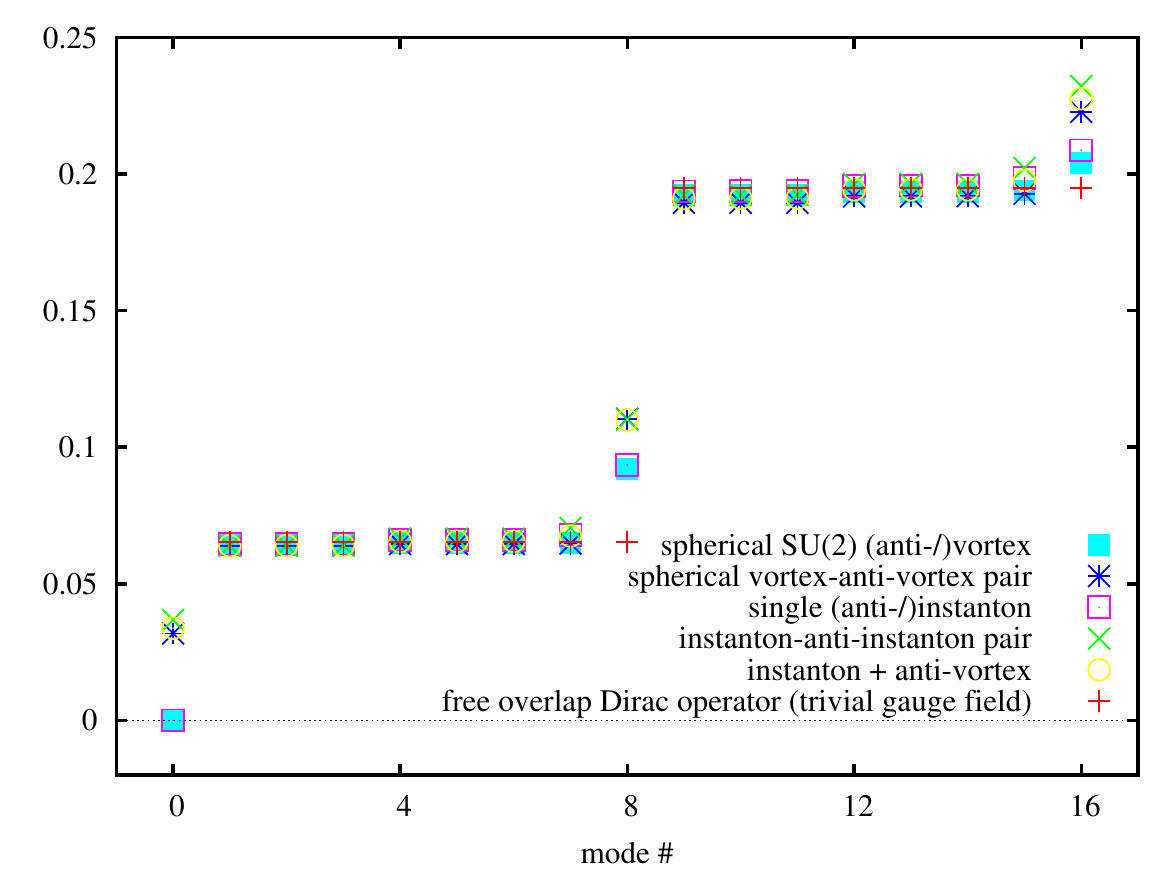}
		b)\includegraphics[width=.48\columnwidth]{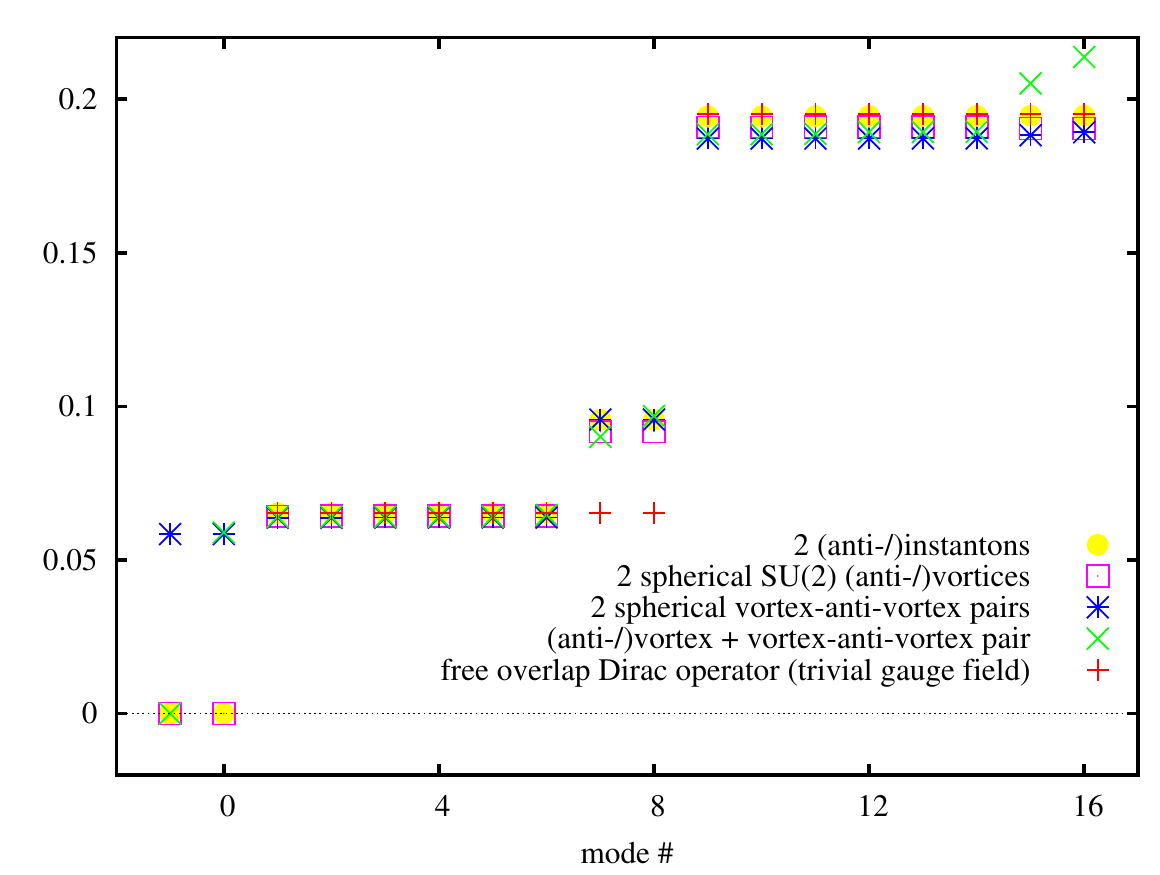}
	\caption{The lowest overlap eigenvalues for instanton and spherical
	vortex configurations compared to the eigenvalues of the free
	(overlap) Dirac operator.}
	\label{fig:oevlsph}
\end{figure}

\begin{figure}
	\centering
		a)\includegraphics[width=.32\columnwidth]{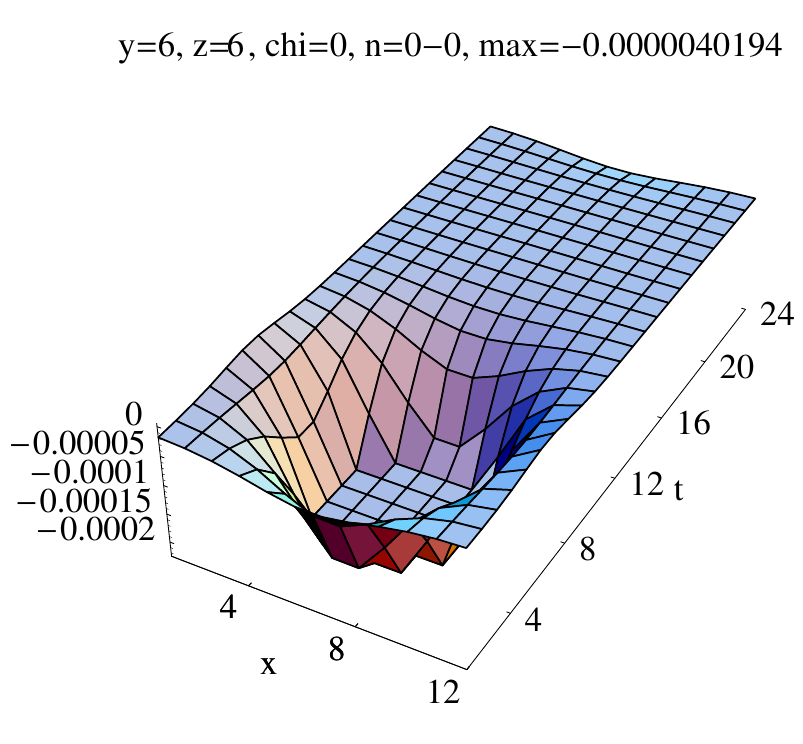}
		\includegraphics[width=.32\columnwidth]{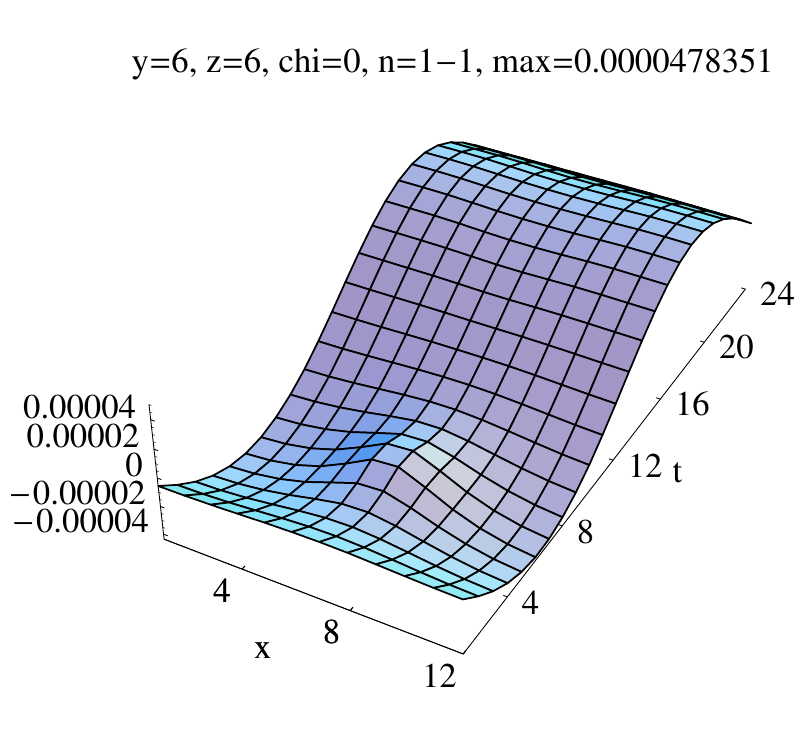}
		\includegraphics[width=.32\columnwidth]{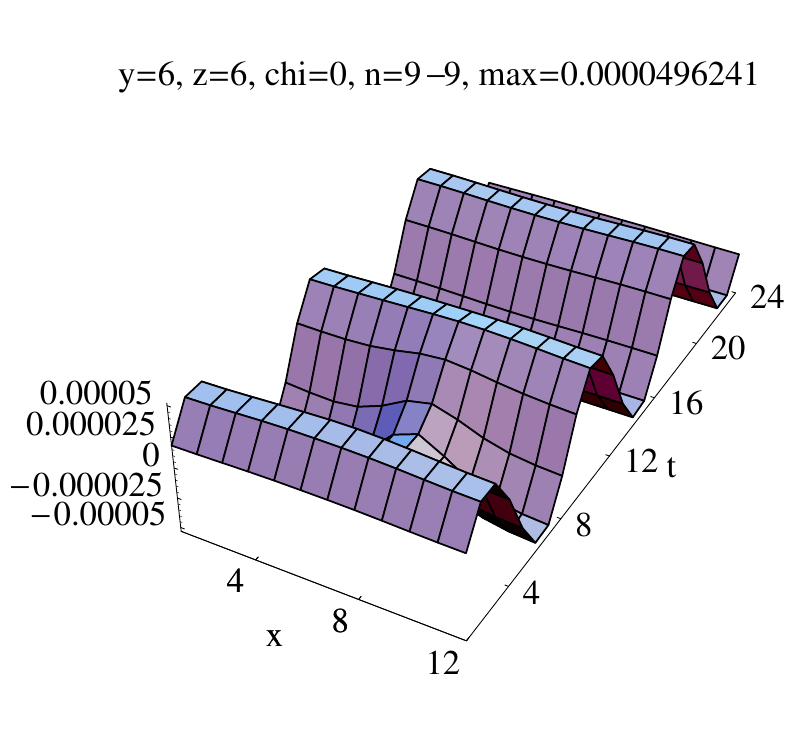}\\
		b)\includegraphics[width=.32\columnwidth]{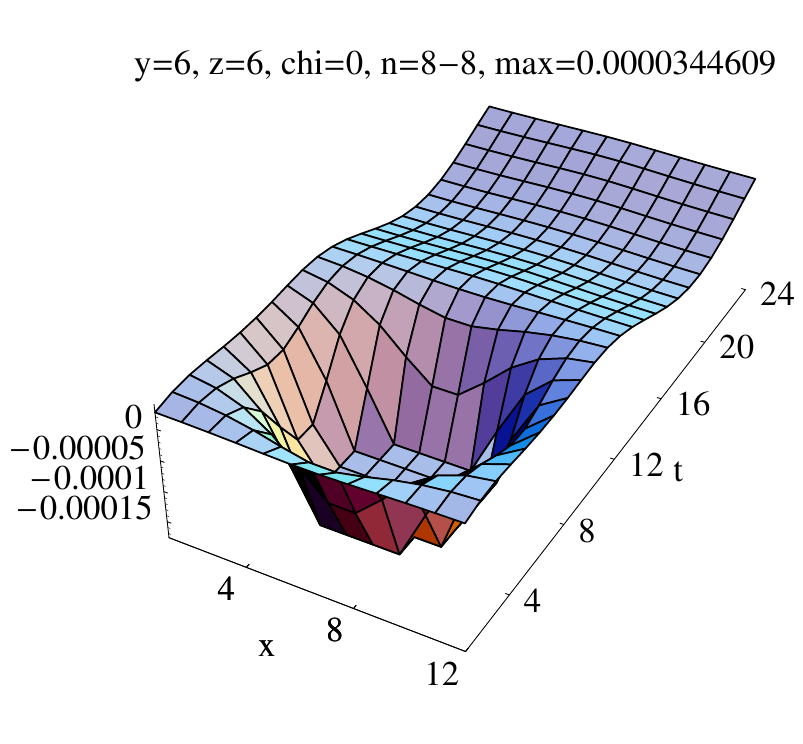}
		\includegraphics[width=.32\columnwidth]{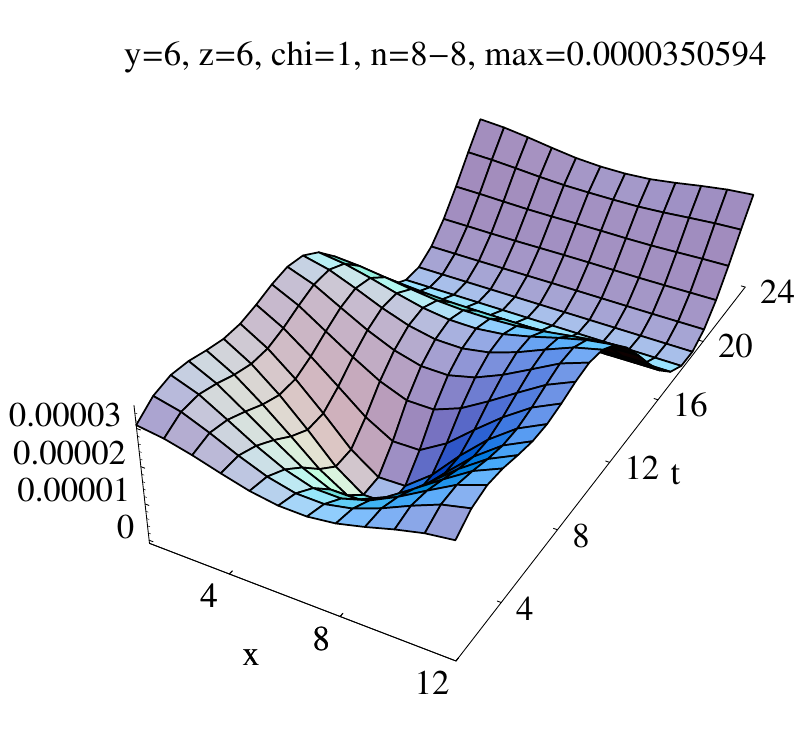}
		\includegraphics[width=.32\columnwidth]{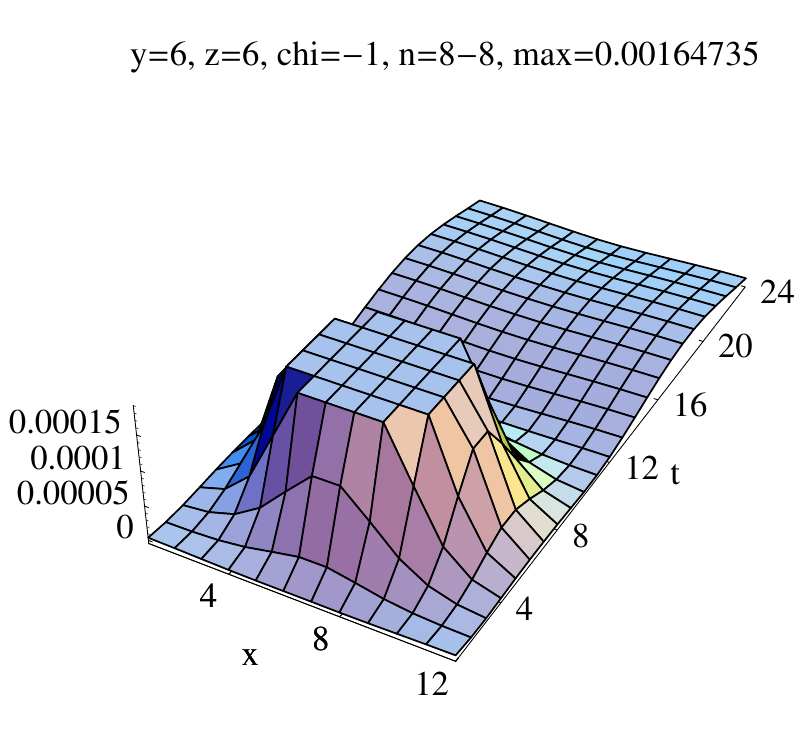}\\
		c)\includegraphics[width=.32\columnwidth]{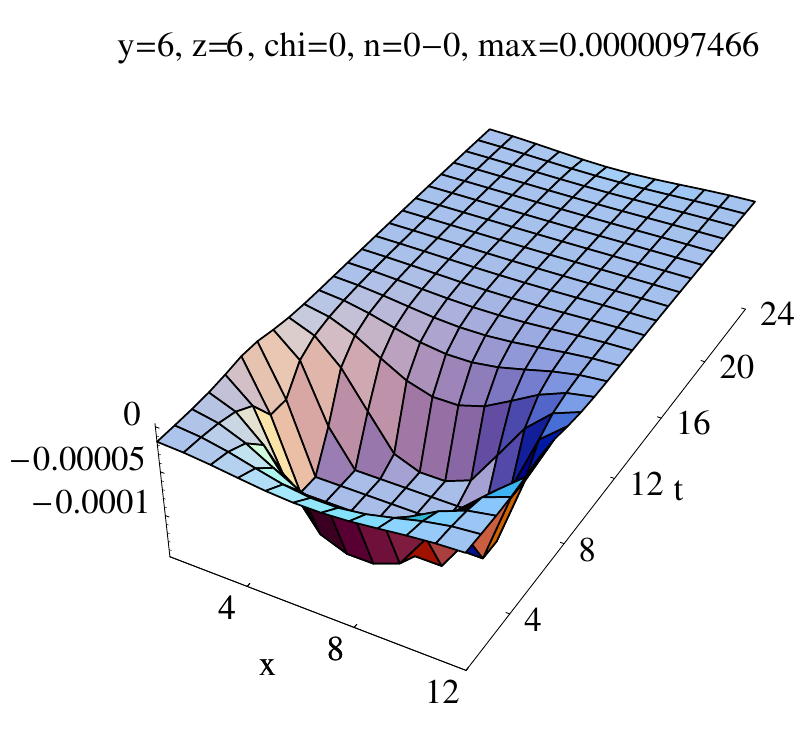}
		\includegraphics[width=.32\columnwidth]{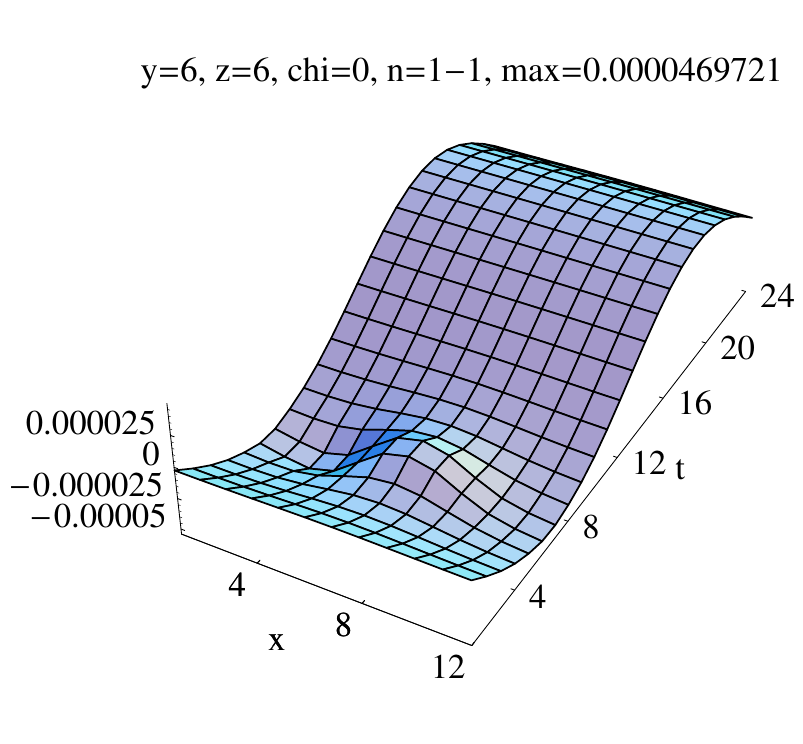}
		\includegraphics[width=.32\columnwidth]{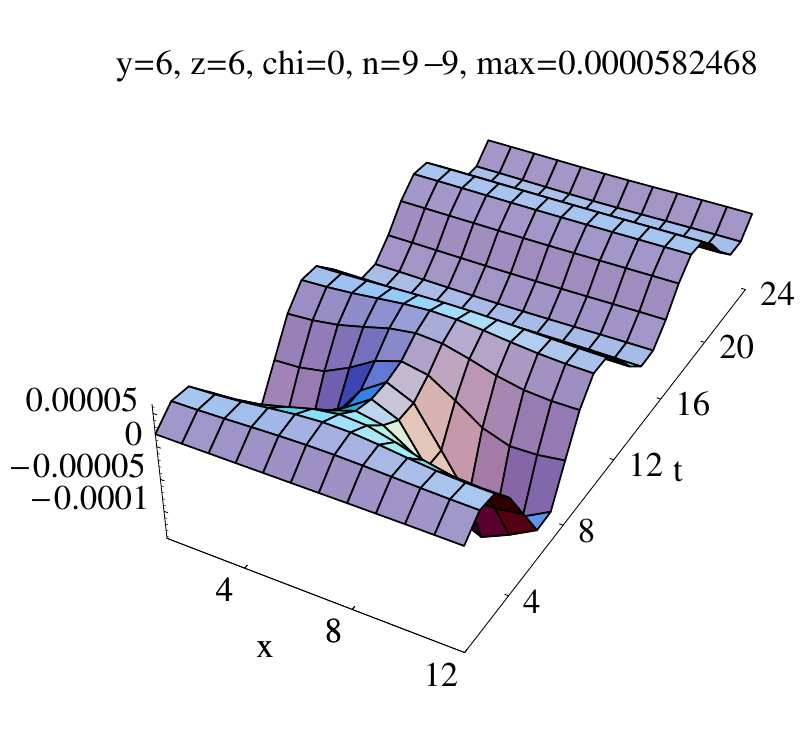}\\
		d)\includegraphics[width=.32\columnwidth]{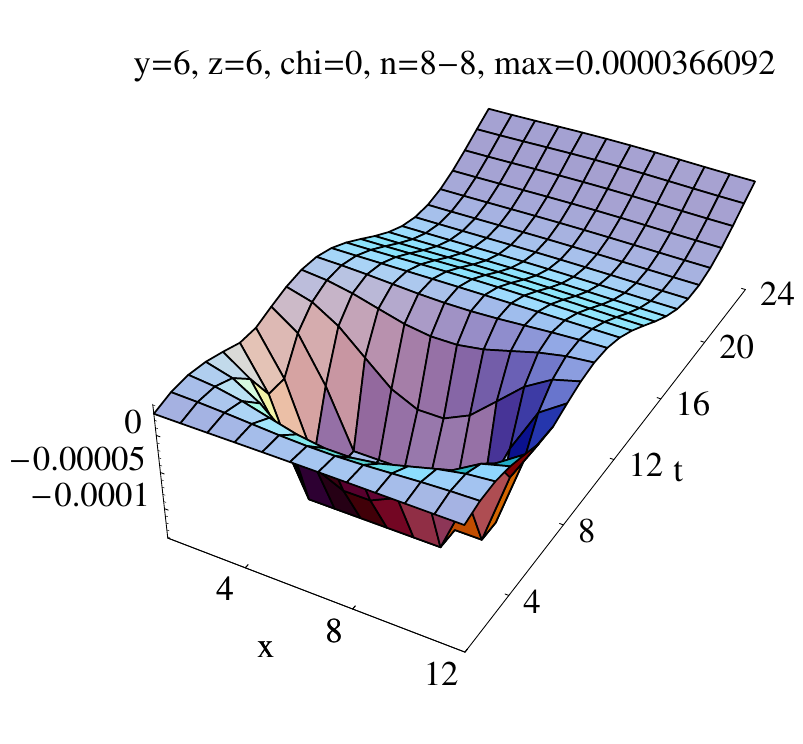}
		\includegraphics[width=.32\columnwidth]{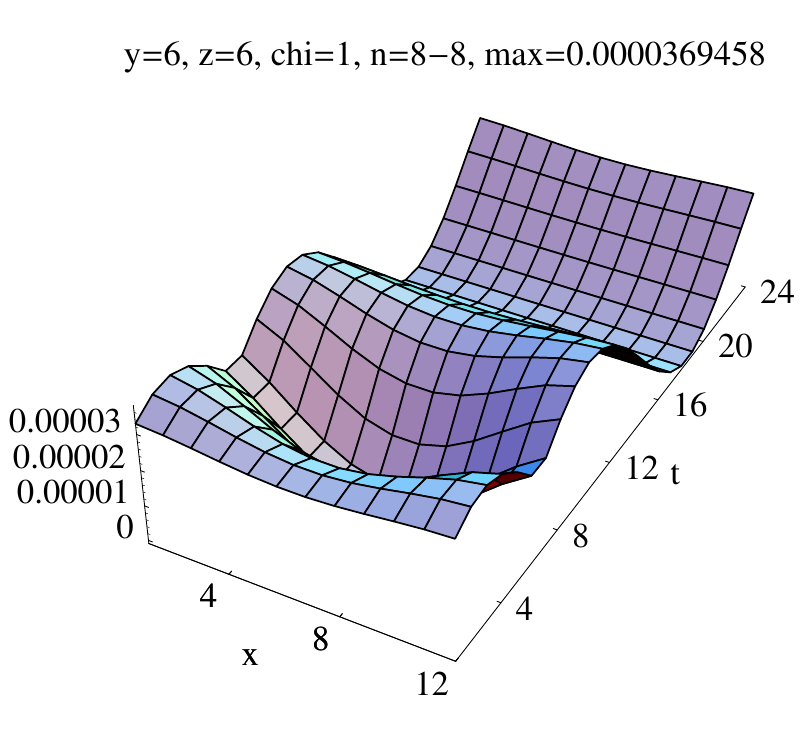}
		\includegraphics[width=.32\columnwidth]{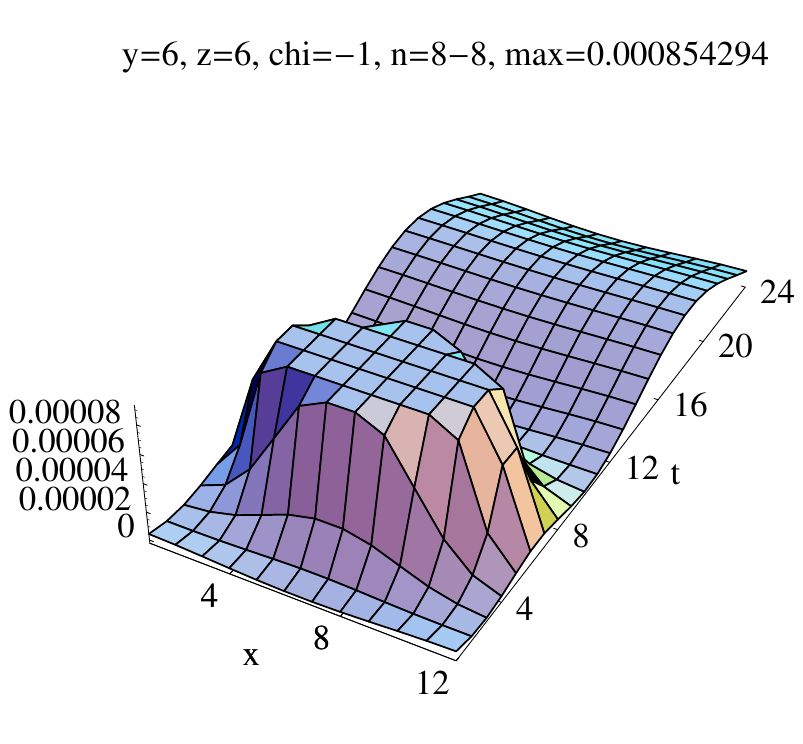}
	\caption{Chiral densities of overlap eigenmodes: a) zero mode
	(left), first (center), ninth (right) and b) eighth ($\rho_5$ left,
	$\rho_+$ center and $\rho_-$ right) nonzero modes for an instanton;
	c) and d) the same as a) and b) but for a spherical vortex.}
	\label{fig:sinsph}
\end{figure}

We start with spherical vortices and show that their effects on fermions is
pretty much the same as those of instantons. We have shown in
Fig.~\ref{fig:finitesizespher} that they attract a zero mode and its scalar
density peaks at the vortex surface. We interpreted the nonzero modes as
eigenmodes of the free Dirac operator, which are shifted slightly because
of their interaction with the nontrivial gauge field content. In
Fig.~\ref{fig:oevlsph}a we see that a single instanton has nearly exactly
the same eigenvalues as a single spherical vortex. In Fig.~\ref{fig:sinsph}
we show that even the chiral densities of the lowest eigenmodes distribute
similarly, except for the fact that the response of the fermions to the
spherical vortex is squeezed in the time direction, since the vortex is
localized in a single time slice ($t=5$). Another interesting issue is that
the nonzero eigenmodes show nice plane wave oscillations, like the free
eigenmodes in Fig.~\ref{fig:freeoverlap}, mode \#8 however shows some
similarity to the zero mode, its eigenvalue is also clearly enhanced
compared to the free spectrum. This is only a side remark however, as we
are not sure how to interpret this and it does not seem to be important for
the creation of near-zero modes since we observe the same effect for
instantons.

\begin{figure}
	\centering
		a)\includegraphics[width=.32\columnwidth]{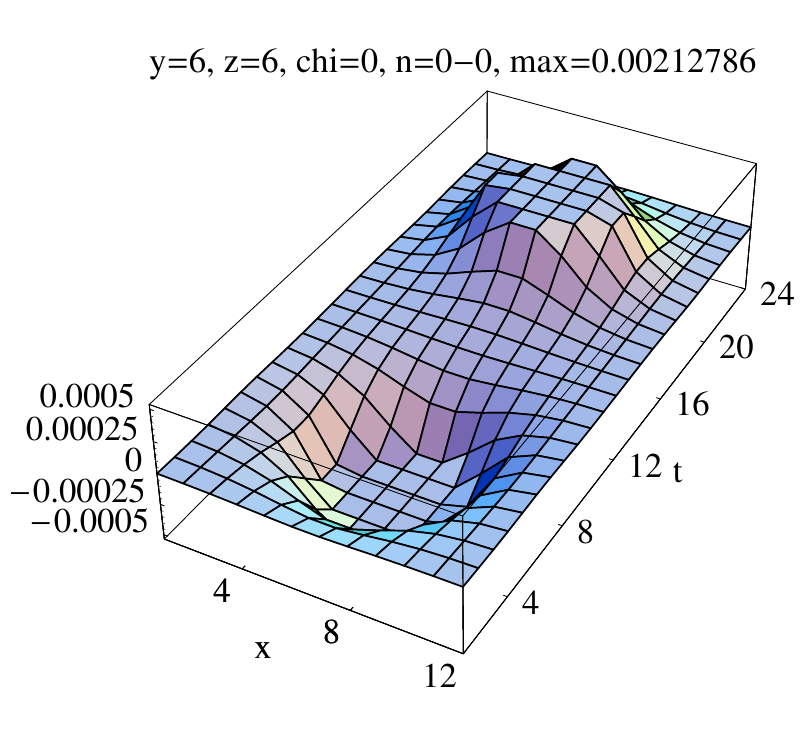}
		\includegraphics[width=.32\columnwidth]{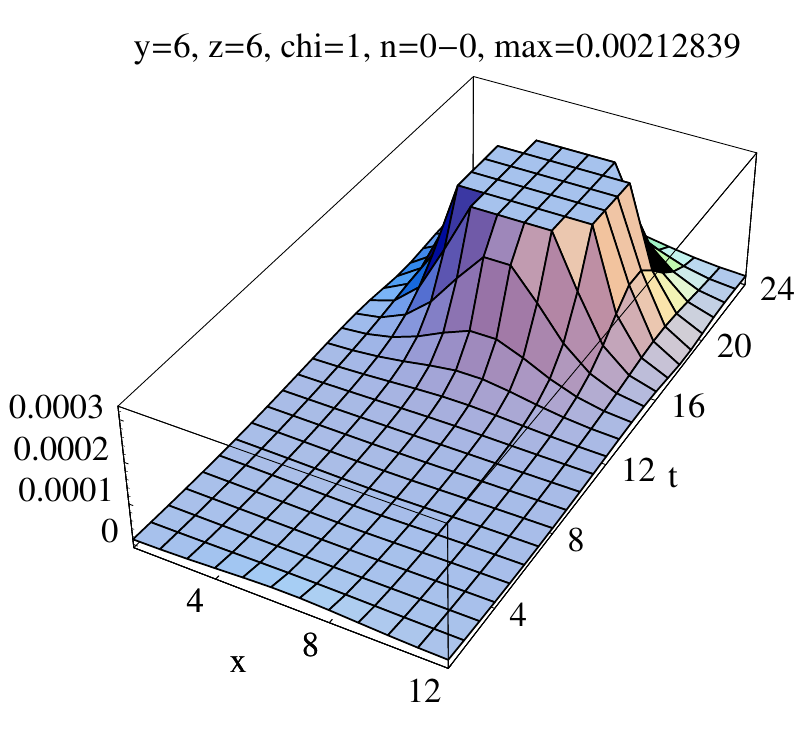}
		\includegraphics[width=.32\columnwidth]{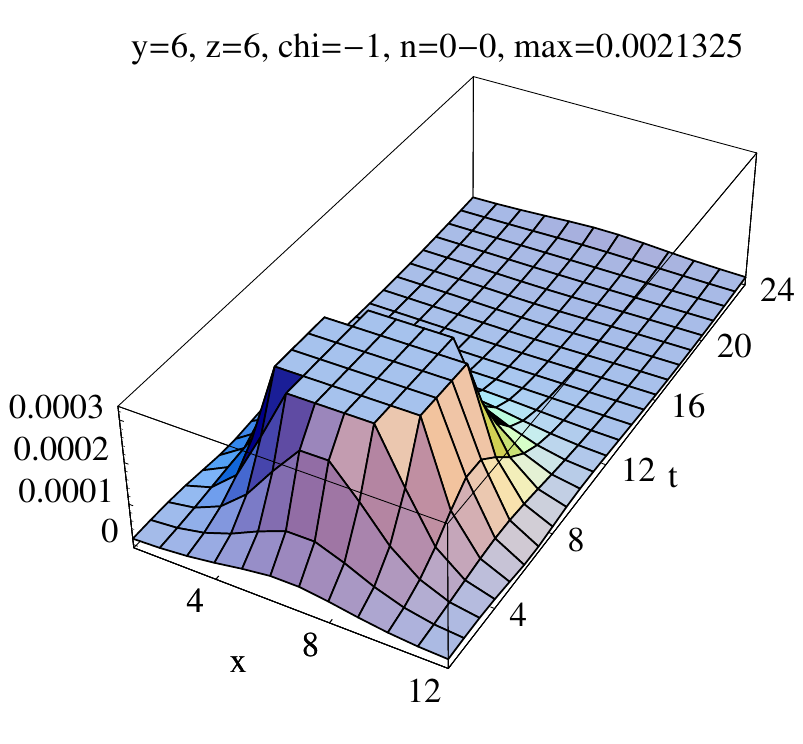}\\
		b)\includegraphics[width=.32\columnwidth]{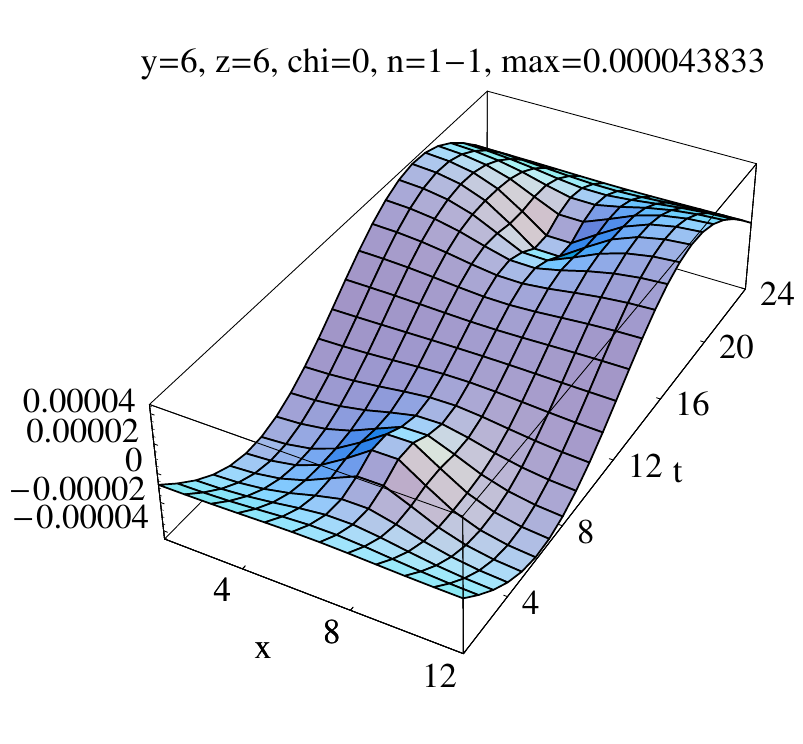}
		\includegraphics[width=.32\columnwidth]{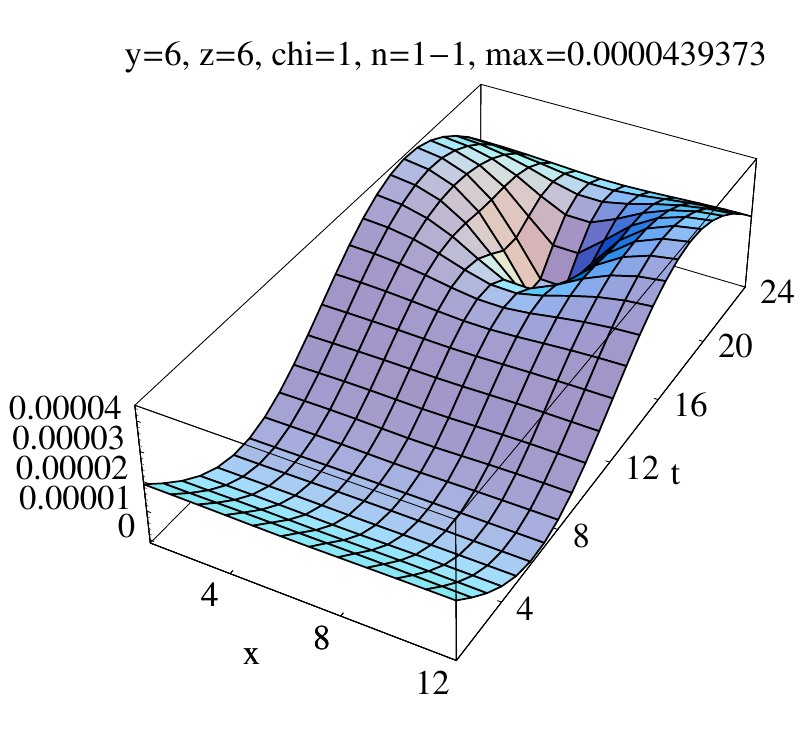}
		\includegraphics[width=.32\columnwidth]{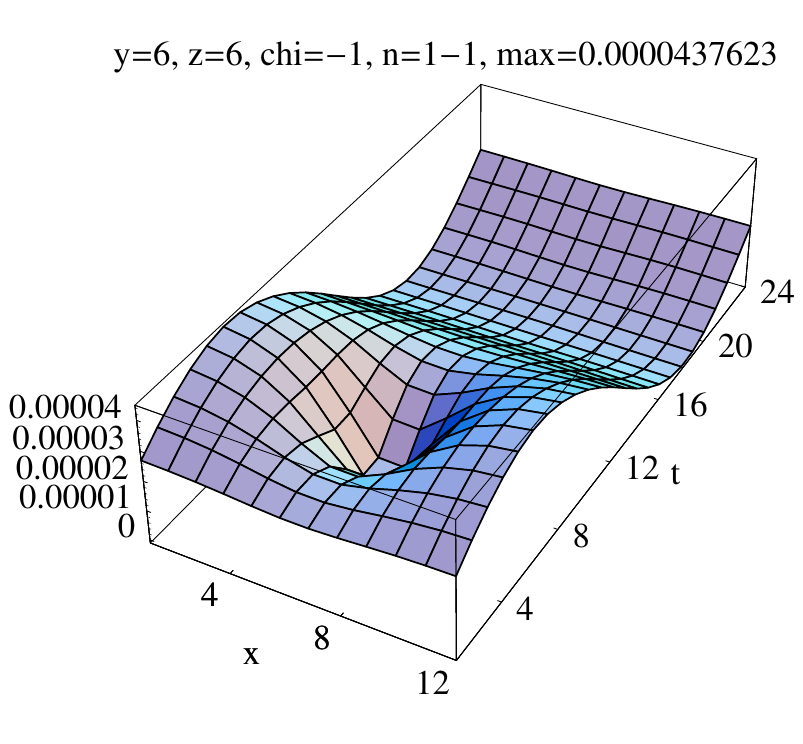}\\
		c)\includegraphics[width=.32\columnwidth]{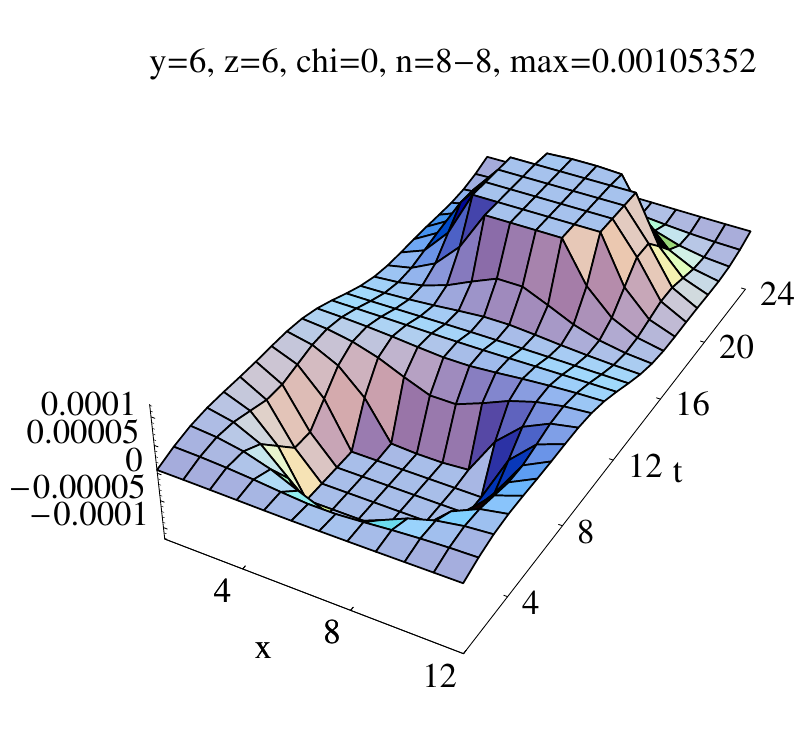}
		\includegraphics[width=.32\columnwidth]{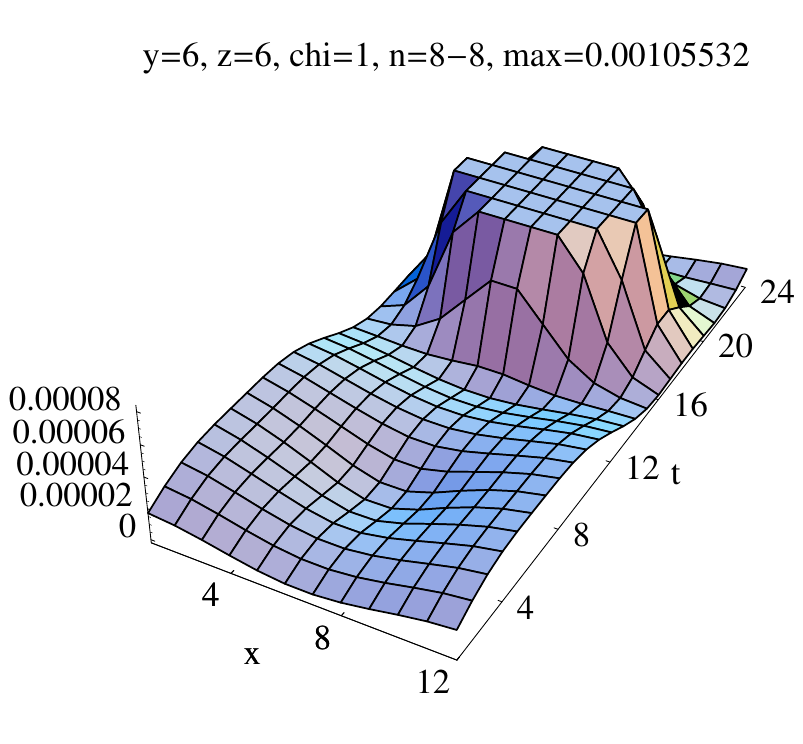}
		\includegraphics[width=.32\columnwidth]{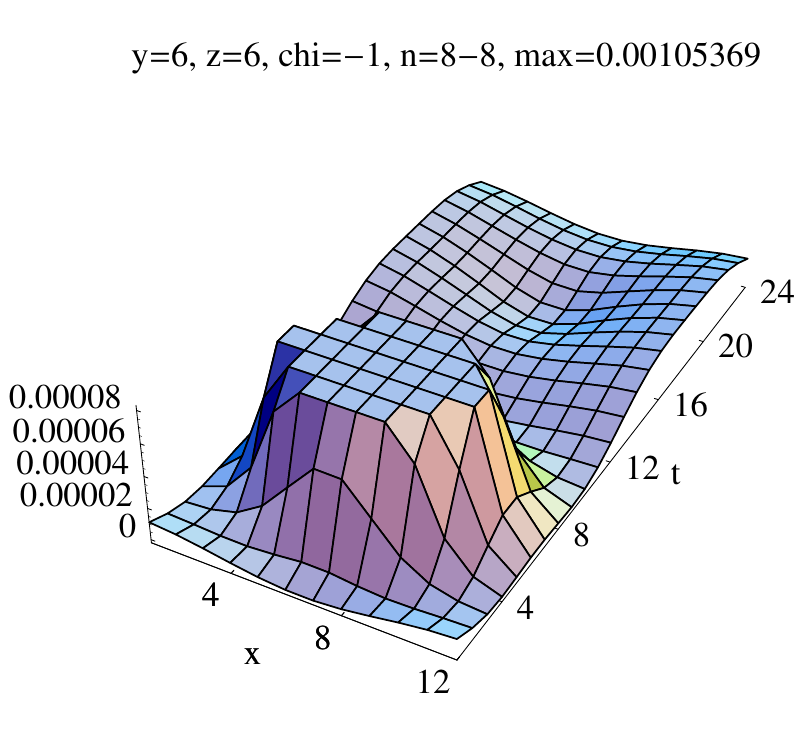}\\
		d)\includegraphics[width=.32\columnwidth]{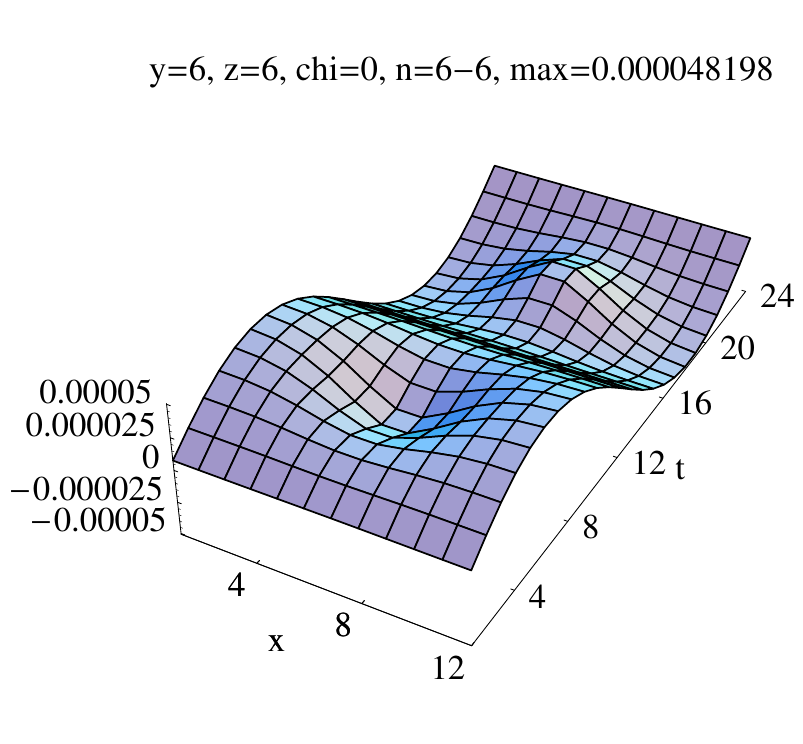}
		\includegraphics[width=.32\columnwidth]{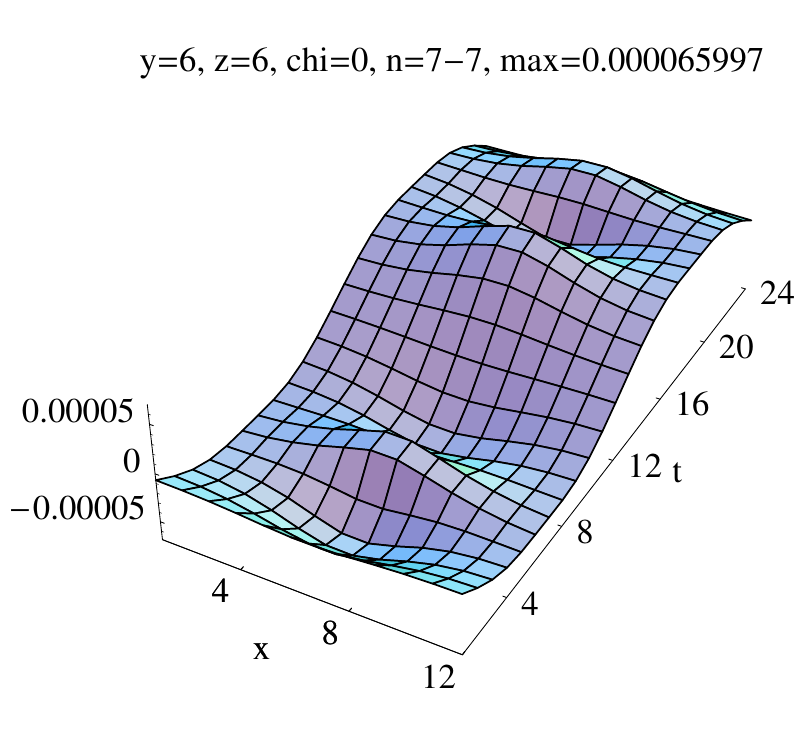}
		\includegraphics[width=.32\columnwidth]{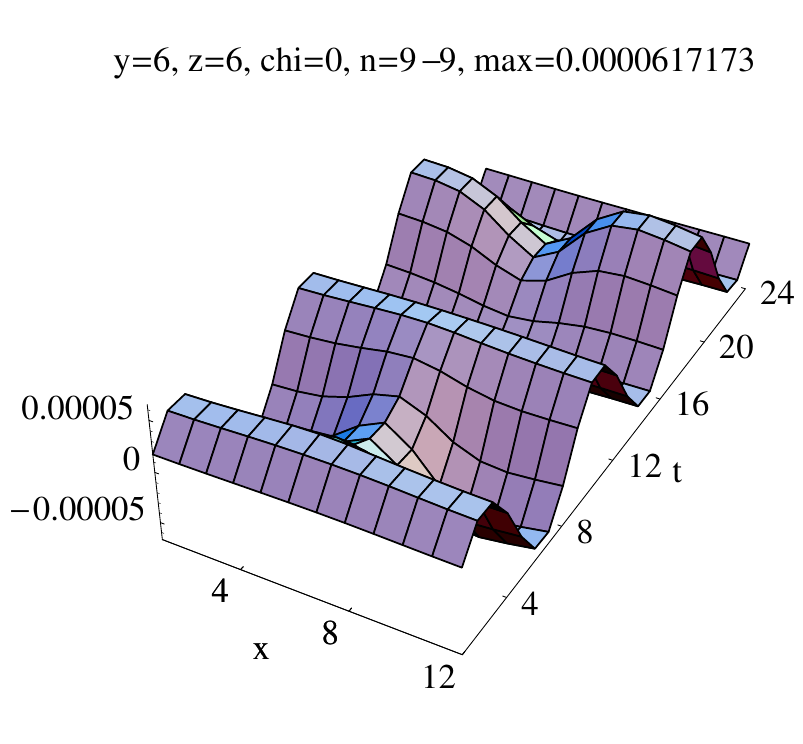}
	\caption{Chiral densities ($\rho_5$ left, $\rho_+$ center and
	$\rho_-$ right column) of the a) lowest (near-zero), b)
	second-lowest (nonzero) and c) eighth (nonzero) eigenmode of the
	overlap Dirac operator for an instanton--anti-instanton pair. d)
	$\rho_5$ of the sixth (left), seventh (center) and ninth (right)
	eigenmode.}
	\label{fig:iai}
\end{figure}

\begin{figure}
	\centering
		a)\includegraphics[width=.32\columnwidth]{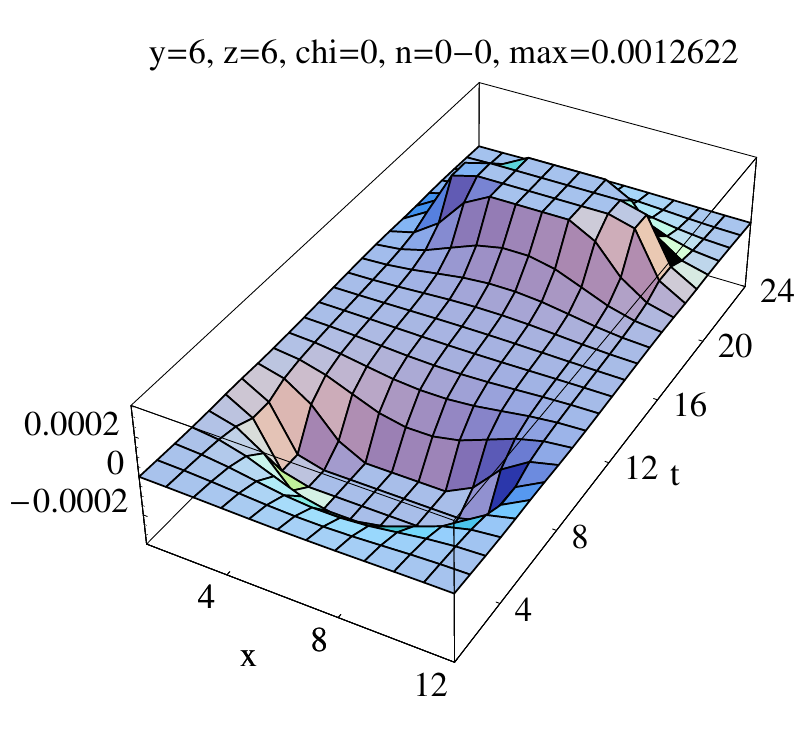}
		\includegraphics[width=.32\columnwidth]{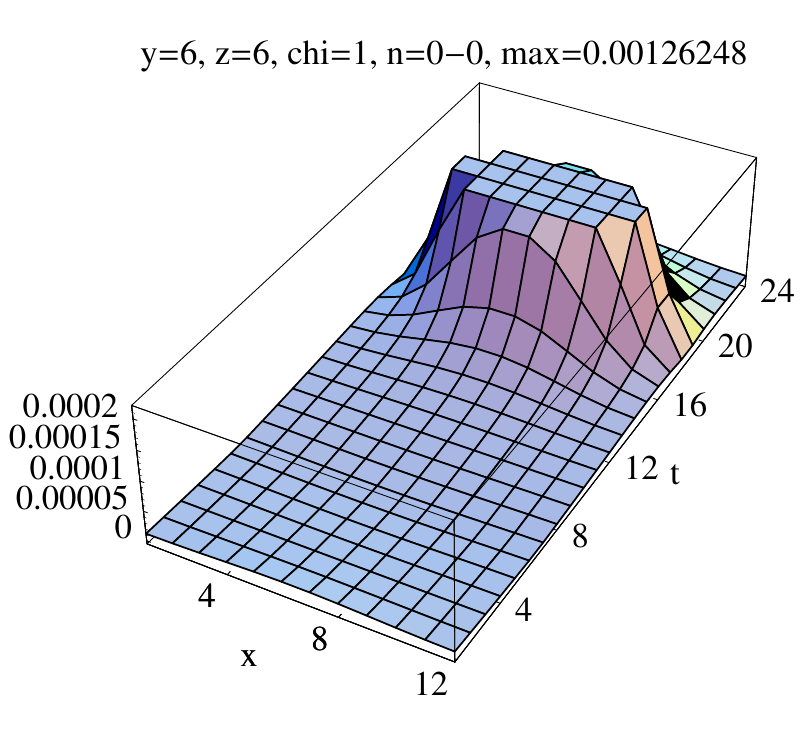}
		\includegraphics[width=.32\columnwidth]{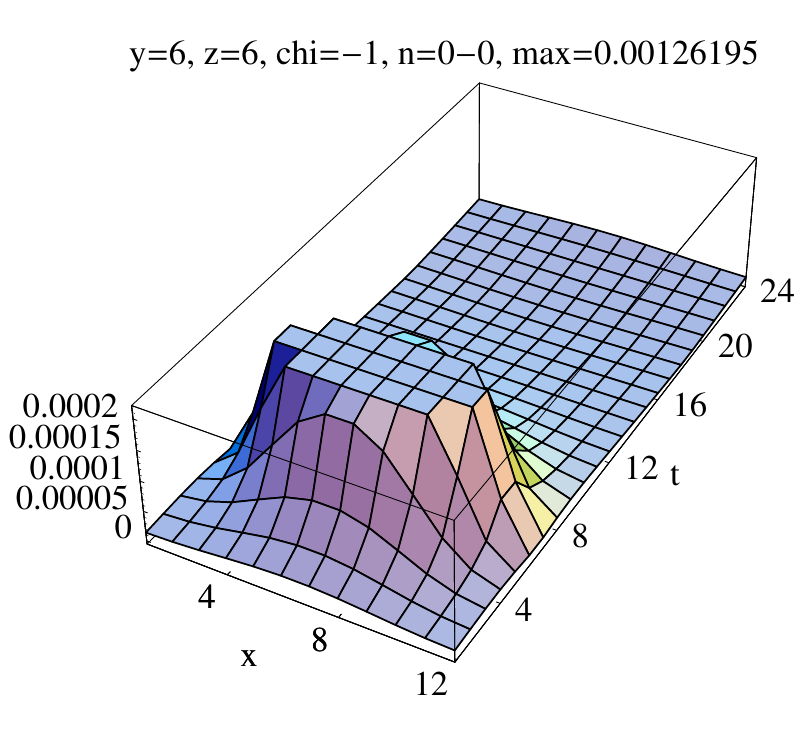}\\
	   b)\includegraphics[width=.32\columnwidth]{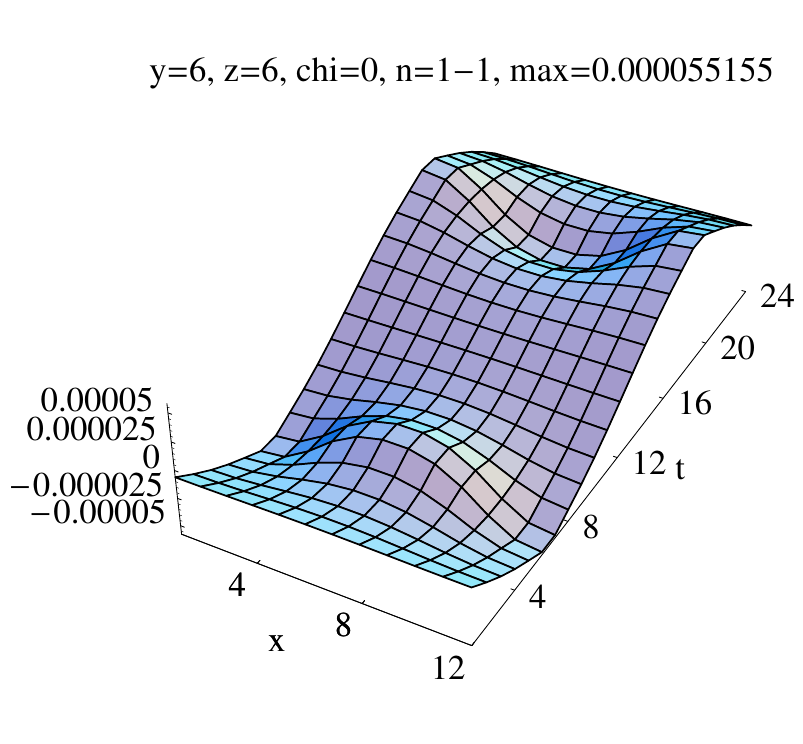}
		\includegraphics[width=.32\columnwidth]{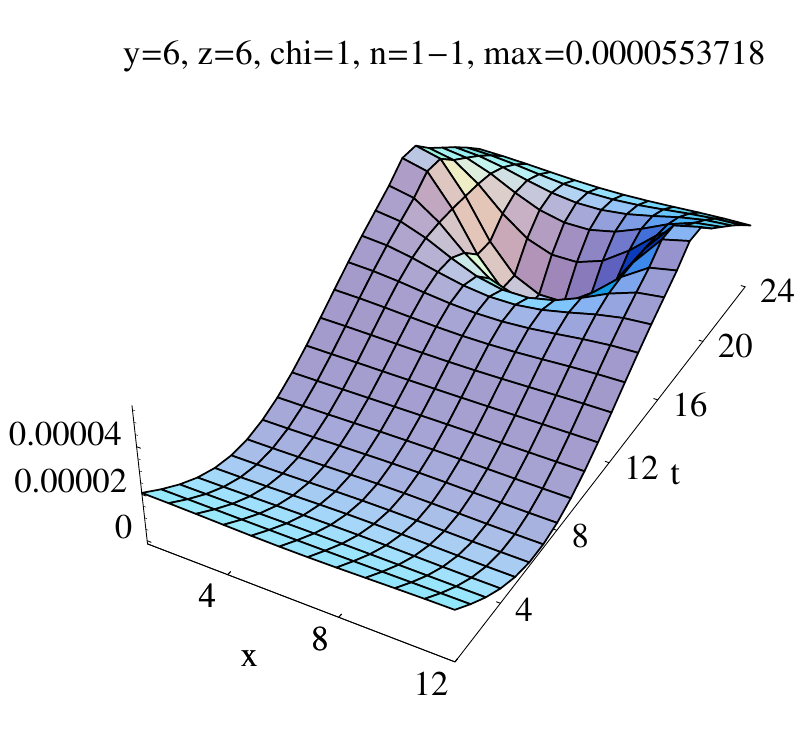}
		\includegraphics[width=.32\columnwidth]{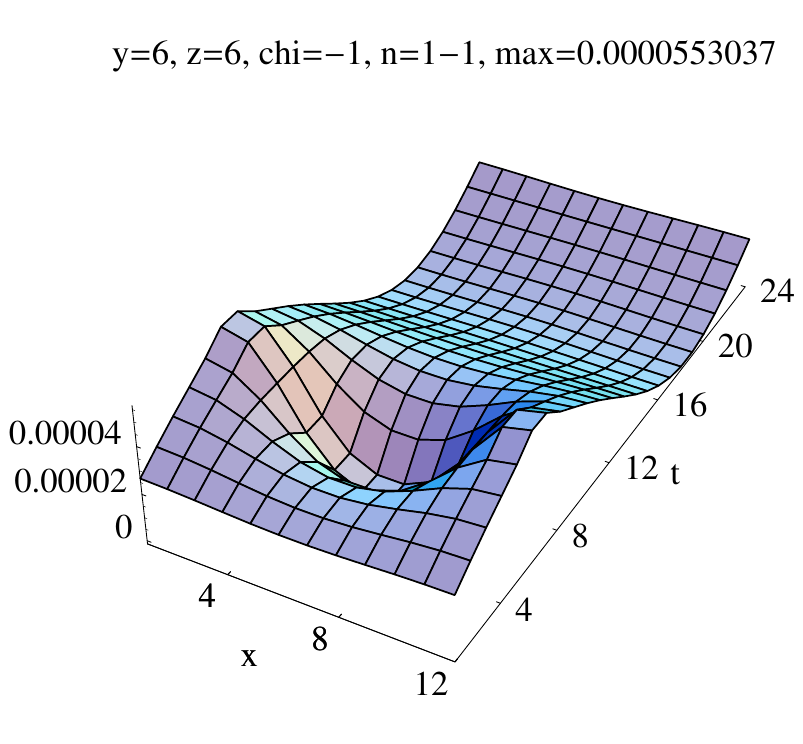}\\
		c)\includegraphics[width=.32\columnwidth]{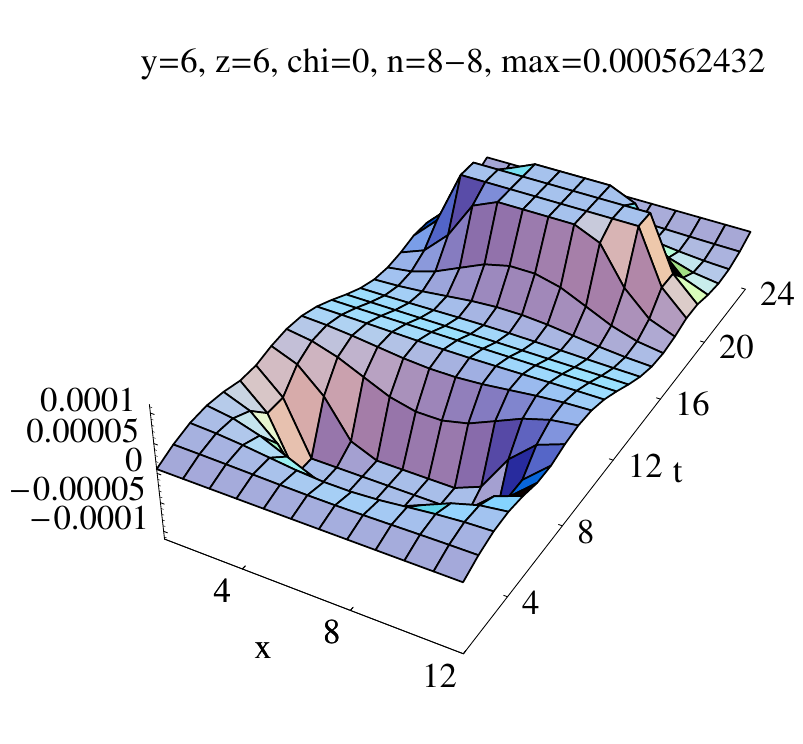}
		\includegraphics[width=.32\columnwidth]{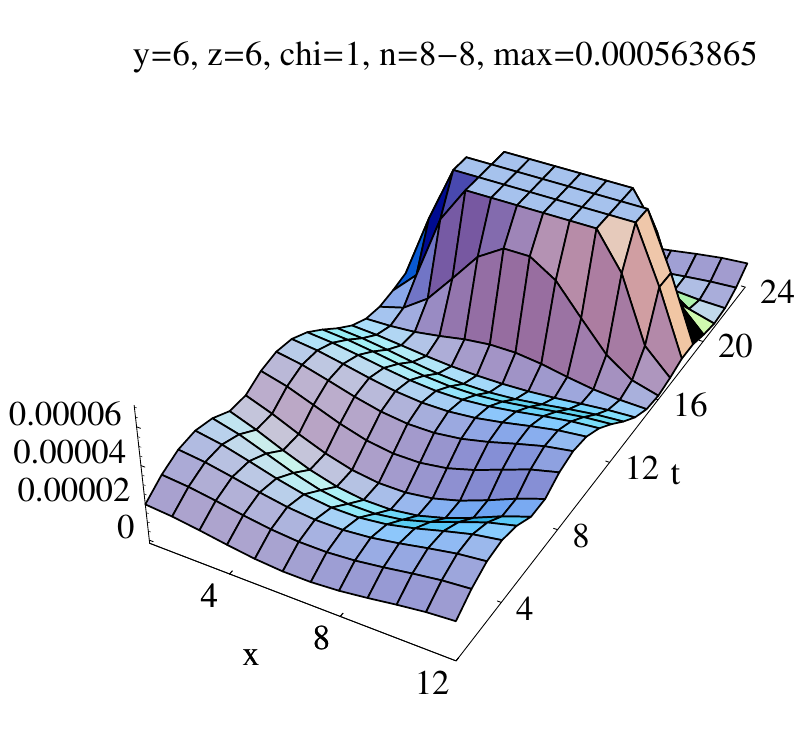}
		\includegraphics[width=.32\columnwidth]{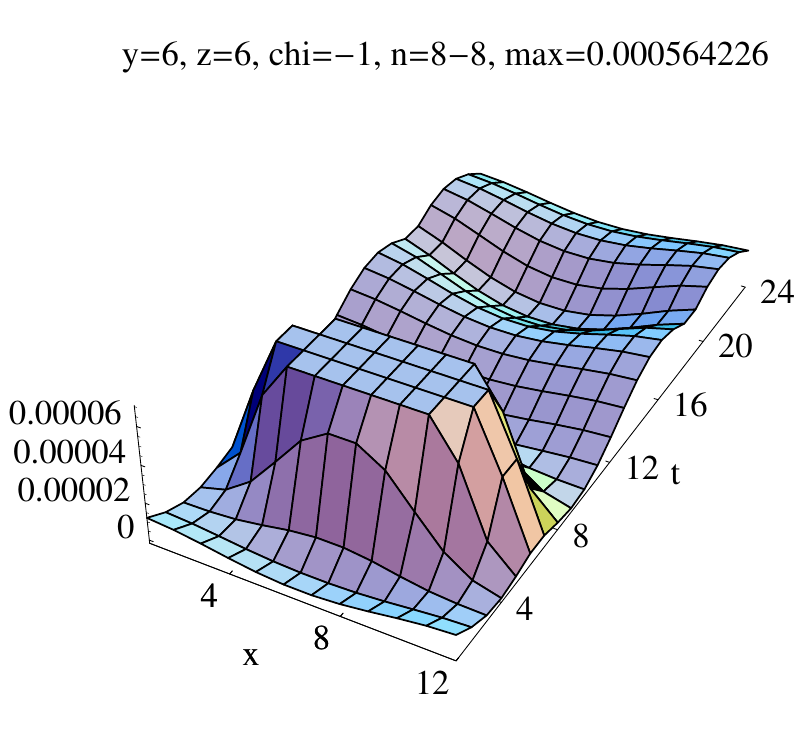}\\
	   d)\includegraphics[width=.32\columnwidth]{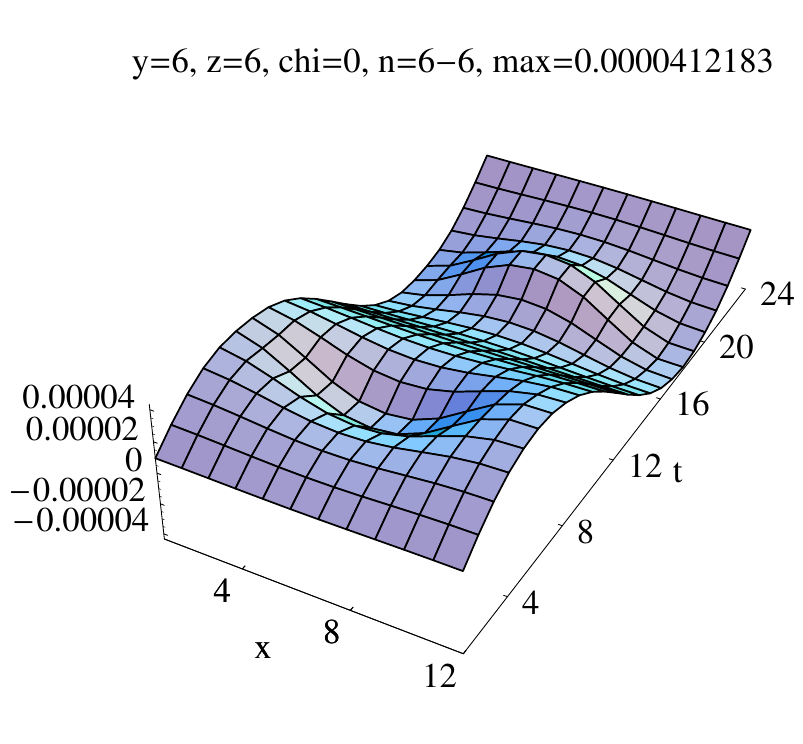}
		\includegraphics[width=.32\columnwidth]{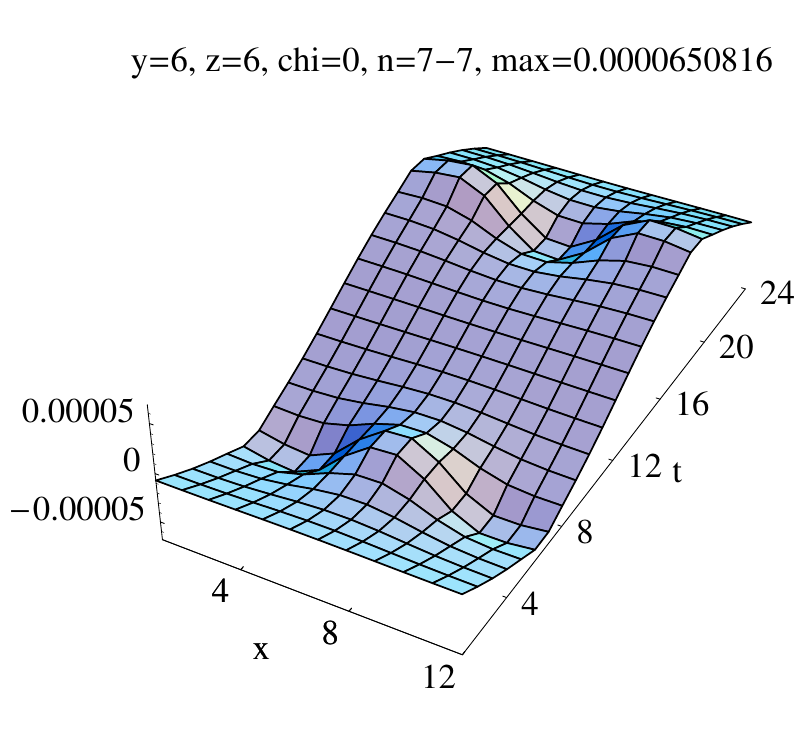}
		\includegraphics[width=.32\columnwidth]{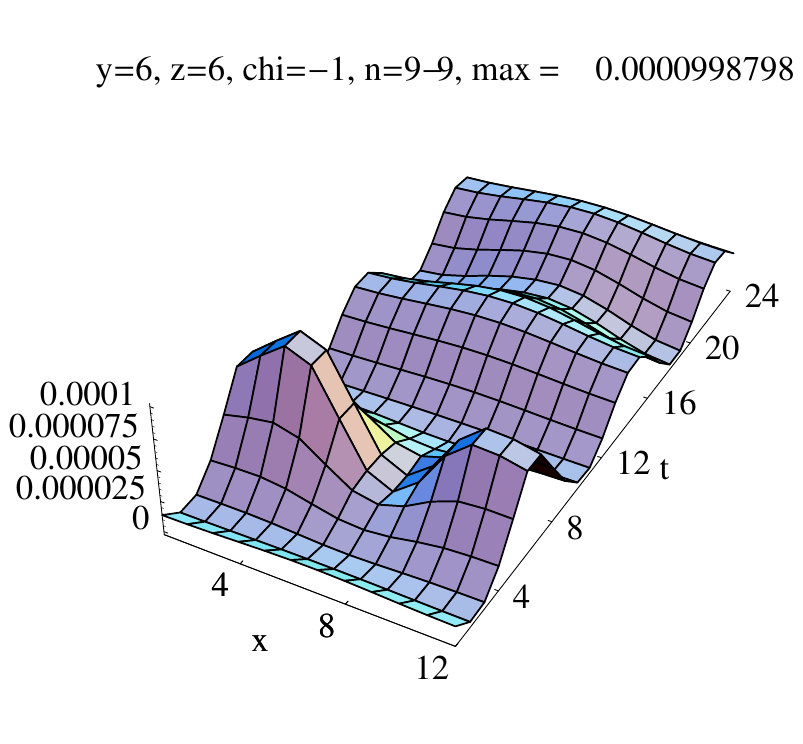}
	\caption{Same as Fig.~\protect\ref{fig:iai} but for a spherical
	vortex--anti-vortex pair.
	Chiral densities ($\rho_5$ left, $\rho_+$ center and
	$\rho_-$ right column) of the a) lowest (near-zero), b)
	second-lowest (nonzero) and c) eighth (nonzero) eigenmode.
	d) $\rho_5$ of the sixth (left), seventh (center) and ninth (right)
	eigenmode.}
	\label{fig:tat}
\end{figure}

\begin{figure}
	\centering
		a)\includegraphics[width=.32\columnwidth]{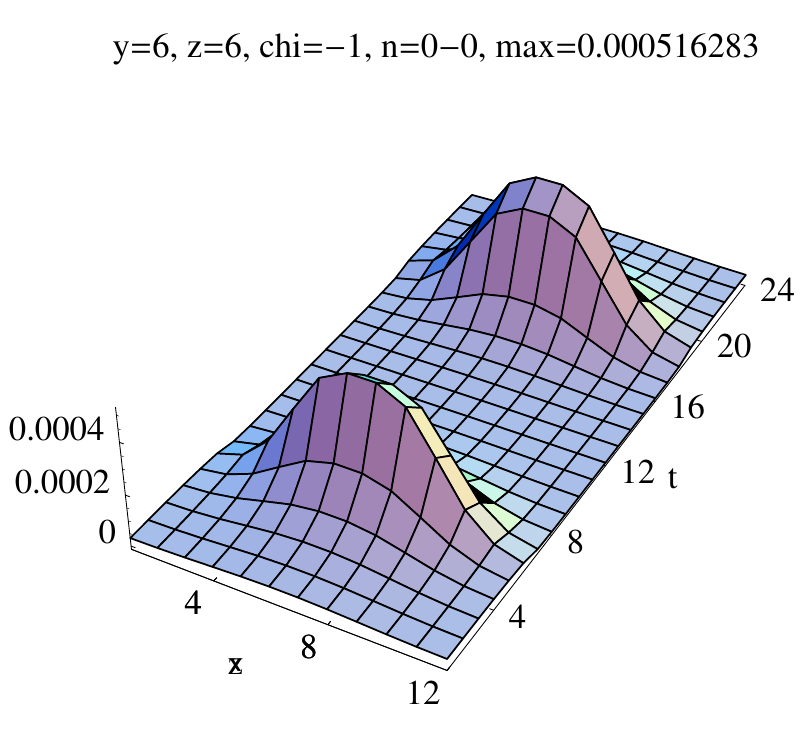}
		\includegraphics[width=.32\columnwidth]{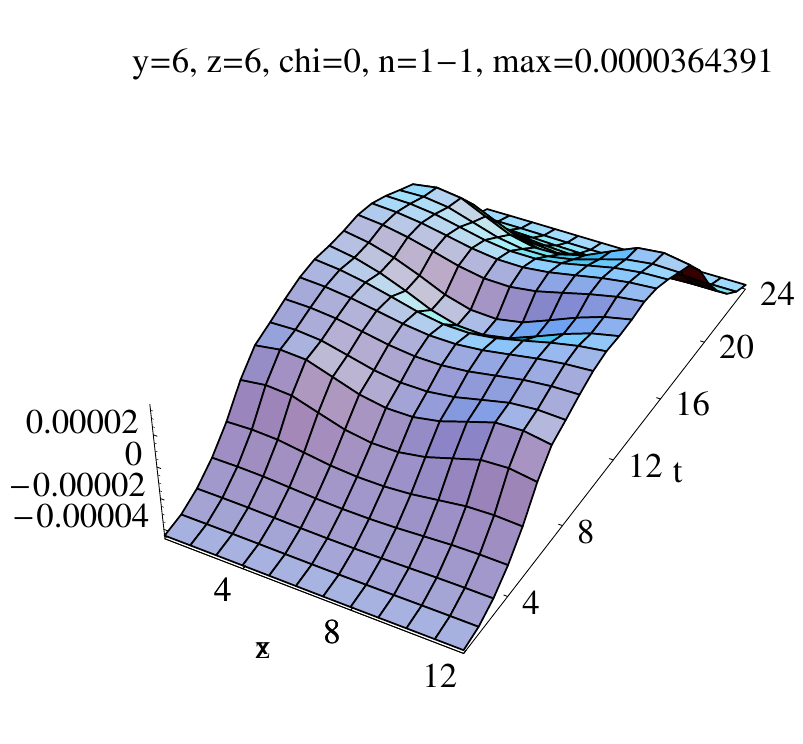}
		\includegraphics[width=.32\columnwidth]{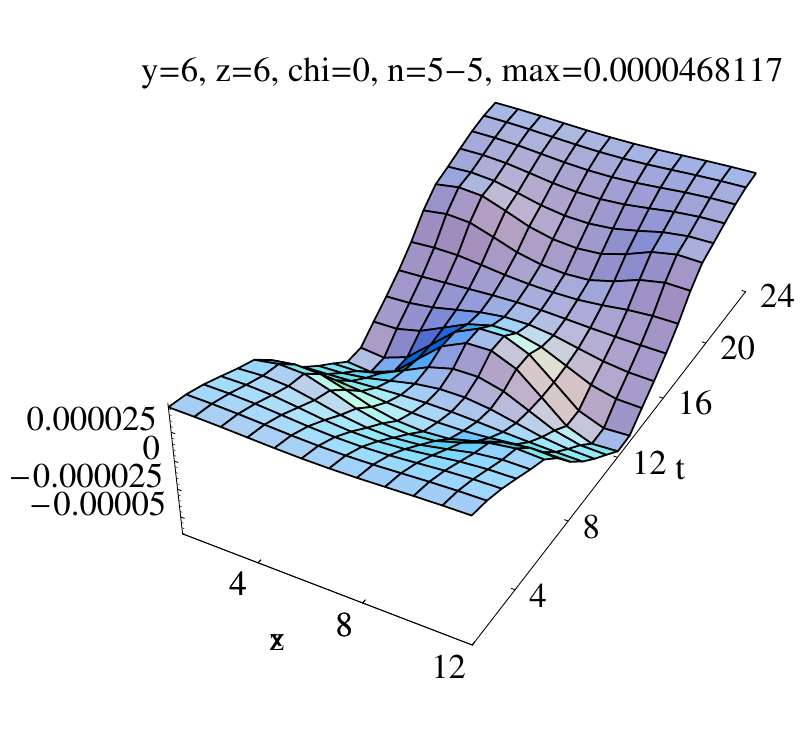}\\
	   b)\includegraphics[width=.32\columnwidth]{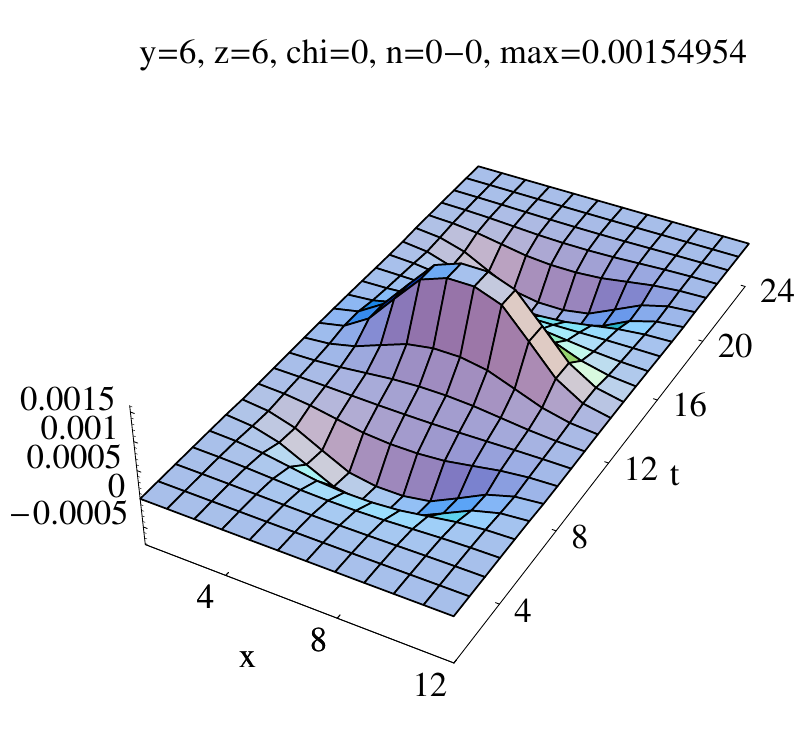}
		\includegraphics[width=.32\columnwidth]{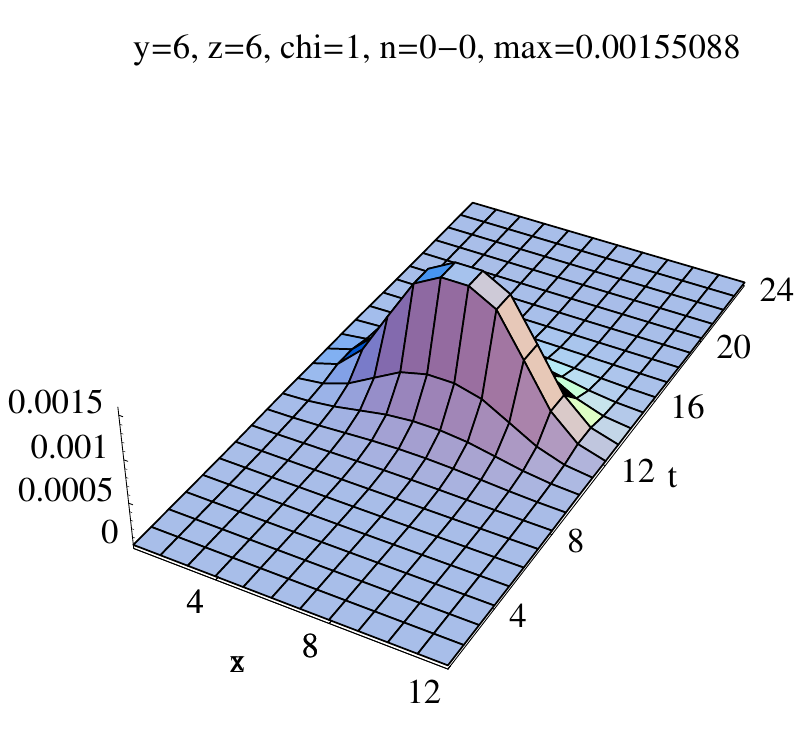}
		\includegraphics[width=.32\columnwidth]{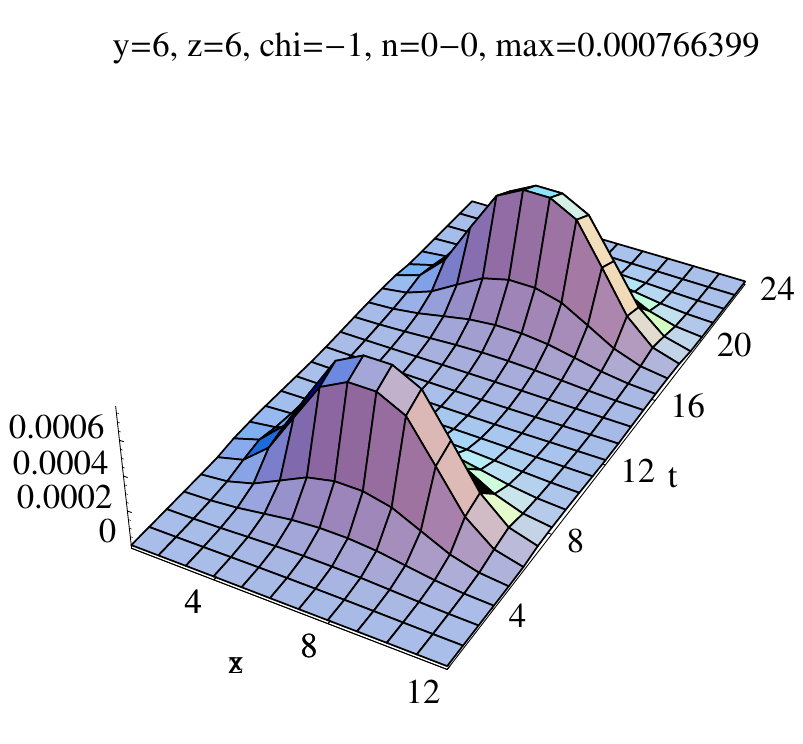}\\
		c)\includegraphics[width=.32\columnwidth]{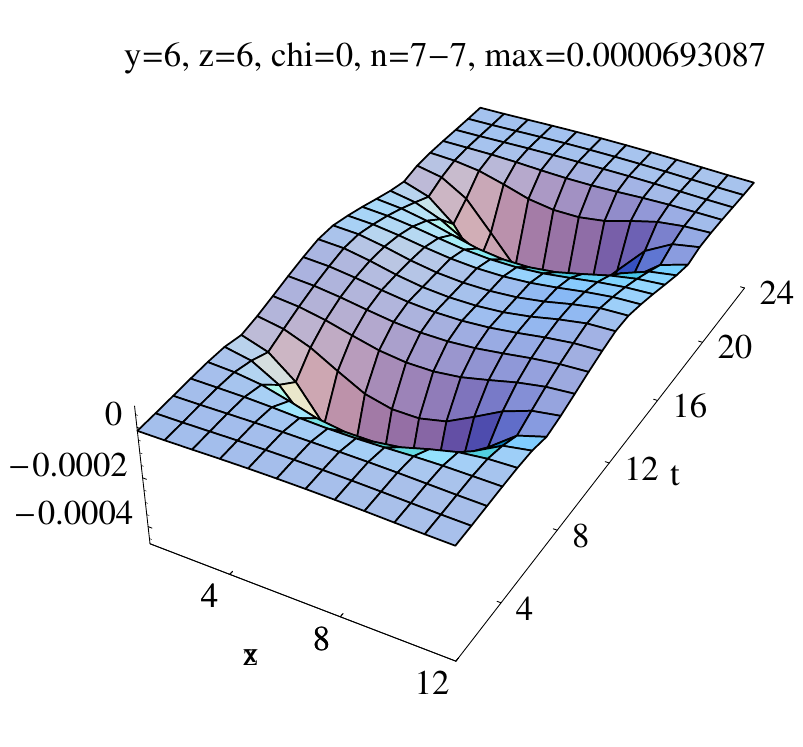}
		\includegraphics[width=.32\columnwidth]{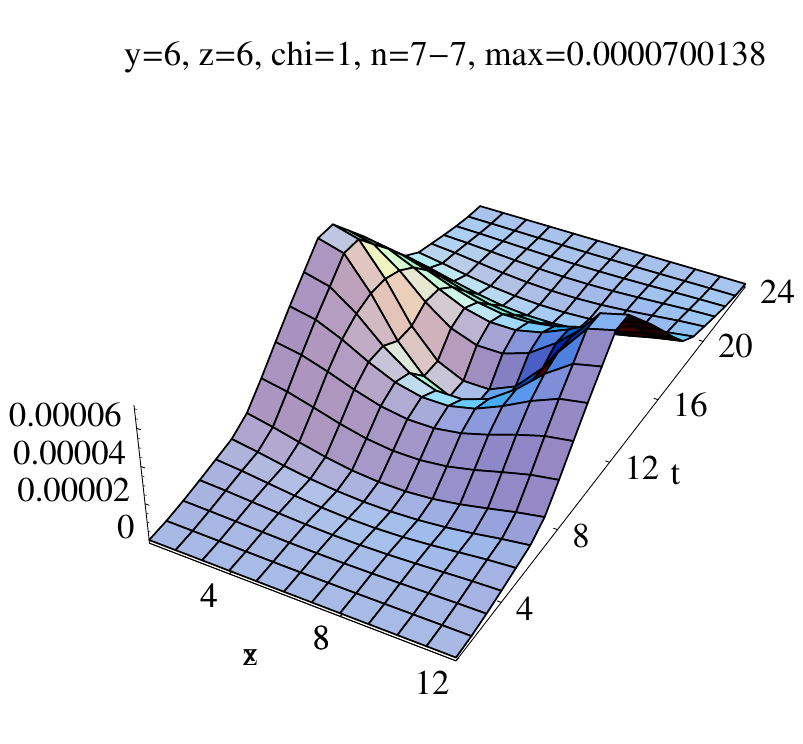}
		\includegraphics[width=.32\columnwidth]{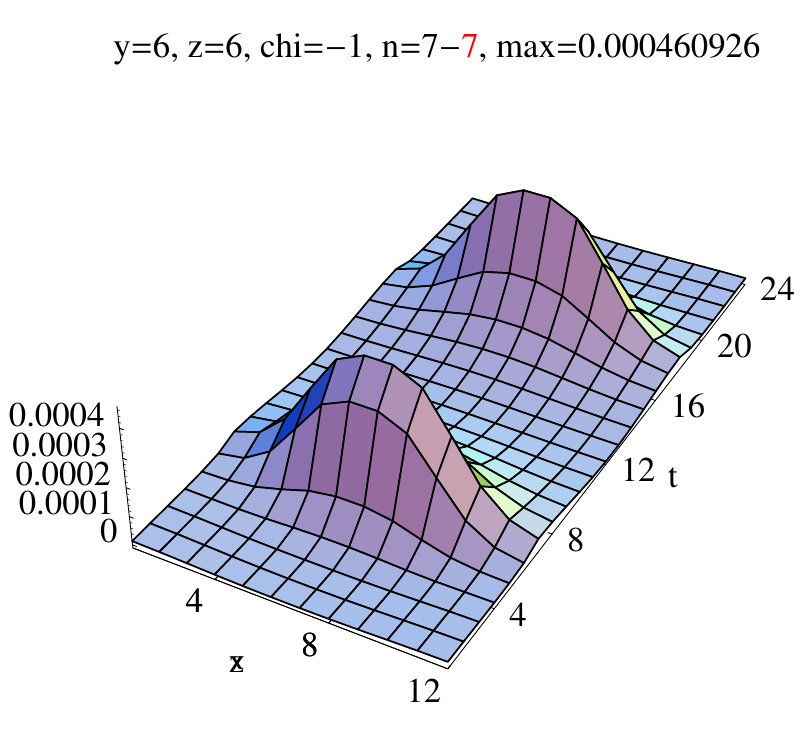}\\
	   d)\includegraphics[width=.32\columnwidth]{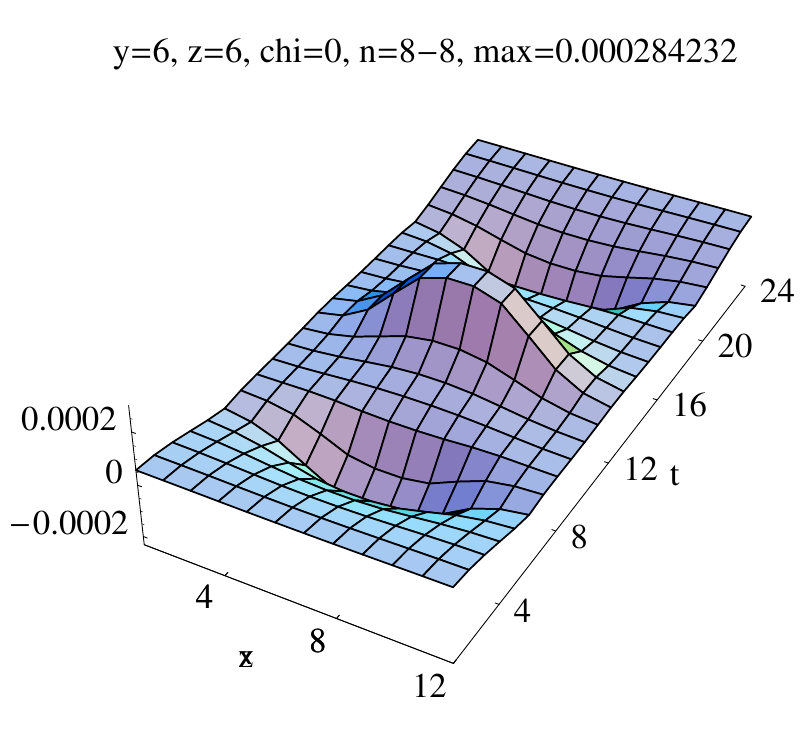}
		\includegraphics[width=.32\columnwidth]{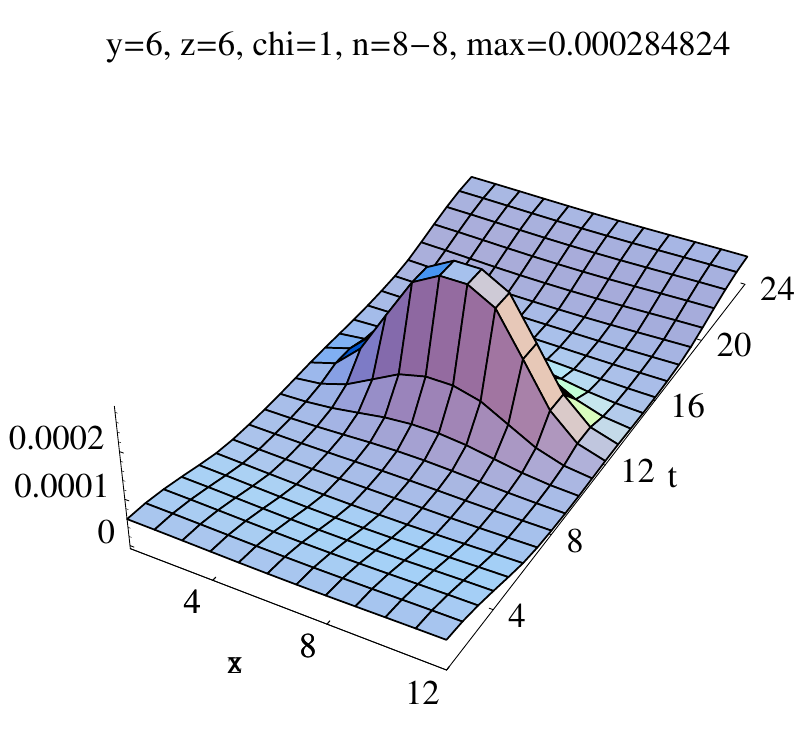}
		\includegraphics[width=.32\columnwidth]{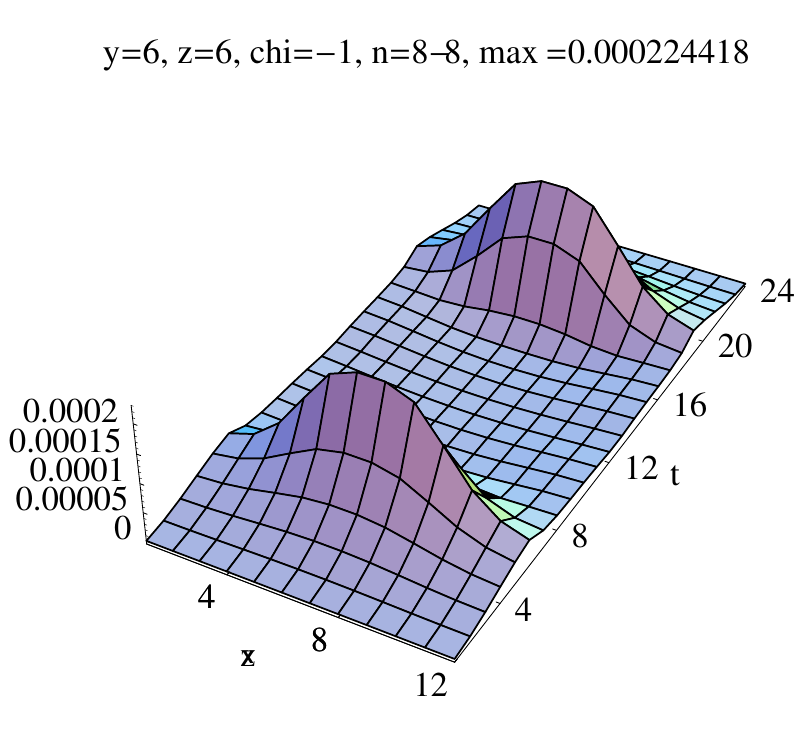}
	\caption{Chiral densities of overlap eigenmodes for spherical
	vortices at $t=4$ and $t=20$ and an anti-vortex at $t=12$: a) zero
	mode $\rho_-\#0-$ (left), $\rho_5\#1$ (center) and $\rho_5\#5$
	(right). b) near-zero (second-lowest) mode \#0, c) \#7
	($\approx\#0-$) and d) \#8 ($\approx\#0$) with $\rho_5$ (left),
	$\rho_+$ (center) and $\rho_-$ (right).}
	\label{fig:3sph}
\end{figure}

We further plot the spectra of instanton--anti-instanton, spherical
vortex--anti-vortex and instanton--anti-vortex pairs in
Fig.~\ref{fig:oevlsph}a. We again see nearly exactly the same eigenvalues
for instanton or spherical vortex pairs, but now we get instead of two
would-be zero modes a pair of near-zero modes for each pair. The chiral density
plots in Fig.~\ref{fig:iai} for the instanton--anti-instanton pair and
Fig.~\ref{fig:tat} for the spherical vortex--anti-vortex pair show, besides
the similar densities, that the near-zero mode is a result of two chiral
parts corresponding to the two constituents of the pairs. The nonzero modes
can again be identified with the free overlap modes with the same side
remark for mode \#8.
In Fig.~\ref{fig:oevlsph}b we plot the eigenvalues of two
(anti-)instantons and two spherical (anti-)vortices giving topological
charge $Q=2$ ($Q=-2$) and therefore two zero modes, two vortex--anti-vortex
pairs with two near-zero modes and a configuration with two vortices and an
anti-vortex ({\it i.e.}, a single vortex plus one vortex--anti-vortex pair)
giving one zero mode ($Q=1$) and one near-zero mode. The chiral densities
for the last configuration in Fig.~\ref{fig:3sph} show that the zero mode
peaks at both spherical vortices, the near-zero mode again consists of two
chiral parts from the (second/would-be) zero mode of the two vortices and
the (would-be) zero mode of the anti-vortex. Now the modes \#7 and \#8
clearly deviate from the free eigenvalues showing similar densities as the
zero and the near-zero mode respectively. 

\begin{figure}[t]
	\centering
		a)\includegraphics[width=.48\columnwidth]{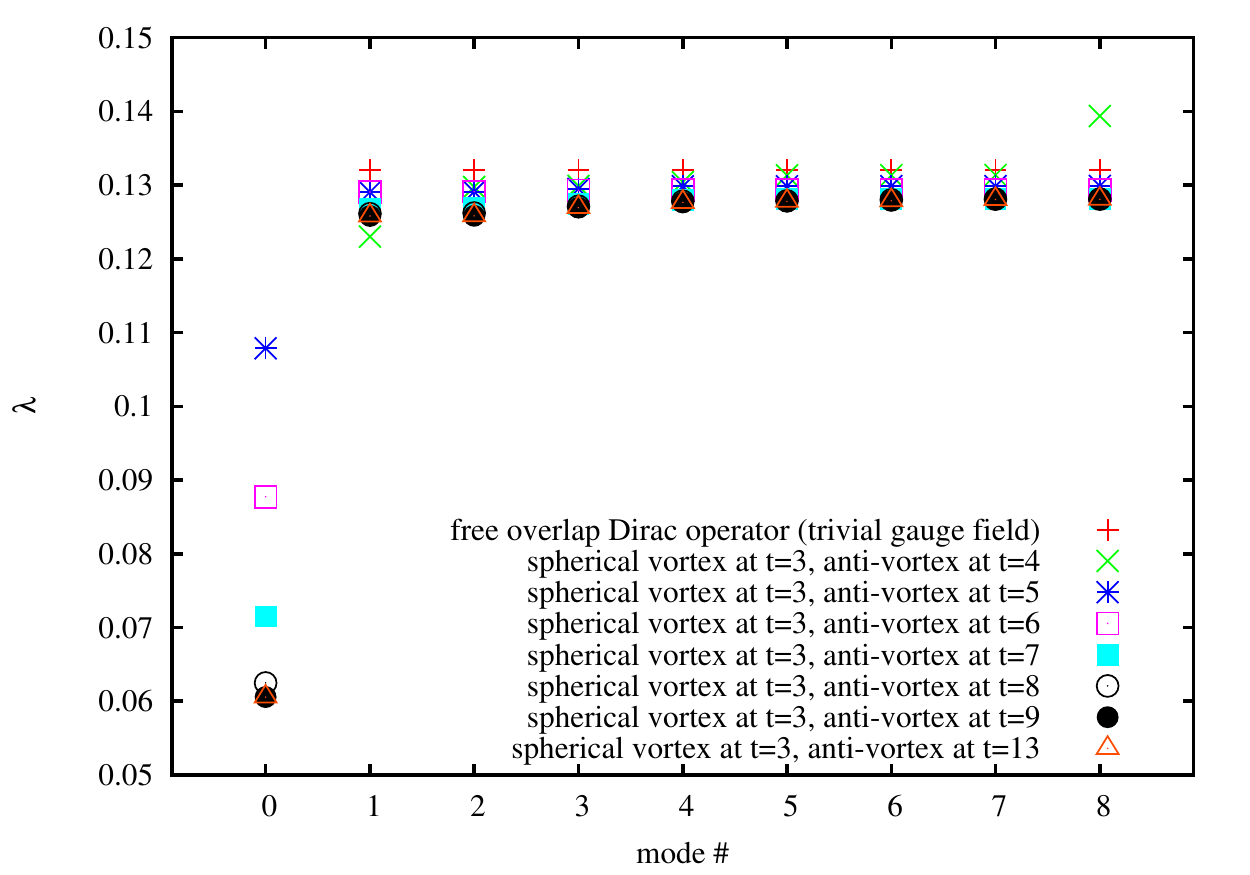}
		b)\includegraphics[width=.38\columnwidth]{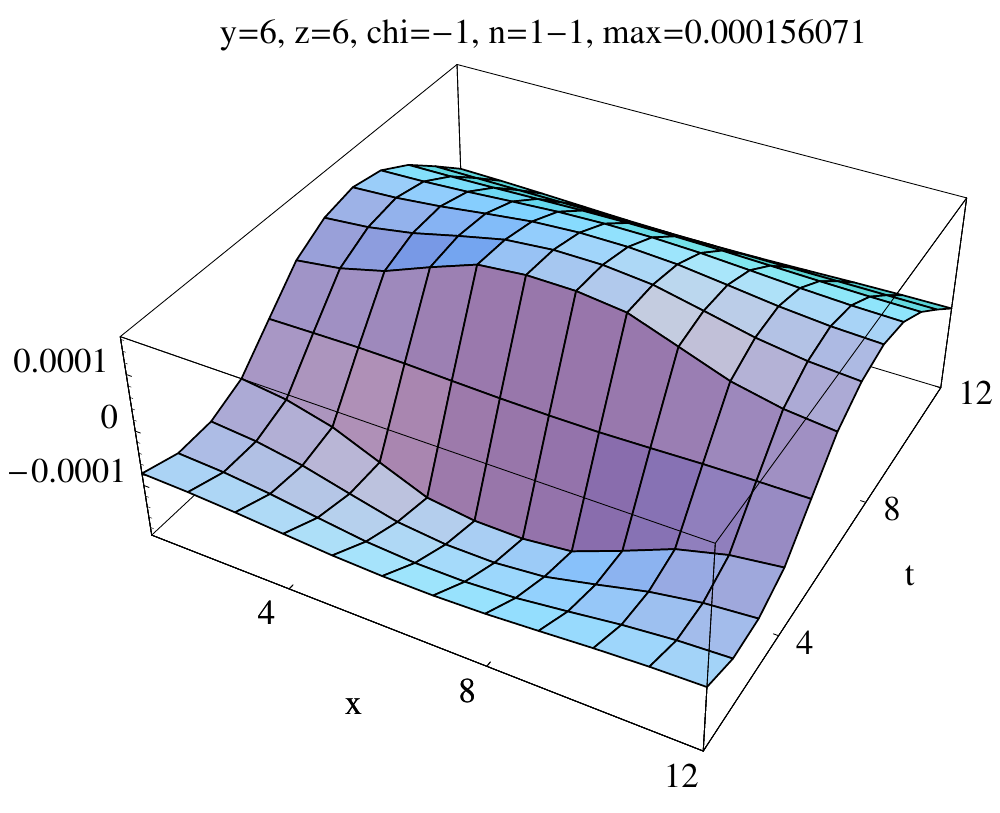}
		\caption{a) The lowest overlap eigenvalues for spherical vortex--anti-vortex
		pairs with varying distance compared to the eigenvalues of the free
	(overlap) Dirac operator. b) Chiral density of the lowest overlap eigenmode for
	spherical vortex and anti-vortex in neighboring time slices (t=3 and t=4).
Its plane wave behavior shows that this eigenmode is not a near-zero mode and hence it
appears as mode \#1 in a).} 
	\label{fig:ovdist1}
\end{figure}

\begin{figure}[t]
	\centering
		\includegraphics[width=.32\columnwidth]{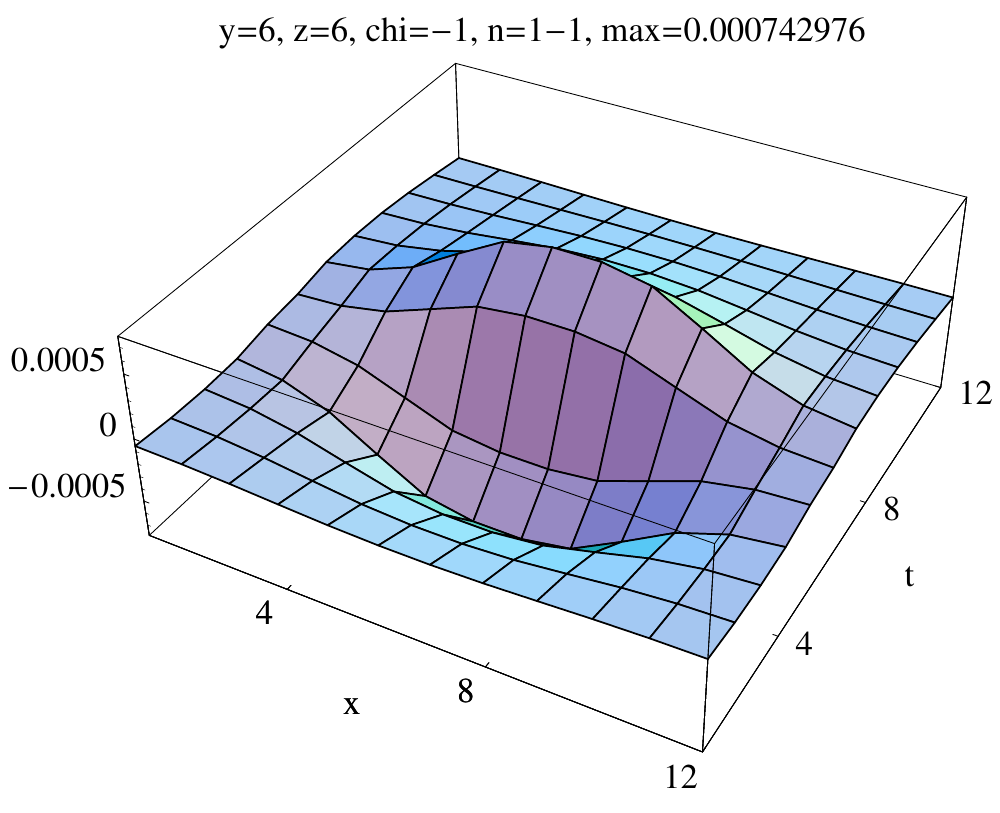}
		\includegraphics[width=.32\columnwidth]{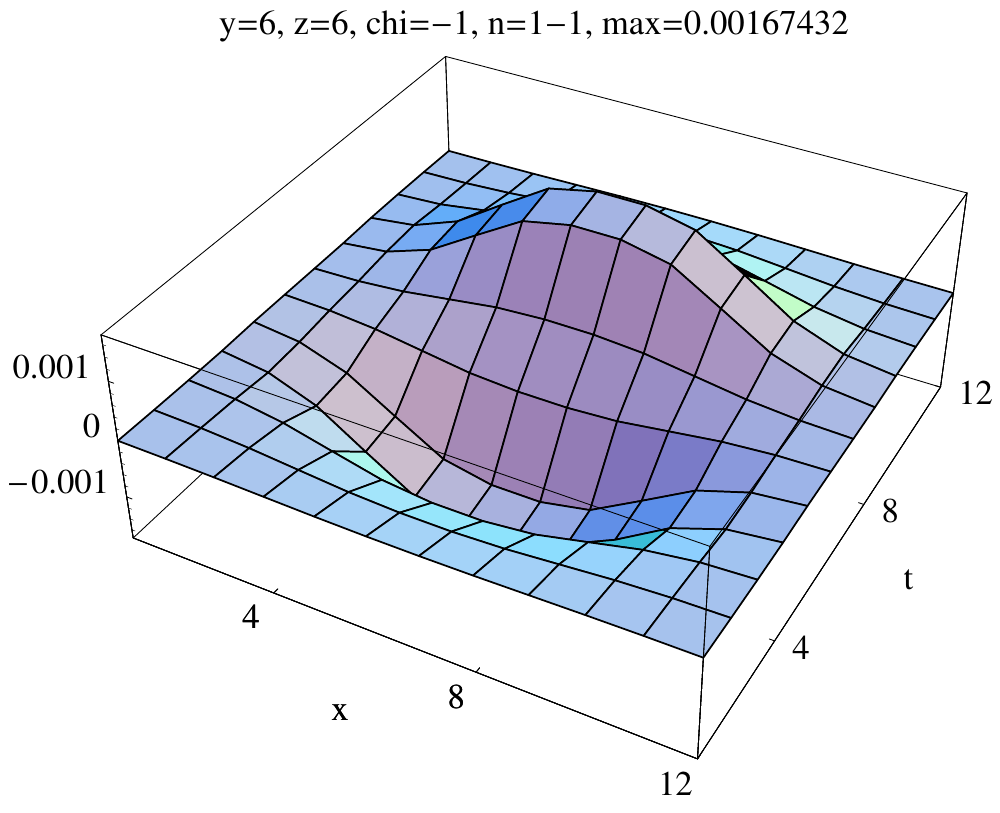}
		\includegraphics[width=.32\columnwidth]{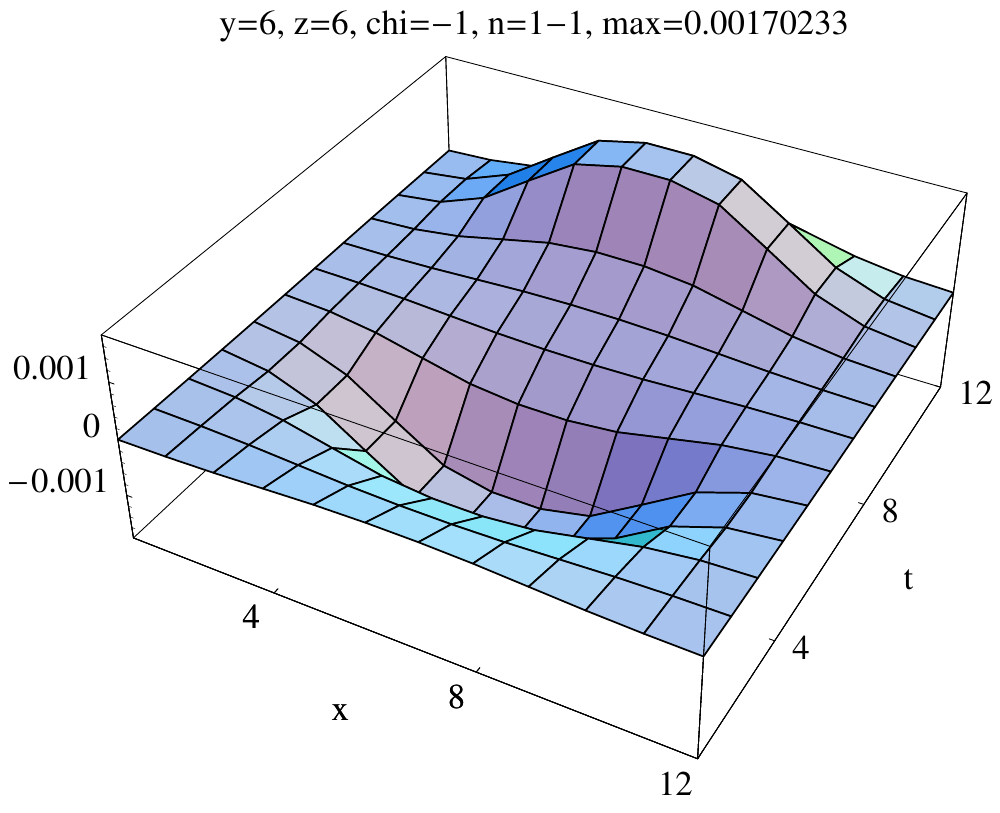}
	\caption{Chiral density of the overlap near-zero mode for a spherical vortex
	at t=3 and an anti-vortex at t=5, t=7 and t=9.}
	\label{fig:ovdist2}
\end{figure}

Finally we want to analyze the effect of the distance between vortex and
anti-vortex. In Fig.~\ref{fig:ovdist1}a) we plot the low-lying eigenvalues for a
spherical vortex--anti-vortex pair with varying distance in time direction. The near-zero mode is clearly shifted away from zero as the vortex and anti-vortex approach each
other. If they lie in neighboring time slices ({\it i.e.} t=3 and t=4) the
Dirac operator cannot resolve them as individual objects, {\it i.e.}
the eigenmodes overlap heavily and no near-zero mode is produced. This can
also be seen from the plane wave behavior in the chiral density of the lowest
eigenmode in Fig.~\ref{fig:ovdist1}b). The chiral densities of the near-zero modes for vortex--anti-vortex pairs with distances 2, 3, and 4 are shown in Fig.~\ref{fig:ovdist2}, they show no plane wave behavior and an increasing degree of local 
chirality. From the density plots we conclude that the eigenmode peaks at the spherical vortices extend over 3-4 time slices. Hence we expect the overlap of the modes to vanish if the vortex and anti-vortex are separated by 5-6 time slices and indeed we see that the eigenvalues do not change significantly if we increase the distance further. Thus, the spherical vortices reproduce all characteristic properties of low-lying modes for chiral symmetry breaking given in~\cite{Horvath:2002gk} and summarized at the end of section~\ref{sec:zms:inst}. The results clearly show that
we may draw the same conclusions for spherical vortices as for instantons concerning the creation of near-zero modes.

\subsection{Planar vortex configurations}\label{sec:plan}

\begin{figure}[h]
	\centering
		a)\includegraphics[width=.48\columnwidth]{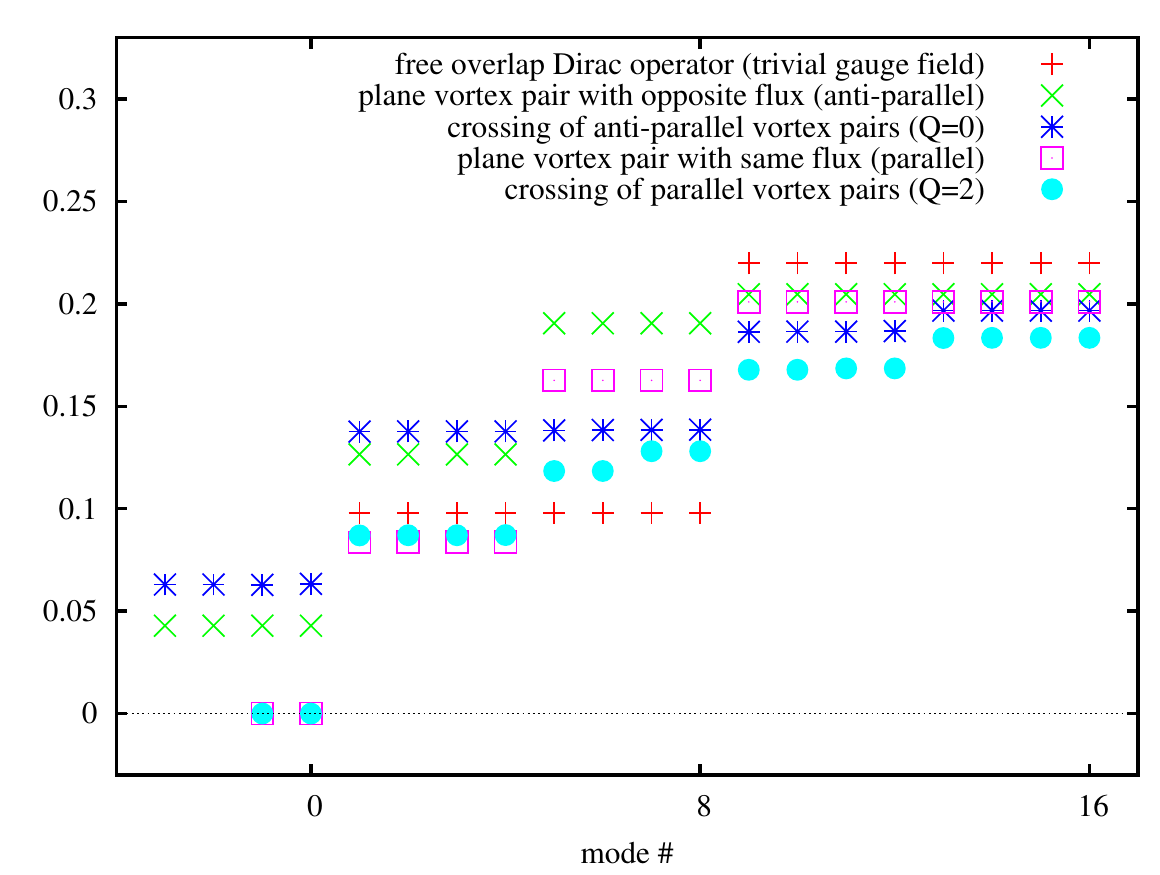}
		b)\includegraphics[width=.44\columnwidth]{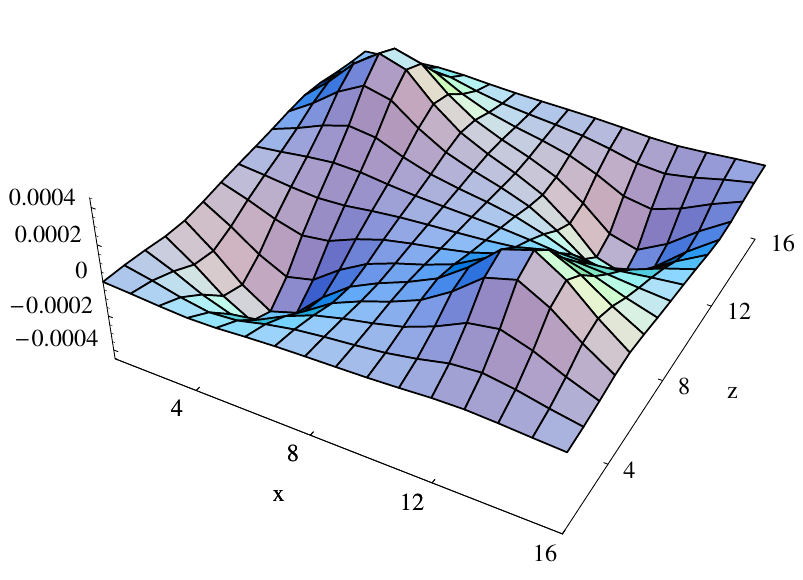}
	\caption{a) The lowest overlap eigenvalues for plane vortex
	configurations compared to the eigenvalues of the free (overlap)
	Dirac operator (red crosses). Note that the low-lying eigenvalues for single
	vortex pairs (green crosses and magenta boxes) only occur for appropriate
	boundary conditions (see text for details).
	b) Chiral density $\rho_5\#0$ in the
	intersection plane of all four near-zero modes of crossing flat vortex
	pairs with opposite flux direction ($Q=0$).}
	\label{fig:ovlart}
\end{figure}

\begin{figure}[h]
	\centering
		a)\includegraphics[width=.22\columnwidth]{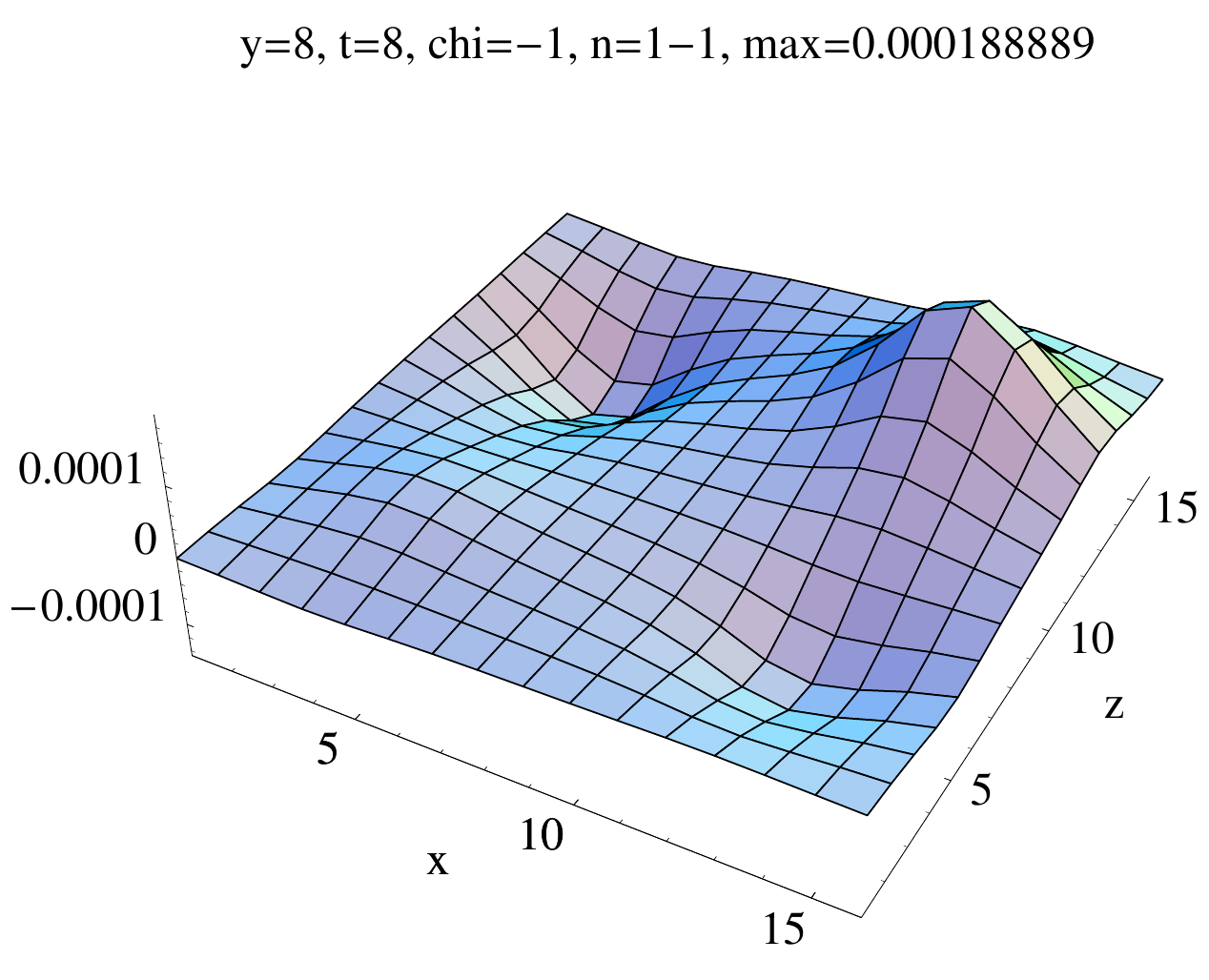}
		b)\includegraphics[width=.22\columnwidth]{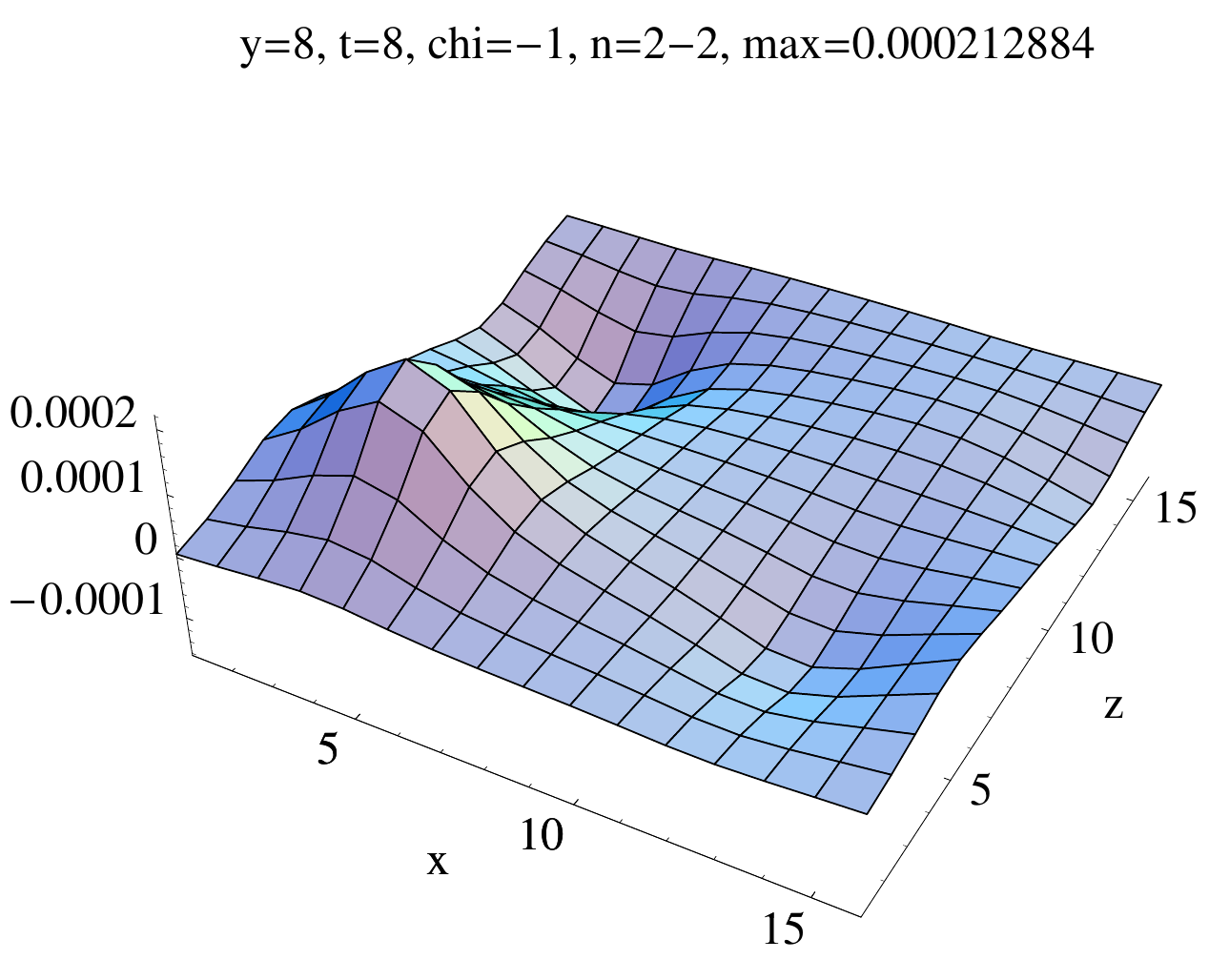}
		c)\includegraphics[width=.22\columnwidth]{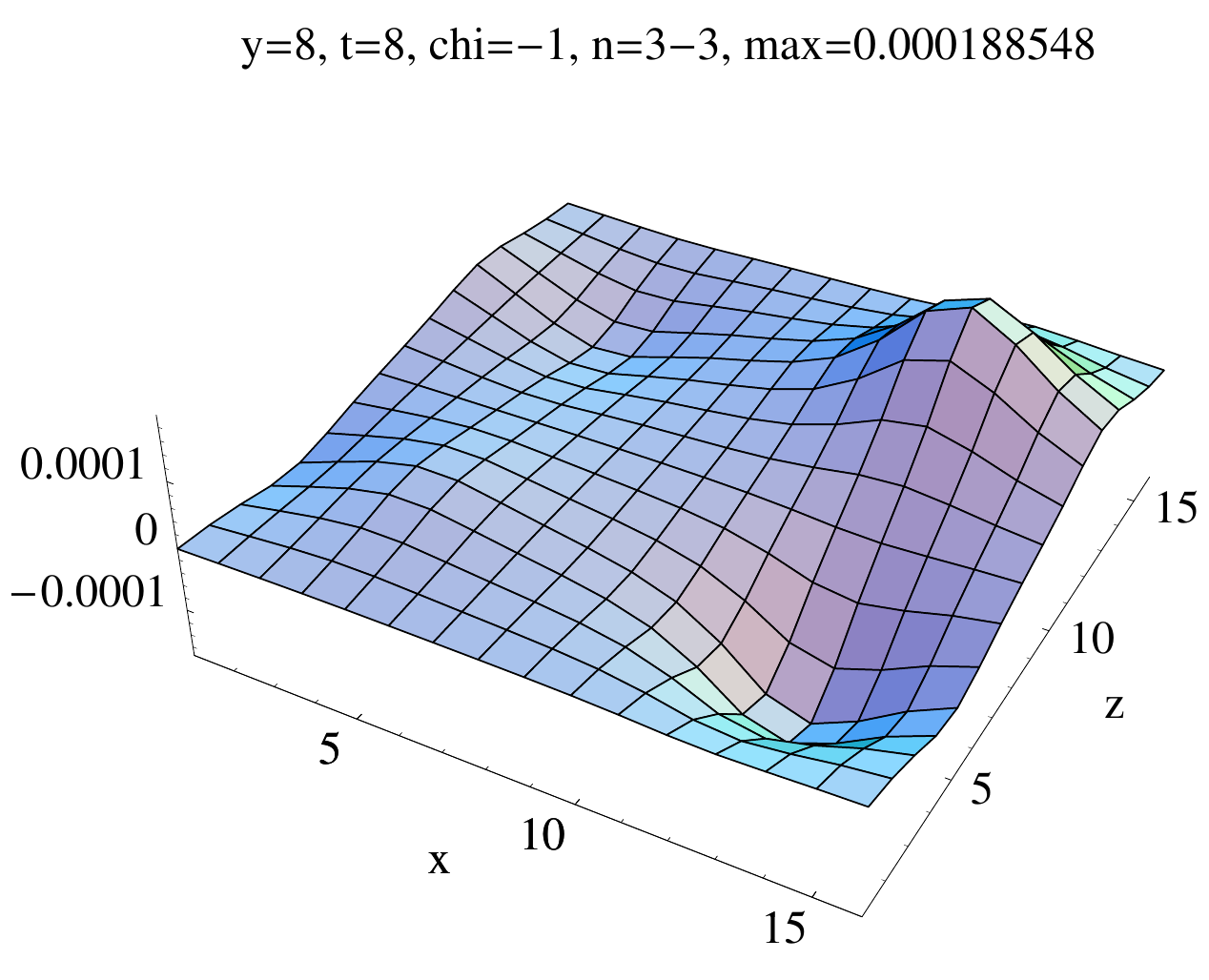}
		d)\includegraphics[width=.22\columnwidth]{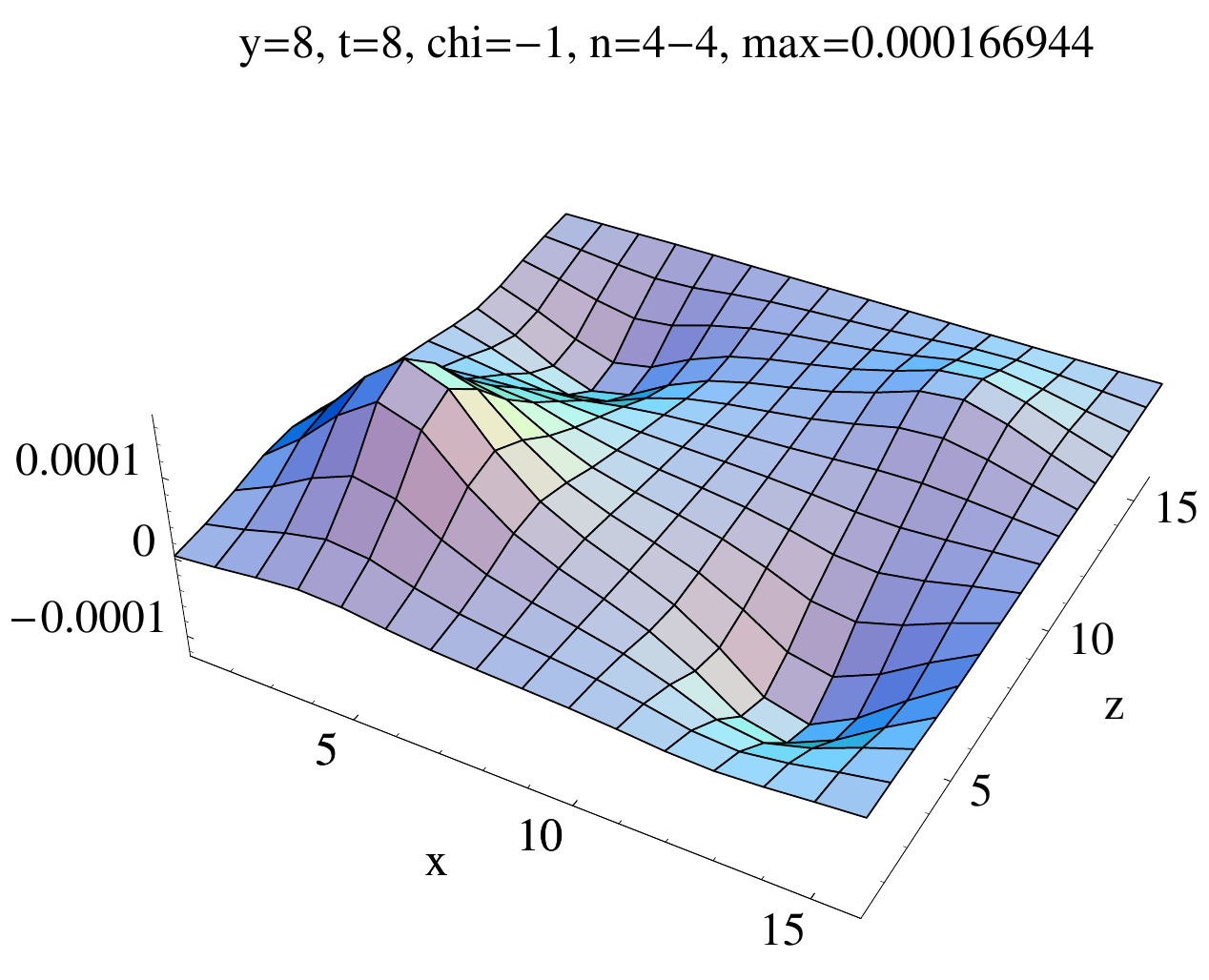}
	\caption{The individual modes of Fig.~\ref{fig:ovlart}b are mainly
localized at two neighboring intersection points with opposite topological charge
contributions: a) mode \#1 mainly peaks at the intersections in the back at $z=12$, b)
mode \#2 is mainly localized at the intersections to the left ($x=4$), c) mode \#3
at $x=12$ (right) and d) mode \#4 at $z=4$ (front).} 
	\label{fig:plq0s}
\end{figure}

For plane vortices the situation is more complicated, as we do not get single,
localized lumps of topological charge $Q=\pm1$, which would attract single zero
modes. We rather deal with vortex intersection points each contributing with
$Q=\pm1/2$ and only attracting zero modes as combinations of topological charge
contributions. Recall from Fig.~\ref{fig:3sph}a, that even for two spherical
vortices and one anti-vortex the corresponding zero mode cannot be matched to a
single vortex, but the zero mode belongs to both topological charge contributions.
As shown in Fig.~\ref{fig:vortcross3d}a and c one can get two different values
of topological charge $Q$ for two pairs of planar vortex surfaces, intersecting
in four points. For parallel flux direction we get two real zero modes, according to the total topological charge $Q=\pm2$. These modes we analyzed in~\cite{Hollwieser:2011uj}, they peak at least at two of the four topological charge contributions of $Q=1/2$. For pairs of planar vortex surfaces with opposite (anti-parallel) flux
directions we get $Q=0$ according to the topological charge contributions in
Fig.~\ref{fig:vortcross3d}c and four real near-zero modes. The sum of local
chiralities of these four modes peaks at the intersection points according to the
sign of the local topological charge, as shown in Fig.~\ref{fig:ovlart}b.
Fig.~\ref{fig:plq0s} shows that every near-zero mode is concentrated on two
intersections with opposite topological charge contribution. The number of
near-zero modes seems to be related to the four possible combinations to get
topological charge $Q=0$ from two of the four $Q=\pm1/2$ contributions. The
non-zero modes again show plane wave oscillations. 

\begin{figure}[h]
	\centering
		a)\includegraphics[width=.31\columnwidth]{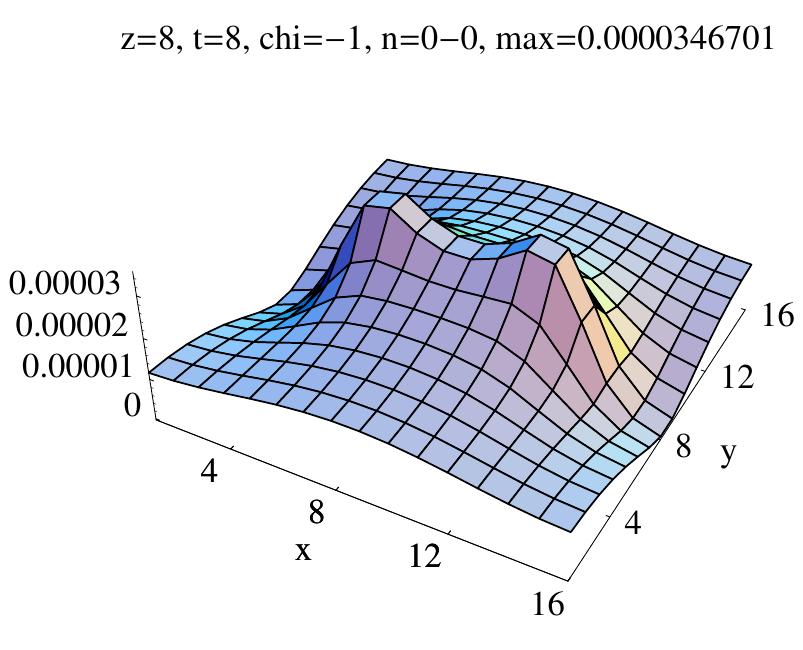}
		b)\includegraphics[width=.31\columnwidth]{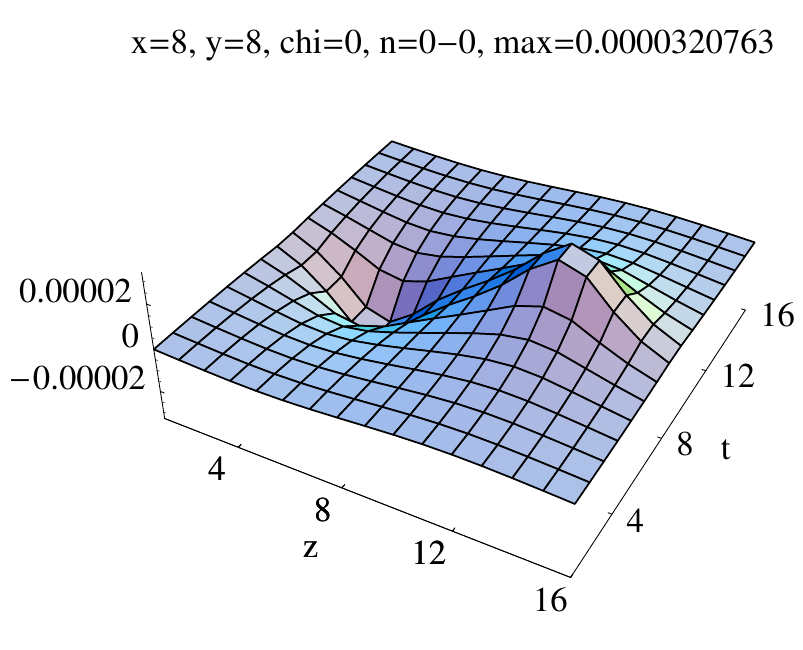}
		c)\includegraphics[width=.31\columnwidth]{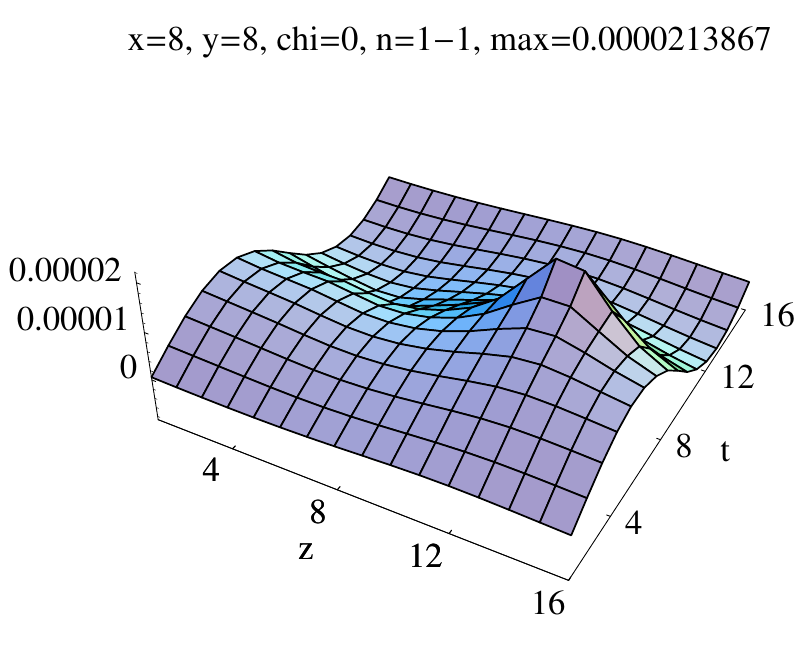}
	\caption{Chiral densities of the low-lying eigenmodes of the
	overlap Dirac operator for plane vortices: a) two zero modes of one parallel vortex pair, {\it i.e.}, two vortices with the same flux direction,
	b) four near-zero modes and c) first nonzero mode of one anti-parallel vortex pair,
	{\it i.e.}, two vortices with opposite flux direction. Note that the modes
in a) and b) are only present for periodic boundary conditions in directions parallel to the vortex flux (see text for details).}
	\label{fig:planes}
\end{figure}

The presented results are obtained for the usual anti-periodic boundary conditions in the time-direction and periodic boundary conditions in the spatial directions. However, we want to emphasize that the number of (near-) zero modes does not change with respect to boundary conditions, even if we impose anti-periodic boundary conditions in all directions. 
The reason it is important to ascertain this is that plane vortex pairs by
themselves are somewhat special, as they can attract zero modes already on their
own according to their magnetic flux, due to their essentially two-dimensional
nature (see Sec.~\ref{sec:zms:vort} and Reinhardt {\it et
al.}~\cite{Reinhardt:2002cm}). For periodic boundary conditions we get for a
single pair of planar vortex surfaces with the same (parallel) flux direction
two non-chiral zero modes. They have the same chirality peaking at the two
vortices, see Fig.~\ref{fig:planes}a and compare to Fig. 1 in
~\cite{Reinhardt:2002cm}. In fact, these modes are remnants of the trivial gauge
fields, where one gets four non-chiral zero modes for periodic boundary
conditions. The two "missing" zero modes, which would of course have opposite
chirality are suppressed by the vortex structure. For a single vortex pair with
opposite flux direction (anti-parallel vortices) we get four non-chiral near-zero modes
peaking at the two vortices with opposite (local) chiralities (two left-handed
and two right-handed), see Fig.~\ref{fig:planes}b. The non-zero modes
again show plane wave oscillations, see Fig.~\ref{fig:planes}c. In four
dimensions these non-chiral modes can be removed by anti-periodic boundary conditions in at least one of the directions parallel to the vortex flux, {\it i.e.}, for usual
anti-periodic boundary conditions in the time direction the near-zero mode results in this paragraph are only valid for "spatial" plane vortices, e.g. xy-vortices. For
zt-vortices there are no (near-) zero modes, as there are none for anti-periodic
boundary conditions in the z-direction and in the case of xy-vortices with
anti-periodic boundary conditions in the x or y-direction. Since the 
(near-) zero modes induced by single pairs of planar vortices can be removed by
appropriate boundary conditions, but they persist for two intersecting vortex
pairs regardless of boundary conditions, we conclude that indeed the intersections by
themselves can cause near-zero modes. 

Now, the mechanism of Sec.~\ref{sec:sphvort} or the analog instanton liquid
model does not directly apply to the case of planar vortices, since there are no
localized lumps of topological charge $Q=\pm 1$. Nevertheless the vortices
attract chiral (near-)zero modes via their intersections with topological charge
$Q=\pm1/2$, similar to the case of merons~\cite{Reinhardt:2001hb} and
calorons~\cite{Bruckmann:2009pa}. We conclude that the color structure of
vortices and their intersection points are able to create a finite density of near-zero modes and break chiral symmetry via the Banks-Casher relation.


\subsection{Asqtad Staggered Modes}\label{sec:asqtag}

\begin{figure}[h]
	\centering
		a)\includegraphics[width=.48\columnwidth]{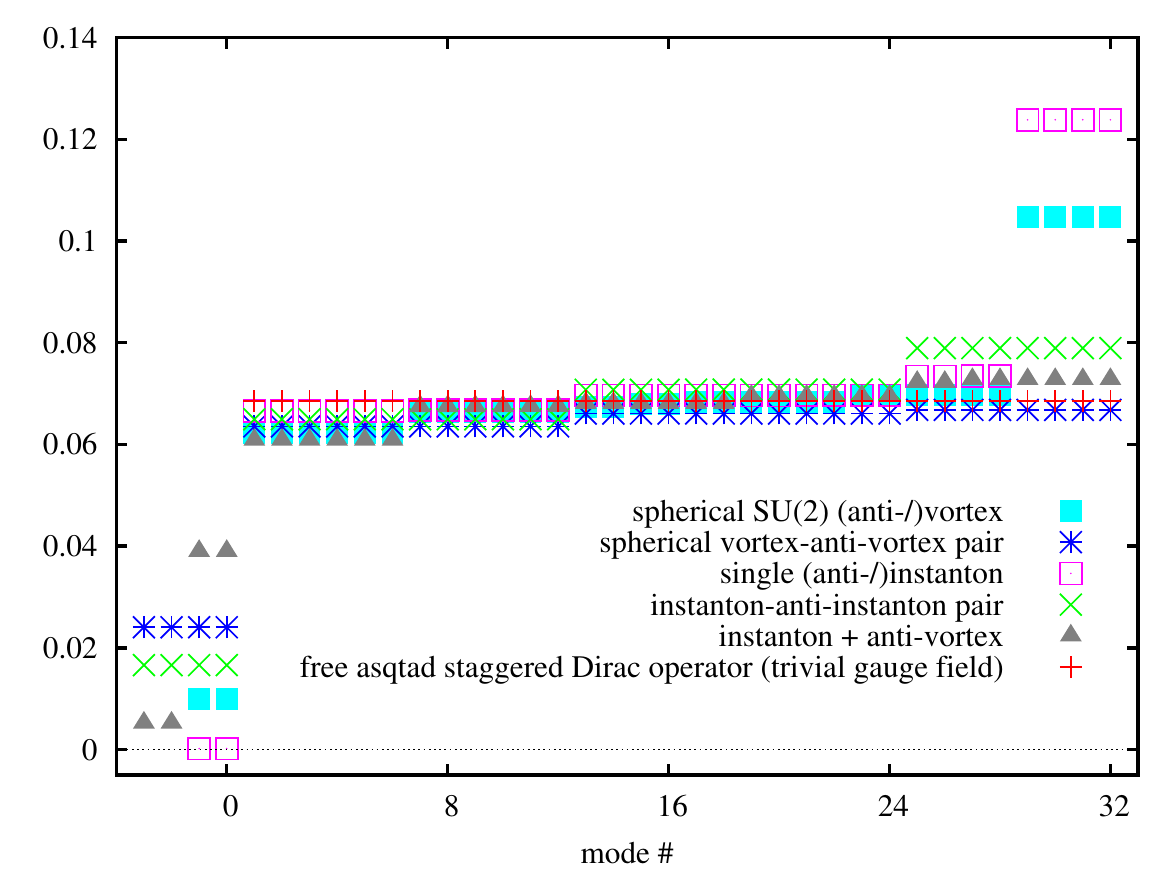}
		b)\includegraphics[width=.48\columnwidth]{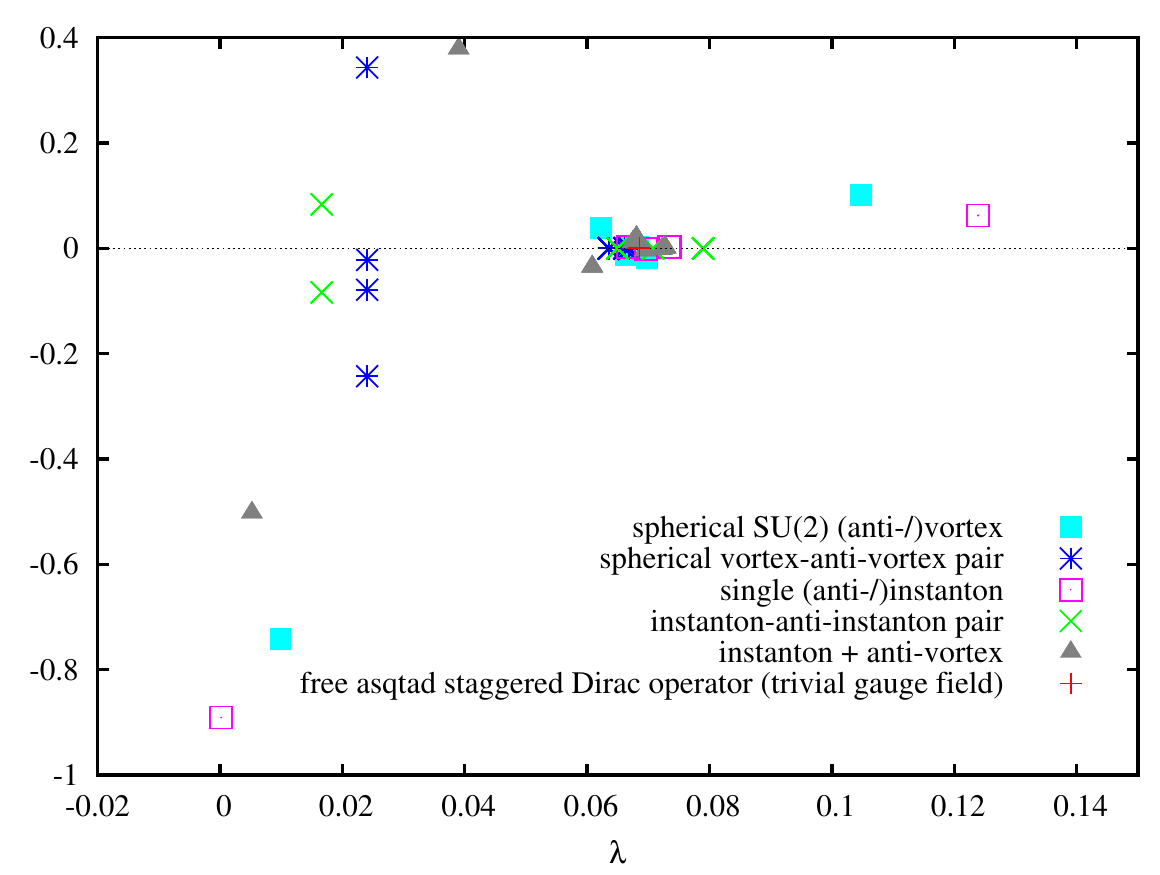}
	\caption{The lowest asqtad staggered eigenvalues for instanton and
	spherical vortex configurations compared to the eigenvalues of the
	free Dirac operator. b) Chirality of the corresponding eigenmodes.}
	\label{fig:aevsphr}
\end{figure}

\begin{figure}[h]
	\centering
		a)\includegraphics[width=.48\columnwidth]{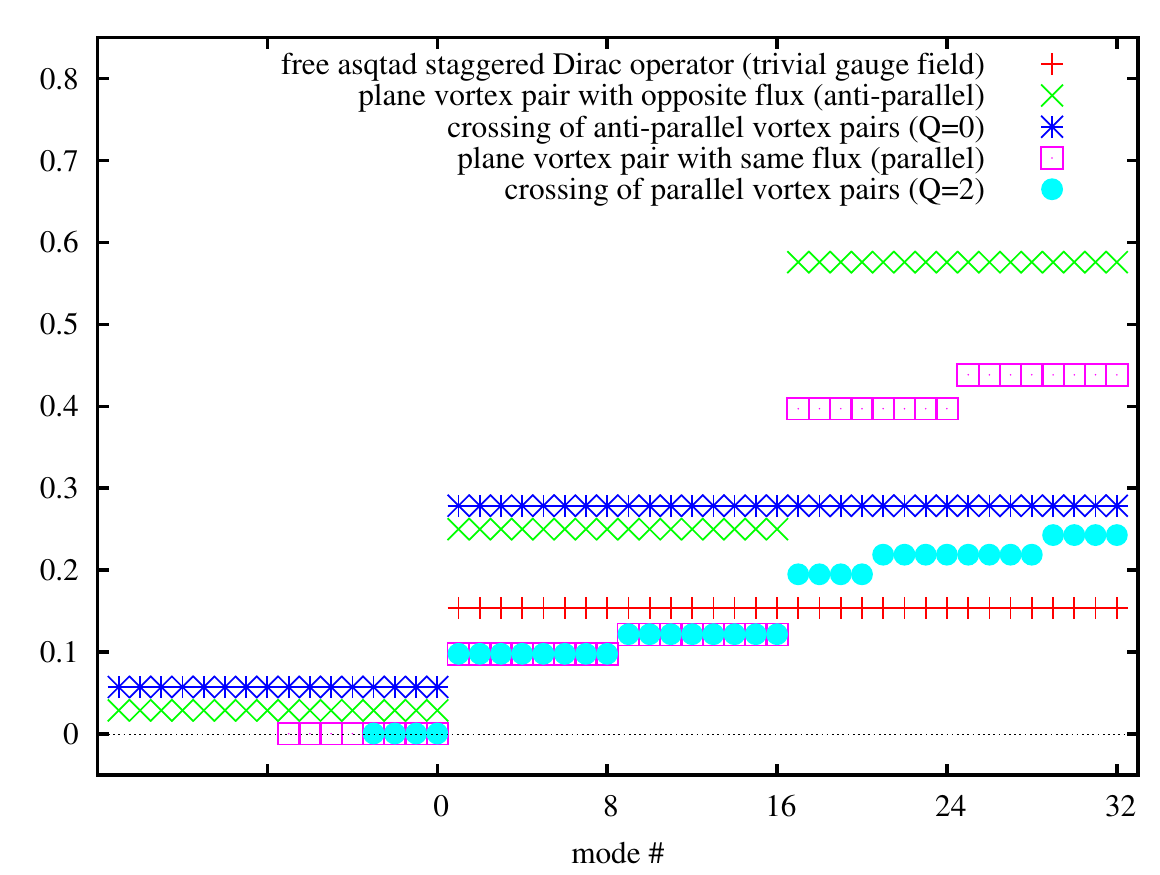}
		b)\includegraphics[width=.48\columnwidth]{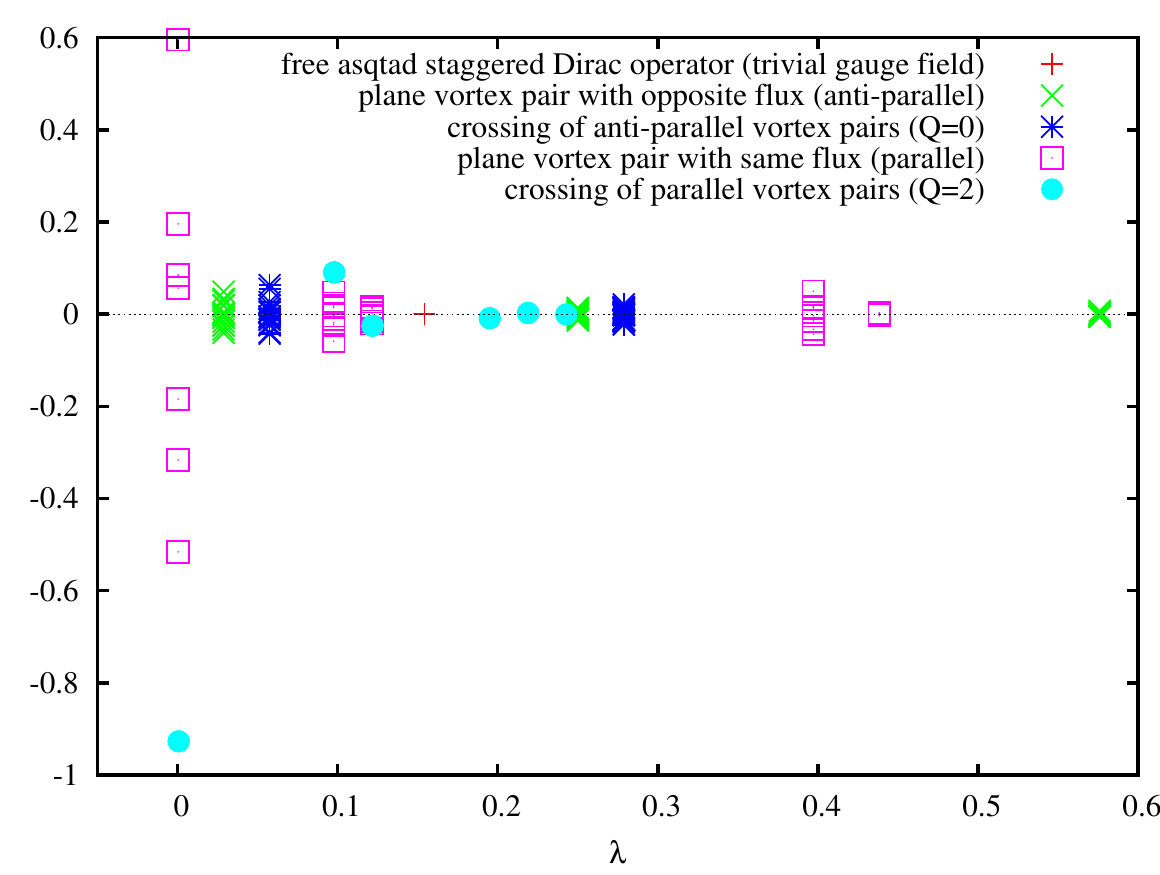}
	\caption{a) The lowest asqtad staggered eigenvalues for plane
	vortex configurations compared to the eigenvalues of the free Dirac
	operator. b) Chirality of the corresponding eigenmodes. Again, the low-lying
	eigenmodes for single vortex pairs (green crosses and magenta boxes) only occur for appropriate boundary conditions (see text for details).}
	\label{fig:aevsart}
\end{figure}

\begin{figure}[h]
	\centering
		a)\includegraphics[width=.48\columnwidth]{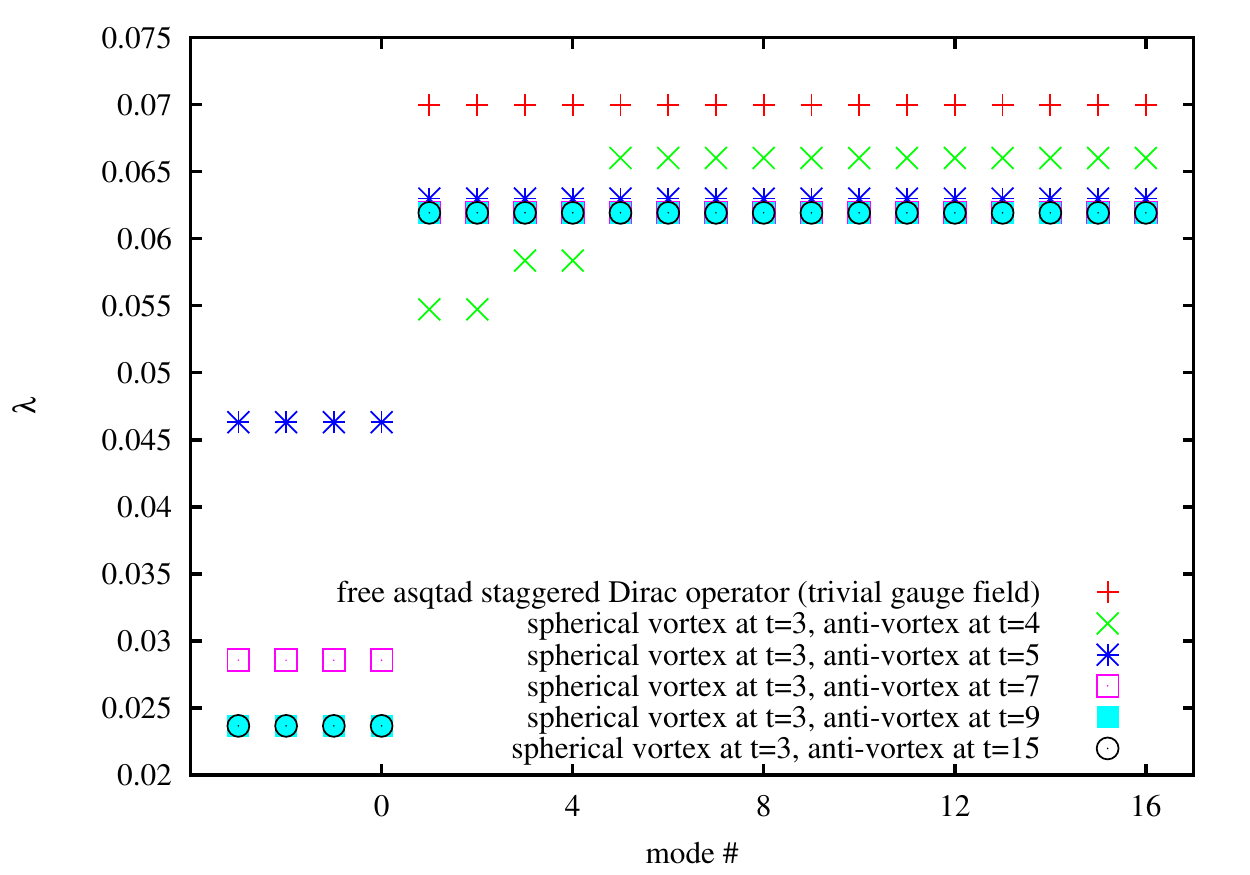}
		b)\includegraphics[width=.38\columnwidth]{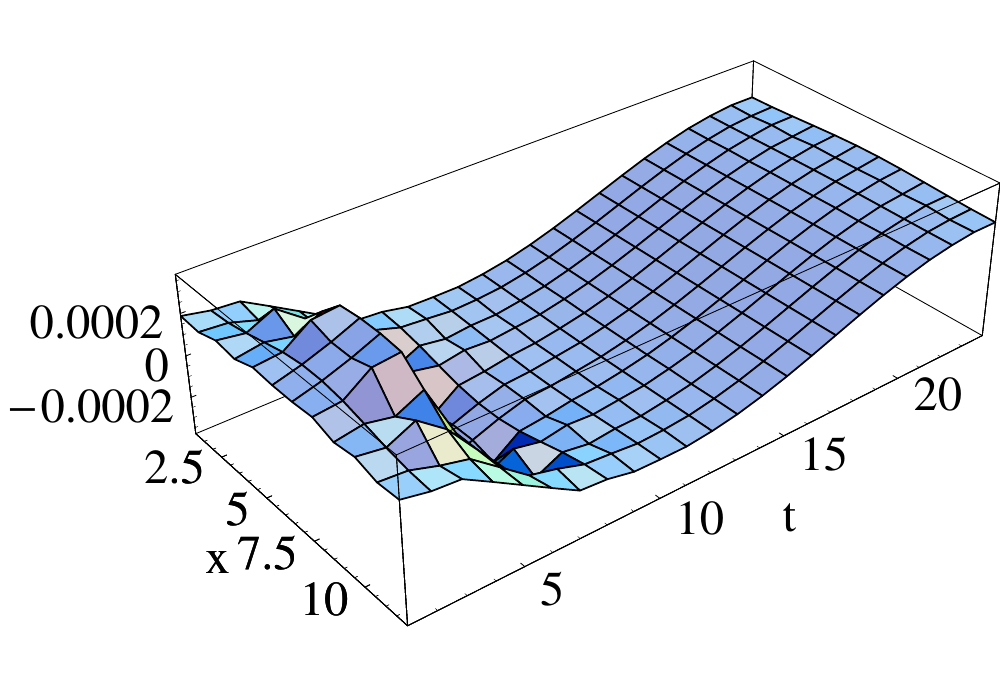}
		\caption{a) The lowest asqtad staggered eigenvalues for spherical vortex--anti-vortex pairs with varying distance compared to the eigenvalues of the free (overlap) Dirac operator. b) Chiral density of the lowest asqtad staggered eigenmode for
	spherical vortex and anti-vortex in neighboring time slices (t=3 and t=4).}
	\label{fig:avdist1}
\end{figure}

\begin{figure}[h]
	\centering
		\includegraphics[width=.32\columnwidth]{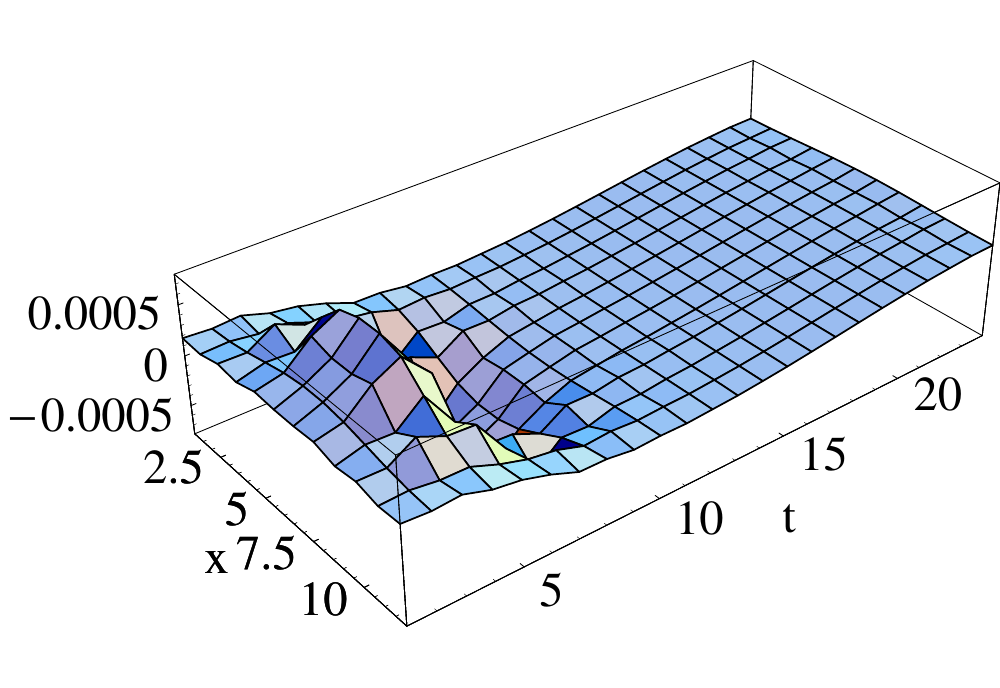}
		\includegraphics[width=.32\columnwidth]{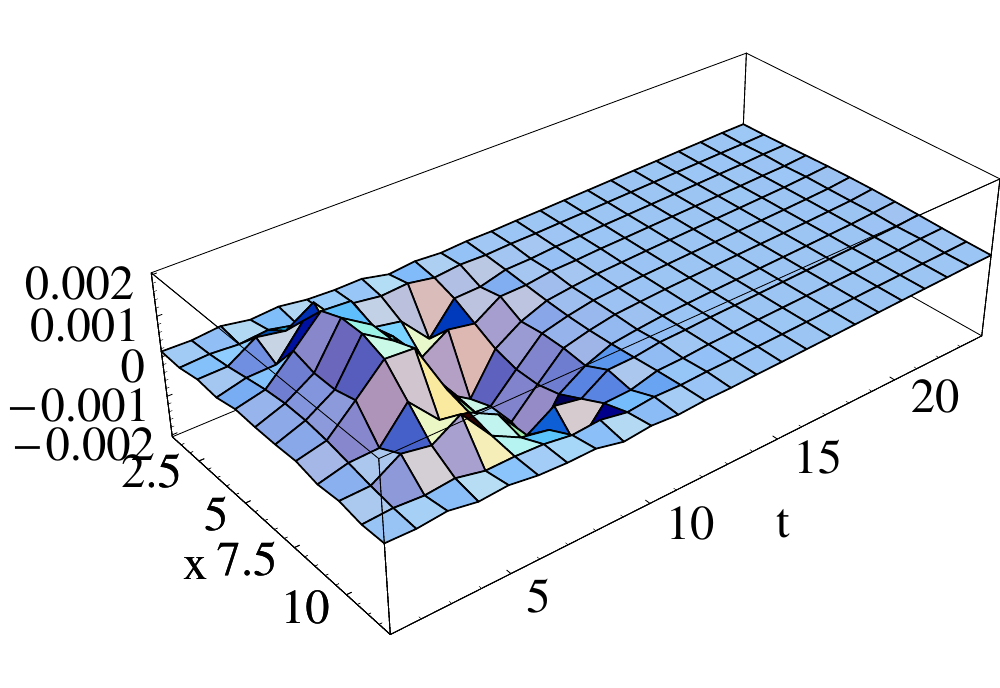}
		\includegraphics[width=.32\columnwidth]{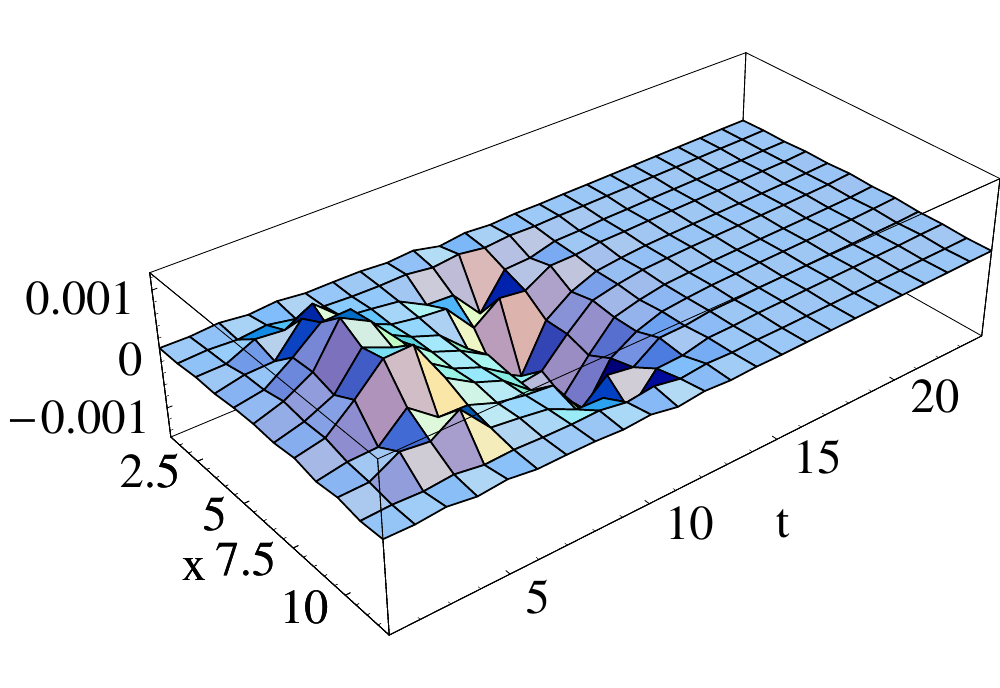}
	\caption{Chiral density of the asqtad staggered near-zero mode for a spherical vortex in time slice t=3 and an anti-vortex in a) t=5, b) t=7 and c) t=9.}
	\label{fig:avdist2}
\end{figure}

\begin{figure}[h]
	\centering
		a)\includegraphics[width=.31\columnwidth]{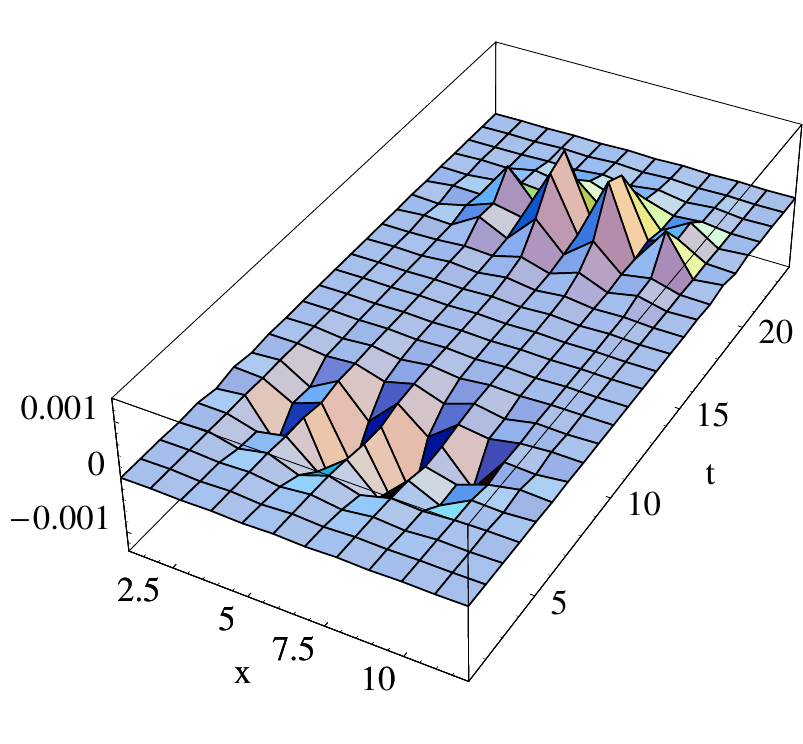}
		b)\includegraphics[width=.31\columnwidth]{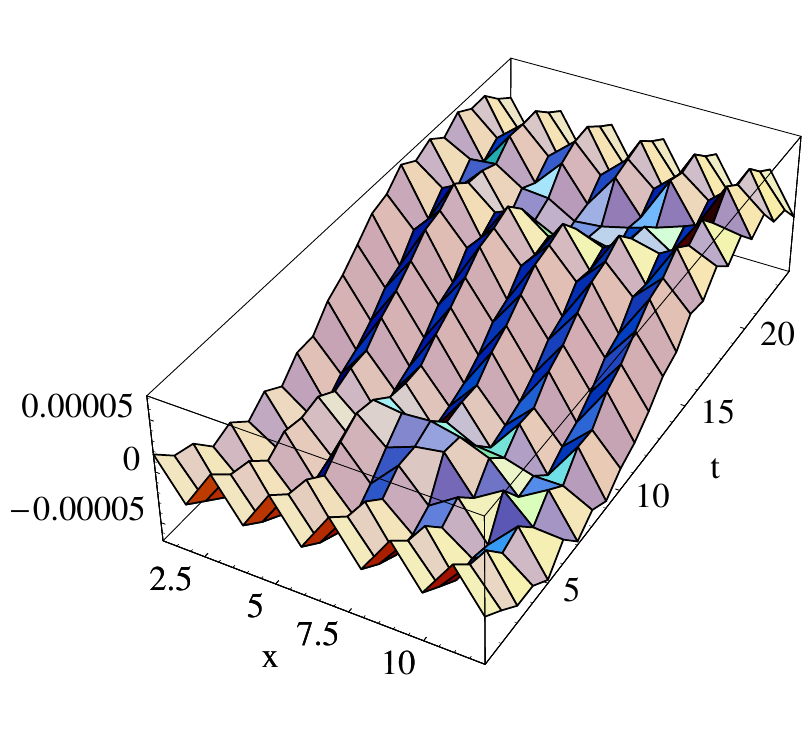}
		c)\includegraphics[width=.31\columnwidth]{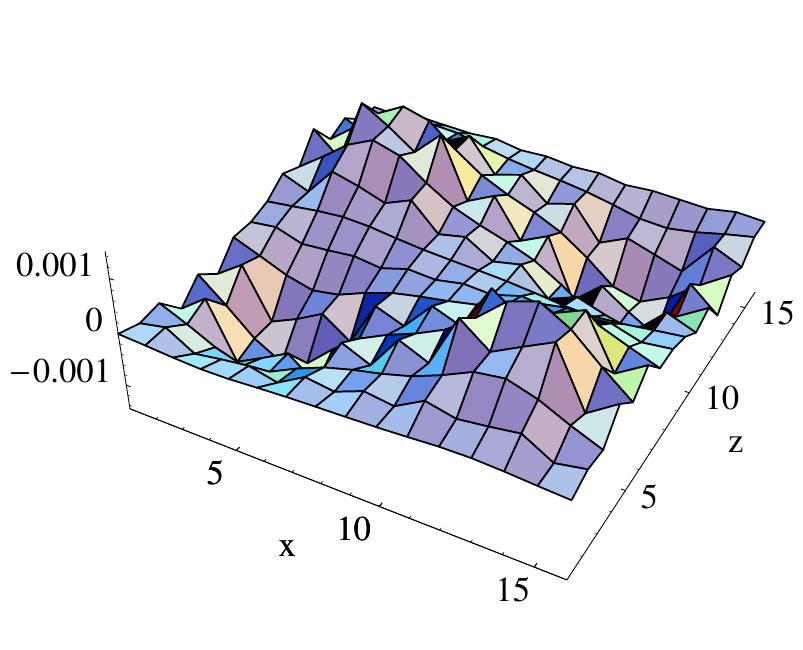}
	\caption{Chiral densities of the low-lying eigenmodes of the
	asqtad staggered Dirac operator: a) four near-zero and b) nonzero modes of a spherical vortex--anti-vortex pair, and c) 16 near-zero modes of the crossing flat vortex pairs in Fig.~\ref{fig:ovlart}b).}
	\label{fig:asqtad}
\end{figure}

For completeness we shortly discuss the asqtad staggered eigenmodes for the
presented configurations (see
Figs.~\ref{fig:aevsphr},~\ref{fig:aevsart},~\ref{fig:avdist1},~\ref{fig:avdist2},~\ref{fig:asqtad}). In principle the same conclusions as for overlap
modes apply if we consider the double degeneracy of asqtad staggered modes
due to charge conjugation. Hence we have two (would-be) zero modes for
$Q=\pm1$ and there are four times as many near-zero modes since two
would-be zero modes do result in two pairs of near-zero modes instead of one for
overlap modes. Remember that for a single pair of plane vortices with parallel
flux the overlap operator finds two non-chiral zero modes. It is interesting to
note, that the staggered operator identifies them as non-chiral near-zero modes as
indicated by their number and chiralities (magenta boxes in
Fig.~\ref{fig:aevsart}). The non-chiral modes for single vortex pairs can again
be removed by anti-periodic boundary conditions in directions parallel to the vortex flux. The chirality of the asqtad staggered eigenmodes is given by $\langle\psi\gamma_5\psi\rangle$, where $\gamma_5$ corresponds to a displacement along the diagonal of a hypercube. 
Staggered fermions do not have exact zero modes, but a separation between
would-be and non-zero modes is observed for improved staggered quark actions~\cite{Wong:2004ai}. The plots show that the would-be and near-zero modes have enhanced chirality compared to nonzero modes and we
even observe the local chiral density properties for the near-zero modes
which we discussed for the overlap modes.

\clearpage

\section{Conclusions}
The instanton liquid model provides a physical picture of chiral symmetry
breaking via the idea of quarks “hopping” between random instantons and
anti-instantons, changing their helicity each time. This process can be
described by quarks propagating between quark-instanton vertices. As
fermions do not seem to make much of a difference between instantons and
spherical vortices this picture can be extended to colorful spherical
center vortices. In fact, the spherical vortices reproduce all characteristic
properties of low-lying modes for chiral symmetry breaking~\cite{Horvath:2002gk}:
\begin{itemize}
\item Their probability density is clearly peaked at the location of the
	vortices. 
\item The local chirality at the peaks exactly matches the sign of the
	topological charge and the size of the chiral lump is correlated to the extension of the topological structure.
\item As a spherical vortex and an anti-vortex approach each other, the
	eigenvalues are shifted further away from zero and the localization and local chirality properties fade.
\end{itemize}

In the vortex picture the model of chiral symmetry breaking can be formulated even more generally, as we have shown that various shapes of vortices attract (would-be) zero modes which contribute via interactions to a finite density of near-zero modes with local chiral
properties, {\it i.e.}, local chirality peaks at corresponding topological
charge contributions. The simple picture of localized would-be zero modes
from the instanton liquid model or spherical vortex configurations as
discussed in Sec.~\ref{sec:inter} and \ref{sec:sphvort} does not apply
directly to general vortex structures as there are not only topological
charge contributions of $Q=\pm1$. In Monte Carlo configurations we do not,
of course, find perfectly flat or spherical vortices, as one does not find
perfect instantons. The general picture of topological charge from vortex
intersections, writhing points and even color structure contributions or
instantons can provide a general picture of $\chi$SB: any source of
topological charge can attract (would-be) zero modes and produce a finite
density of near-zero modes leading to chiral symmetry breaking via the
Banks-Casher relation. Here one also has to ask what could be the dynamical
explanation of $\chi$SB. We can try the conjecture that only a combination
of color electric and magnetic fields leads to $\chi$SB, electric fields
accelerating color charges and magnetic fields trying permanently to
reverse the momentum directions on spiral shaped paths. Such reversals of
momentum keeping the spin of the particles should especially happen for
very slowly moving color charges. Alternatively we could argue that
magnetic color charges are able to flip the spin of slow quarks, {\it i.e.}
when they interact long enough with the vortex structures.

Finally, it seems that vortices not only confine quarks into bound states 
but also change their helicity in analogy to the instanton liquid model. 
We therefore emphasize that the center vortex model of quark confinement is indeed capable of describing chiral symmetry breaking.  
While we can not give a conclusive answer to the question of a dynamical explanation for the mechanism of $\chi$SB, we can speculate that the generation of near-zero
modes demonstrated for artificial configuration in this paper, carries over
to vortices present in Monte Carlo generated configurations. As the
near-zero modes are located around intersection and writhing points of vortices that
carry topological charge, the behavior away from these points
would seem to be far less important.

We conclude by remarking that other mechanisms of chiral symmetry breaking,
in addition to the instanton liquid paradigm or the vortex picture
described in this paper, may be operative in the Yang-Mills vacuum. For
instance, it also seems possible that, even in the absence of would-be zero
modes, the random interactions of quarks with the vortex background may be
strong enough to smear the free dispersion relation such that a finite
Dirac operator spectral density at zero virtuality is generated.  In fact,
a confining interaction by itself generates chiral symmetry breaking,
independent of any particular consideration of would-be zero modes
connected to topological charge~\cite{Casher:1979vw}. However, this effect on its own is not sufficiently strong for a quantitative explanation of the chiral
condensate; other effects, among them possibly the ones considered in this
article, must play a role. Also, the importance of the
long-range nature of low-dimensional topological structures for the
understanding of the mechanism of $\chi$SB in QCD was underlined by various
results of different groups~\cite{Horvath:2002zy,Horvath:2003is,Bowman:2010zr,Braguta:2010ej,OMalley:2011aa,Buividovich:2011mj,O'Malley:2011zz,Braguta:2013kpa}, 
and agrees well with a vortex picture of $\chi$SB.

\acknowledgments{We thank {\v S}tefan Olejn\'{\i}k and Michael Engelhardt for helpful discussions. This research was supported by the Austrian Science Fund FWF (``Fonds zur F\"orderung der wissenschaftlichen Forschung'') under Contract No. P22270-N16 (R.H.) and by an Erwin Schr\"odinger Fellowship under Contract No. J3425-N27 (R.H.).}

\bibliographystyle{utphys}
\bibliography{literatur}

\end{document}